\documentclass[iop]{emulateapj}

\usepackage{url}
\usepackage{graphicx}

\def\aj{AJ}%
\def\araa{ARA\&A}%
\def\apj{ApJ}%
\def\apjl{ApJ}%
\def\apjs{ApJS}%
\def\aap{A\&A}%
\def\aaps{A\&AS}%
\def\mnras{MNRAS}%
\def\pasp{PASP}%
\def\nat{Nature}%
\def\physrep{Phys.~Rep.}%
\def\s4g{S$^4$G}

\def\m20{$\rm M_{20}$}
\def\gm{$\rm G_{M}$}
\def\mum{$\rm \mu$m}

\def\Aonetwo{$\langle A_1 \rangle_i$}
\def\Atwotwo{$\langle A_2 \rangle_i$}
\def\Aonethree{$\langle A_1 \rangle_o$}
\def\Atwothree{$\langle A_2 \rangle_o$}

\def\nsamp{2349}

\slugcomment{}
\shorttitle{Morphological Parameters of S$^4$G Galaxies}
\shortauthors{Holwerda et al.}

\begin{document}

\title{Morphological Parameters of Spitzer Survey of Stellar Structure in Galaxies}

\author{B.W. Holwerda\altaffilmark{1}
J--C. {Mu{\~n}oz-Mateos}\altaffilmark{2}, S. {Comer\'{o}n}\altaffilmark{3}, S. Meidt\altaffilmark{4}, K. Sheth\altaffilmark{2,5,6}, 
\and
S. Laine\altaffilmark{6}, J.~ L. Hinz\altaffilmark{7}, M.~ W. Regan\altaffilmark{8}, A. Gil de Paz\altaffilmark{9},  
\and
K. Men\'{e}ndez--Delmestre\altaffilmark{10}, M. Seibert\altaffilmark{10}, T. Kim\altaffilmark{2,11}, T. Mizusawa\altaffilmark{4}, 
\and
E. Laurikainen\altaffilmark{3,12}, H. Salo\altaffilmark{3}, J. Laine\altaffilmark{3}, D. A. Gadotti\altaffilmark{13}, D. Zaritsky\altaffilmark{7},
\and
 S. Erroz-Ferrer\altaffilmark{14,15}, L. C. Ho\altaffilmark{10}, J. H. Knapen\altaffilmark{14,15}, E. Athanassoula\altaffilmark{16}, A. Bosma\altaffilmark{16}, and N. Pirzkal\altaffilmark{8} }

\altaffiltext{1}{European Space Agency Research Fellow (ESTEC), Keplerlaan 1, 2200 AG Noordwijk, The Netherlands, email: \url{benne.holwerda@esa.int} or \url{benne.holwerda@gmail.com}, twitter: @benneholwerda}
\altaffiltext{2}{National Radio Astronomfy Observatory/NAASC, 520 Edgemont Road, Charlottesville, VA 22903, USA}
\altaffiltext{3}{Astronomy Division, Department of Physical Sciences, FIN-90014 University of Oulu, P.O. Box 3000, Oulu, Finland}
\altaffiltext{4}{Max-Planck-Institut f\"{u}r Astronomie / K\"{o}nigst\"{u}hl 17 D-69117, Heidelberg, Germany}
\altaffiltext{5}{Spitzer Science Center, 1200 East California Boulevard, Pasadena, CA 91125, USA}
\altaffiltext{6}{Spitzer Science Center, Mail Stop 220-6, 1200 East California Boulevard, Pasadena, CA 91125, USA}
\altaffiltext{7}{Steward Observatory, University of Arizona, 933 North Cherry Avenue, Tucson, AZ 85721, USA}
\altaffiltext{8}{Space Telescope Science Institute, 3700 San Martin Drive, Baltimore, MD 21218, USA}
\altaffiltext{9}{Departamento de Astrof\'{i}sica, Universidad Complutense de Madrid, 28040 Madrid, Spain}
\altaffiltext{10}{The Observatories of the Carnegie Institution of Washington, 813 Santa Barbara Street, Pasadena, CA 91101, USA}
\altaffiltext{11}{Astronomy Program, Department of Physics and Astronomy, Seoul National University, Seoul 151-742, Republic of Korea}
\altaffiltext{12}{Finnish Centre of Astronomy with ESO (FINCA), University of Turku, Vislntie 20, FI-21500, Piikki, Finland}
\altaffiltext{13}{European Southern Observatory, Casilla 19001, Santiago 19, Chile}
\altaffiltext{14}{Instituto de Astrof\'{i}sica de Canarias, V\'{i}a L\'{a}ctea s/n 38205 La Laguna, Spain}
\altaffiltext{15}{Departamento de Astrof\'{i}sica, Universidad de La Laguna, Avda. Astrof\'{i}sico Francisco S\'{a}nchez s/n 38206 La Laguna, Spain}
\altaffiltext{16}{Aix Marseille Universit\'e, CNRS, LAM (Laboratoire d'Astrophysique de Marseille) UMR 7326, 13388, Marseille, France}

\begin{abstract}
The morphology of galaxies can be quantified to some degree using a set of scale-invariant parameters. 
Concentration (C), Asymmetry (A), Smoothness (S), the Gini index (G), relative contribution of the brightest pixels to the 
second order moment of the flux (\m20), ellipticity (E), and the Gini index of the second order moment ($G_M$) have all been 
applied to morphologically classify galaxies at various wavelengths. Here we present a catalog of these 
parameters for the {\em Spitzer} Survey of Stellar Structure in Galaxies (\s4g), a volume-limited near-infrared 
imaging survey of nearby galaxies using the 3.6 and 4.5 \mum\ channels of the IRAC camera onboard 
{\em Spitzer Space Telescope}. Our goal is to provide a reference catalog of near-infrared quantified morphology
for high-redshift studies and galaxy evolution models with enough detail to resolve stellar mass morphology.
We explore where normal, non-interacting galaxies --those typically found on the Hubble tuning fork-- lie in this parameter 
space and show that there is a tight relation between Concentration ($C_{82}$) and \m20\ for normal galaxies.  
\m20~ can be used to classify galaxies into earlier and later types (e.g., to separate spirals from irregulars).
Several criteria using these parameters exist to select systems with a disturbed morphology, i.e., those that 
appear to be undergoing a tidal interaction. We examine the applicability of these criteria to {\em Spitzer} 
near-infrared imaging. We find that four relations, based on the parameters A and S, G and \m20, \gm, 
and C and \m20, respectively, select outliers in the 
morphological parameter space, but each selects different subsets of galaxies. Two criteria 
($G_M > 0.6$, $G > -0.115 \times M_{20} + 0.384$) seem most appropriate to identify possible mergers and the 
merger fraction in near-infrared surveys.
We find no strong relation between lopsidedness and most of these morphological parameters, except for 
a weak dependence of lopsidedness on Concentration and $M_{20}$. 

\end{abstract}


\section{\label{s:intro}Introduction}

The formation and evolution of galaxies leave an imprint on their light profile and morphology. 
Classification of their appearance has mostly been done by eye \citep[see][for a review]{Sandage05} 
but in recent years there has been a concerted effort to move the classification of visible light images 
of galaxies to a quantifiable basis. 
Classical morphological classification has long been done visually. Visual morphological classification
is impractical for the very large surveys now underway \citep[an interesting exception is the Galaxy Zoo 
project,][]{galaxyzoo}. Thus the goal is to find a quantifiable morphological classifier. 

An obvious classifier is the radial light profile of the galaxy, starting with the distinct profiles of ellipticals and 
spirals, later more generalized to the S\'ersic profile with the power of the profile ($n$) as the single 
identifier \citep{Sersic68}. However, these profiles ignore much of the small-scale detail in a galaxy image on 
which human classifiers rely for more subtle distinctions. 

Consequently, a range of scale-invariant parameters have been proposed, starting from various 
concentration indices \citep{Abraham00, CAS}, asymmetry \citep{Abraham00, Bershady00}, 
some indicator of smoothness of small-scale structure \citep{CAS}, and later including the Gini 
inequality parameter \citep{Abraham03}, the second order moment of the light distribution \citep{Lotz04}, 
ellipticity \citep{Scarlata07}, and \gm, the equality of the second order moment distribution \citep{Holwerda11c}.
The resulting parameter space is hardly mathematically orthogonal or complete but it has seen extensive as well as very  
specific use. There are clear advantages of simple parameterizations of galaxy morphology: no human 
biases, practical to implement on millions of objects, the possibility to directly and qualitatively compare 
across wavelength and redshift or to other characteristics. For example, at higher redshift, there are many 
galaxies that do not conform to the classical Hubble morphological classification but these can still be 
qualified using this system.
Based on a choice of parameter space and training sample, one can subsequently try to classify  
galaxies along the Hubble Tuning fork through a machine-learning algorithm \citep[e.g.,][]{Lahav96, 
Molinari98, Ball04, Scarlata07, Kormendy12}.

Disturbed morphology can be used to identify ongoing galaxy major mergers, and morphological
classification parameters have seen much use on galaxy samples observed at low and high redshift to 
infer the fraction and rate at which galaxies merge \citep{Conselice03b, Conselice05a, Conselice08b, 
Conselice09b, Yan05, Bundy05, Cassata05,Ravindranath06, Scarlata07, Trujillo07, Lotz08b, Jogee09, 
Darg09,  Lopez-Sanjuan09b, Lopez-Sanjuan09a}.
Concurrently, these parameters have shown promise in classifying objects along the Hubble 
tuning fork, both locally \citep{CAS, Lotz04, Taylor-Mager07, Munoz-Mateos09a}, and at high redshift 
\citep{Scarlata07, Huertas-Company09a}.
Meanwhile, efforts using visual inspection and classification by single observers or crowds have kept 
apace with quantified, automated classifiers \citep[e.g.,][Holwerda et al. {\em submitted}]{Darg09, 
Fortson12, Hoyle11, Keel13, Skibba09, Skibba11b, Skibba12, Masters10, Masters11, Masters12}. 

In this paper, we report our application of the popular morphological parameters to the data of the 
{\em Spitzer} Survey of Stellar Structure in Galaxies \citep[$\rm S^4G$,][ \protect\url{www.cv.nrao.edu/~ksheth/s4g}]{s4g}.

The IRAC camera \citep{IRAC} on board the {\em Spitzer Space Telescope} \citep{Spitzer} mostly 
maps the older stellar population at 3.6 and 4.5 $\mu$m and hence stellar mass in these systems 
\citep[][2013, {\it in preparation}]{Eskew12, Meidt12a}, and IRAC images are much less encumbered by dust extinction than 
any visible light images\footnote{see also \cite{Holwerda07a, Holwerda07b}}. 
Thus, the \s4g morphological parameterization reveals the  underlying stellar mass Hubble type 
rather than the apparent one, somewhat distorted by dust and star-formation \citep{Buta10}.  
The \s4g sample is one of the largest and uniformly selected and observed in the near-infrared, 
inviting the possibility of a study of the relations between near-infrared morphology parameters 
with each other and Hubble type, tidal disturbance, lopsidedness, etc. 

Strongly disturbed systems occupy a known sub-space of these morphological parameters. 
Those selected from this \s4g sample can be compared to the canonical Arp catalog of disturbed 
galaxies, to illustrate how well the morphological parameterization selects individual disturbed galaxies. 

The aim of \s4g Survey is to be volume-, mass- and luminosity-limited, using a representative sample of galaxies.
It now becomes possible to infer a local volume merger fraction and rate based on the morphological 
selection of disturbed systems. Our goals are to: 
(a) describe the \s4g morphological parameter catalog, 
(b) explore where the ``normal" galaxies lie in this quantified morphology parameter space and explore to what degree these can be morphologically typed based on these parameters, and
(c) examine those galaxies that are selected as ``disturbed'' from this catalog by the various morphological criteria.

Our goal for this paper is to present a uniformly computed catalog of the quantified morphology parameters 
for the \s4g\ survey for which codified morphological classifications from the Third Reference 
Catalogue \citep{RC3} exist. This quantified morphological catalog will subsequently serve as a reference for 
higher-redshift surveys where Hubble Types are unknown as well as results from detailed galaxy evolution modeling.

The paper is organized as follows: \S \ref{s:data} briefly describes the \s4g data-products, \S \ref{s:morph} 
the morphological parameters, and the application of these parameters to \s4g. \S \ref{s:result} presents the resulting 
catalogs. \S \ref{s:type} discusses the morphological parameters' relation to Hubble type, \S \ref{s:disturb} discusses
those systems that show clear signs of disturbance, \S \ref{s:lop} discusses the link with lopsidedness, and \S \ref{s:concl} lists our conclusions.

\section{$S^4G$ Data}
\label{s:data}

The Spitzer Survey of Stellar Structure in Galaxies \citep[\s4g][]{s4g} is a volume-, magnitude-, and size-limited ($D<40$
Mpc, $|b| > 30^\circ$, $m_{Bcorr} < 15.5$, $D_{25} >1'$) survey of \nsamp\ nearby galaxies in 3.6 $\mu$m 
and 4.5 $\mu$m (IRAC channels 1 and 2) of the Infrared Array Camera \citep[IRAC,][]{IRAC} of the 
{\em Spitzer Space Telescope} \citep{Spitzer}, using both archival cryogenic- and ongoing warm-mission 
observations \citep[for a full description and selection criteria, see][]{s4g}. All images have been reprocessed 
by the \s4g pipeline. The reprocessed pixel scale is 0\farcs75; the resolution is 1\farcs7 for 3.6 $\mu$m 
and 1\farcs6 for 4.5 $\mu$m. Data have been made public (\url{http://irsa.ipac.caltech.edu/data/SPITZER/S4G/}). 

For this paper we use the first and second pipeline products (P1 \& P2) of \s4g (Regan et al. {\em in preparation})
available from DR1 (January 2013) for \nsamp \ galaxies: the photometry images ({\em phot}) 
from P1 and foreground and background object masks from P2 for both the 3.6 and 4.5 $\mu$m images \cite[see for more details][]{s4g}. 
Our morphological parameters are in concert with the final \s4g\ data products (Mu\~{n}oz-Mateos et al., {\em in preparation}).

\begin{figure}
\begin{center}
\includegraphics[width=0.49\textwidth]{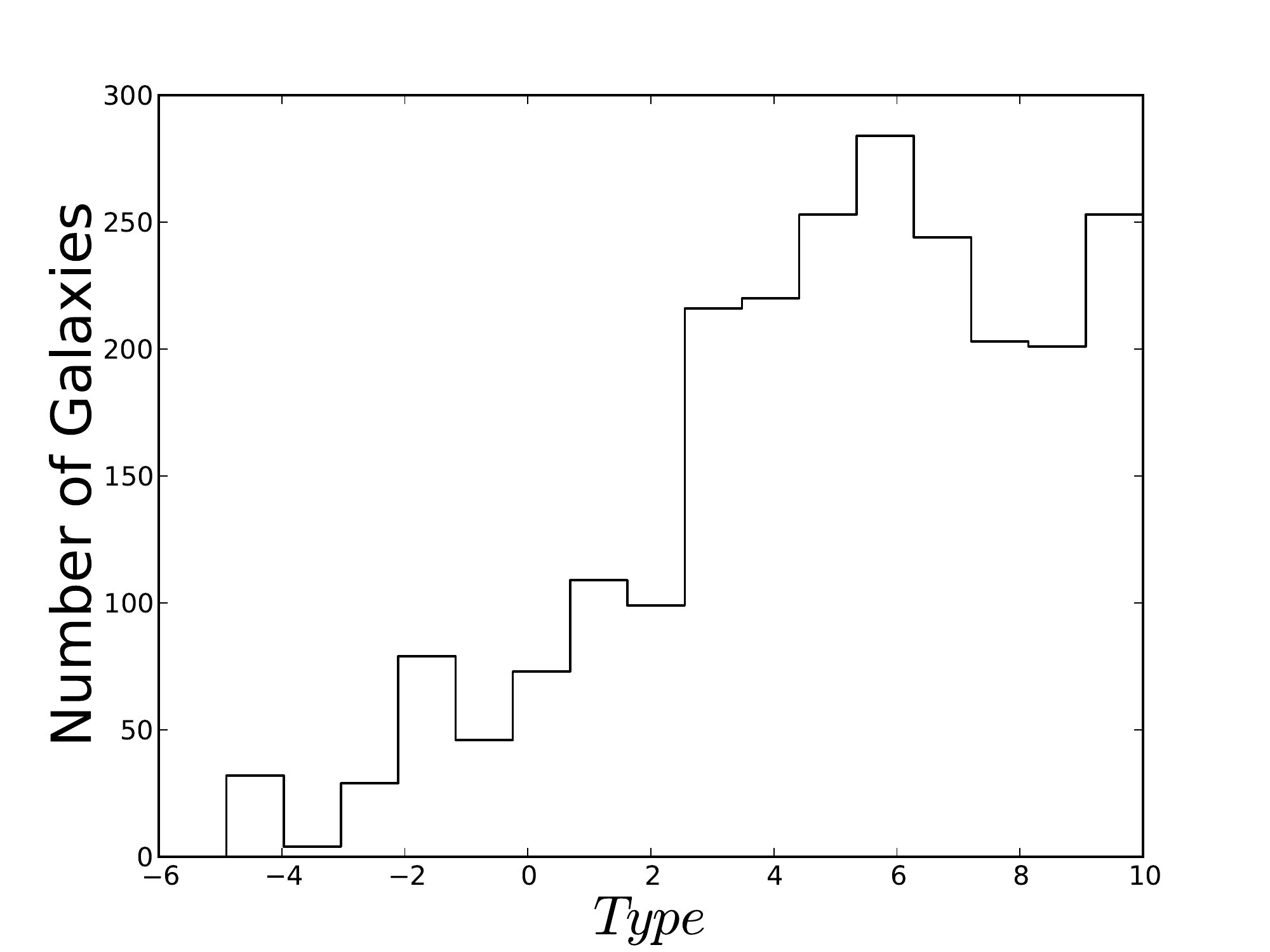}
\caption{The distribution of galaxy types in our sample of \nsamp \ galaxies from \s4g. Early-types ($t<0$) are 
underrepresented in \s4g. }
\label{f:typehist}
\end{center}
\end{figure}

\begin{figure}
\begin{center}
\includegraphics[width=0.49\textwidth]{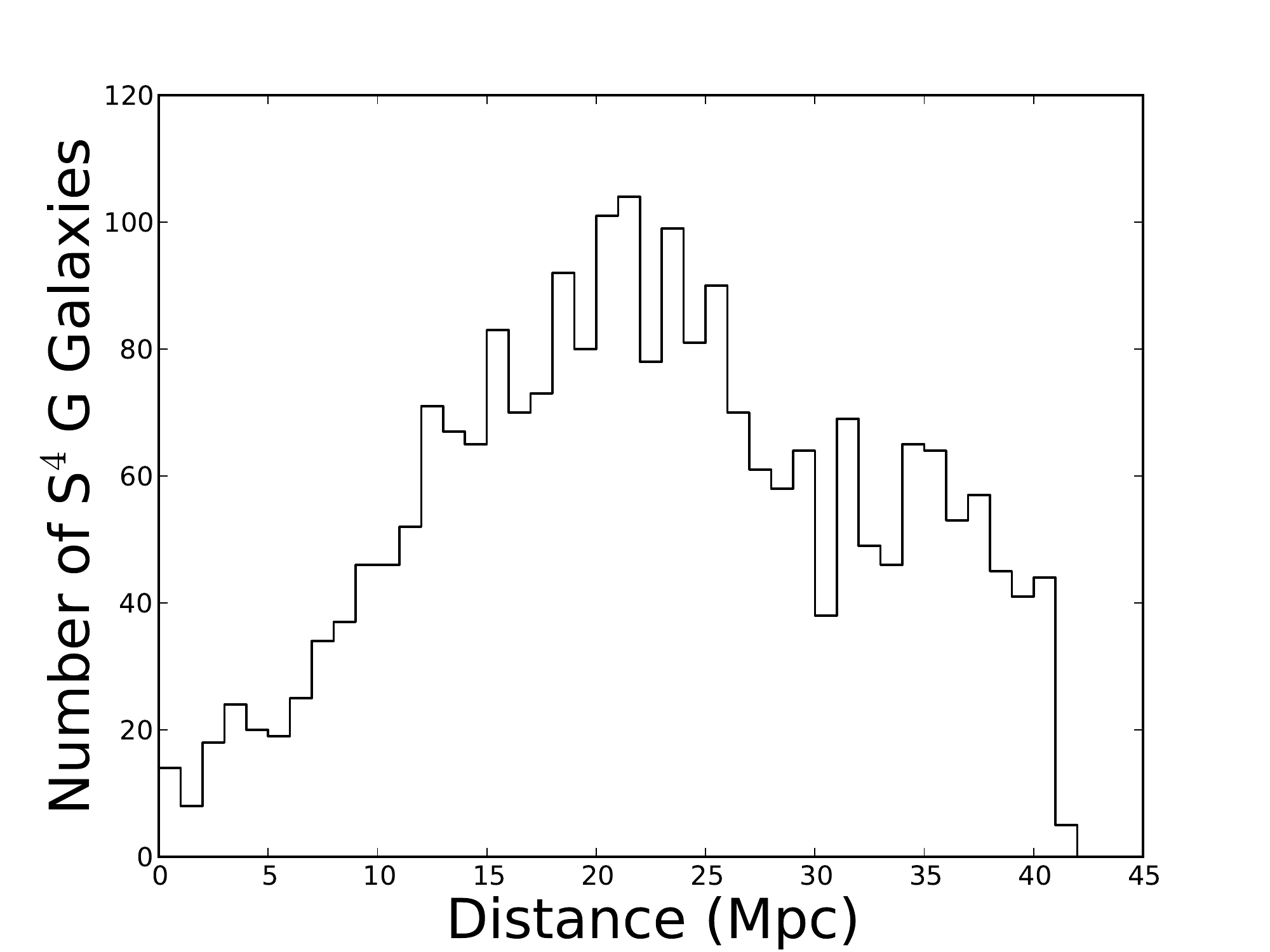}
\caption{The distribution of galaxy distance ($d=v_{rad}/h_0$) of our sample of \nsamp \ galaxies from \s4g. 
Radial velocities from NED where available which means some distances are negative. }
\label{f:disthist}
\end{center}
\end{figure}

\s4g is designed to be a volume-, luminosity- and especially mass-limited representative sample 
Since the initial selection required an H\,I radial velocity from HyperLEDA \citep{hyperleda}, 
early-types are relatively under-represented (Figure \ref{f:typehist}). In addition, early-types are under-represented because they are typically found in denser environments outside the local volume. 
The distribution of distances (based on radial velocities) is shown in Figure \ref{f:disthist}. The majority of our sample is between 10-30 Mpc.
The resolution of the Spitzer/IRAC ($\sim$2") translates to a physical resolution of less than a kpc over this distance range.

\s4g observations ideally trace the stellar mass of galaxies in both the 3.6 and 4.5 \mum\ channels \citep{Pahre04b,Pahre04a}, with the 3.6 \mum\ considered optimal to study the stellar mass \citep[][2013, {\it in preparation}]{Zibetti11,Meidt12a}. However, both have known contaminants such as the 3.3 \mum\ PAH feature in the 3.6 \mum\ channel (see the PAH heating models by \cite{Bakes01b}) and contamination from stochastically heated small dust grains in both \citep{Lu03, Flagey06, Mentuch09, Mentuch10}. We refer the reader to \cite{Meidt12a} and Meidt et al. {\em in preparation} for a comprehensive discussion of the non-stellar and anomalous mass-to-light stellar population contaminating the 3.6 \mum\ channel as a map of stellar mass.


\s4g has seen use on a variety of galaxy phenomena; disk truncation \citep{Comeron12}, bar fraction (Sheth et al. {\em in preparation}), thick disks 
\citep{Comeron11a,Comeron11b,Comeron11c}, visual and automated morphological classification \citep[][Laine et al. {\em in preparation}]{Buta10} and Hinz et al. {\em in preparation}, stellar mass mapping \citep{Meidt12a}, the role of AGB stars in galaxy appearance \citep{Meidt12b}, disk lopsidedness \citep{Zaritsky13},  and spiral structure \citep{Elmegreen11h}, star-formation hidden in spiral arms (Elmegreen et al. {\em submitted}), H$\alpha$ kinematics and the stellar disk \citep{Erroz-Ferrer12}, and Early-type Galaxies with Tidal Debris \citep{Kim12d}.

\section{Morphology Parameters}
\label{s:morph}

In this paper, we use the CAS system (Concentration-Asymmetry-Smoothness) from 
\citep{Bershady00,Conselice00a}, and \cite{CAS}, the Gini and $M_{20}$ system from \cite{Lotz04}, 
and a hybrid parameter $G_M$, the Gini parameter of the second order moment \citep[][]{Holwerda11c}.\\

Concentration is defined as \citep{Kent85}:
\begin{equation}
C_{82} = 5 ~ log \left( {r_{80} \over  r_{20} } \right),
\label{eq:c}
\end{equation}
\noindent where $r_{\%}$ is the radius of the circular aperture which includes that percentage of the 
total light of the object. For example, the SDSS generally uses $\rm C_{42}$, the ratio of the 40\% over 20\% 
radii and \cite{Scarlata07} and \cite{Munoz-Mateos09a} use the 80\% over 20\% ratio ($\rm C_{82}$). 
We opt to use the $\rm C_{82}$ \ definition here\footnote{We must note that the earlier version of our code \citep{Holwerda11a,Holwerda11b,Holwerda11c,Holwerda11d,Holwerda11e}
contained a small error, artificially inflating the concentration values. A quick check revealed this to be 
$C_{new} = 0.38 \times C_{old}$, and the new, correct values are adopted in this paper, \cite{Holwerda12c}, 
Holwerda et al. {\em submitted}, and  Parekh et al. {\em in preparation}, to bring our results in line with the common definition.}. 
This concentration index can be used to quickly discern between light profiles; a de Vaucouleurs 
profile ($I \propto R^{-4}$) has Concentration value of $C_{82}=5.2$, and a purely exponential one 
has a value of $C_{82}=2.7$. It also can be used to identify unique phenomena, for for example H\,I disk 
stripping \citep{Holwerda11e}. In the case of disk galaxies observed in the near-infrared, one can expect 
this parameter to rise in highly inclined disks: there is more light in the line-of-sight in the center of the 
galaxy, less obstructed by dust. \cite{Bendo07} find a smooth increase in Concentration in the 3.6 \mum\ 
channel with inclination (their Figure 2) but \cite{Holwerda11a} find a much more complex relation
for H\,I maps. We derive the inclination from the axes ratio reported in Munoz-Matteos et al. {\em in preparation} 
for the 25 mag/arcsec$^2$ isophote ($cos^2(i) = (q^2-q^2_0)/(1-q^2_0)$) but find no relation between any of the 
morphological parameters and the disk inclination (see Figure \ref{f:incl:6par}). 
We choose not to correct for inclination because (a) we not always know the inclination accurately (typically not 
better than $10^\circ$), (b) any correction would necessarily need to assume a template galaxy to derive the 
inclination from (and by necessity ignore disk thickness) or rely on a 3D galaxy model, and (c) in the case of a 
comparison to high redshift samples, accurate disk inclinations would not be available.
Therefore, we choose not to correct for inclination angle. In effect, we explore apparent rather than intrinsic morphology space, including any effects of viewing angle (e.g., apparent disk ellipticity).
\cite{Scarlata07} adopted such a similar approach, in part because accurate inclinations were not available for their high-redshift sample and the computation of disk inclination are not calibrated with H\,I observations.  \\

In an image with $n$ pixels with intensities $I(i,j)$ at pixel position $(i,j)$, and in which the value of 
the pixel is $I_{180}(i,j)$ in the image rotated by $180^\circ$, Asymmetry is defined as \citep{Schade95, CAS}:
\begin{equation}
A = {\Sigma_{i,j} | I(i,j) - I_{180}(i,j) |  \over 2 \Sigma_{i,j} | I(i,j) |  },
\label{eq:a}
\end{equation}
We chose to ignore the positive background contribution to asymmetry as the {\em Spitzer} data have a very high signal to noise
and the added asymmetry from sky noise is negligible \citep[see also][]{Holwerda11a}. Fully symmetric galaxies (e.g., an elliptical)
would have very low values of Asymmetry. Even a regular spiral would not show a high value of Asymmetry. For example, a grand-design spiral galaxy's spiral arms map onto each other with a 180$^\circ$ rotation \citep[the rotational symmetry of galaxies can be used to infer dust extinction in pairs of galaxies, see][]{kw92,kw00a,kw00b,kw01a,kw01b,Keel13, Holwerda07c,Holwerda09,Holwerda13a,Holwerda13b}.
Small-scale structure (e.g. HII regions) in the arms would however contribute to a higher Asymmetry value for spiral galaxies. 
Flocculant spirals can be expected to be slightly more Asymmetric still. The highest values of Asymmetry can be found in either irregular galaxies (Irr) or galaxies with strong tidal disruptions, provided the tidal structures are included in the calculation and are relatively bright due to recent triggered star-formation. If the wavelength over which the parameter is computed is less sensitive to star-formation, as is the case with the \s4g\ imaging, then the Asymmetry signal of interaction or HII regions in spiral arms can be expected to be lower.\\

\noindent Smoothness \citep[also called Clumpiness in the original][]{CAS} is defined as:
\begin{equation}
S = {\Sigma_{i,j} | I(i,j) - I_{S}(i,j) | \over \Sigma_{i,j} | I(i,j) | },
\label{eq:s}
\end{equation}
\noindent where $I_{S}(i,j)$ is the same pixel in the image after smoothing with a choice of kernel. Smoothness is a parameterized 
version of the unsharp masking technique \cite{Malin78b} used on photographic plates to identify faint 
structures. The various definitions employ different smoothing kernels and sizes, the most recent one 
using a flexible kernel-size of 0.2 Petrosian radius and the  boxcar shape. To simplify matters, we chose 
to use a fixed 3 pixel FWHM Gaussian smoothing for our definition (a 30-300 pc. scale). 
We note that we use the term "Smoothness" for historical reasons as this has become the de facto 
designation of this parameter (the CAS scheme), even though an increase in its value means a more clumpy appearance of the image (hence its original designation ``clumpiness"). Very smooth galaxies (Ellipticals) have very low values of Smoothness 
but in other galaxies, the value of the Smoothness parameter depends on the size of the smoothing kernel used.
If the kernel's size correspond to, for example, the width of spiral arms at the distance of the galaxy, then grand design spirals will have relatively high Smoothness values. Alternatively, if the kernel corresponds to large HII regions (common in the HST surveys),
both spirals and Irregulars will show higher Smoothness values.

\cite{Abraham03} and \cite{Lotz04} introduced the Gini and \m20\ parameters. Both are related to the concentration of the light
but the Gini parameter does not assume the brightest pixel is in the geometric center of the galaxy image 
and the \m20\ parameter is more sensitive to merger signals and does not impose circular symmetry on non-merging galaxies.

The Gini parameter is an economic indicator of equality, i.e., G=1, if all the flux is in one pixel, G=0, if all 
the pixels in the object have equal values
We use the implementation from \cite{Abraham03} and \cite{Lotz04}:
\begin{equation}
G = {1\over \bar{I} n (n-1)} \Sigma_i (2i - n - 1) | I_i |,
\label{eq:g}
\end{equation}
\noindent where $I_i$ is the intensity of pixel $i$ in an increasing flux-ordered list of the $n$ pixels in 
the object, and $\bar{I}$ is the mean pixel intensity. Holwerda et. al. ({\em submitted}) find a weak link with Gini and current star-formation.\\

\cite{Lotz04} introduce the relative second-order moment (\m20) of an object.
The second-order moment of a pixel is: $M_i = I_i \times R_i = I_i \times [(x_i - x_c)^2 + (y_i - y_c)^2]$, where $I_i$ is the value of
pixel $i$ in the image, and $x_i$ and $y_i$ are the $x$ and $y$ coordinates of that pixel and $x_c$ and $y_c$ 
are the position of the galaxyÕs center. Each pixel value is weighted with the projected radius away from the galaxy center.

The total second-order moment of an image is defined as:
\begin{equation}
M_{tot} = \Sigma M_i = \Sigma I_i [(x_i - x_c)^2 + (y_i - y_c)^2]
\end{equation}

When we now rank the pixels by value, we can define the relative second-order moment of the brightest 20\% of the flux:
\begin{equation}
M_{20} = \log \left( {\Sigma_i^k M_i  \over  M_{tot}}\right), ~ {\rm for ~ which} ~ \Sigma_i^k I_i < 0.2 ~ I_{tot} {\rm ~ is ~ true}.\\
\label{eq:m20}
\end{equation}
\noindent where pixel $k$ marks the top 20\% point in the flux-ordered pixel-list. Some authors vary the central position ($x_c$, $y_c$) to minimize this parameter \citep{Lotz04, Bendo07}, but we treat deviations from this value due to variation in the center as a source of uncertainty.

The \m20\ parameter is a parameter that is sensitive to bright structure away from the center of the galaxy; flux is weighted in favor of the outer parts. It therefore is relatively sensitive to tidal structures (provided of course these are included in the calculation), specifically 
star-forming regions formed in the outer spiral or tidal arms. If no such structures are in the image, the 20\% brightest pixels will most likely 
be concentrated in the center of the galaxy, which is weighted lower. Thus, one can expect low values of \m20\ for smooth galaxies with bright nucleus (Ellipticals, S0 or Sa) but much higher values (less negative) for galaxies with extended arms featuring bright HII regions.
For example, \cite{Holwerda12c} show how the combination of \m20\ and Asymmetry can be used to identify extended ultra-violet disks \citep[e.g., those identified by ][]{Thilker07b}. As with the Smoothness parameter, one expects the contributions from star-formation to be much less in \s4g, lowering contrast of HII regions at higher radii; lower values for \m20\ in galaxies that would have a much higher value in bluer wavebands.\\

Instead of the intensity of the pixel ($I_i$) one can use the second order moment of the pixel ($M_i =  I_i [(x_i - x_c)^2 + (y_i - y_c)^2]$) in equation \ref{eq:g}. This is the $G_M$ parameter \citep{Holwerda11c}:
\begin{equation}
G_M = {1\over \bar{M} n (n-1)} \Sigma_i (2i - n - 1) | M_i |,
\label{eq:gm}
\end{equation}
\noindent which is an indication of the spread of pixel values weighted with the projected radial distance to the galaxy center.  
In essence this is the Gini parameter with a different weighting scheme than unity for each pixels. Similar to the \m20\ parameter, it emphases 
the flux from the outer regions of the galaxy. If there is significant flux in the outer parts, this will boost the value of \gm. Contrary to \m20, it does not depend on a somewhat arbitrary delineation of the brightest 20\% flux for the denominator but relies on all pixel values. Unlike the Gini parameter, however, it does rely on a supplied center of the galaxy (to compute $M_i$). For concentrated galaxies, the \gm\ and Gini values will be close together but as relatively more flux is evident in the outer parts of the galaxy, \gm\ will be higher. \cite{Holwerda11b} found this a good {\em single} parameter to identify active mergers (sweeping tidal tails etc.) from atomic hydrogen maps (H\,I).\\

\cite{Scarlata07} added the ellipticity of a galaxy's image to the mix of parameters in order to classify 
galaxies according to type in the COSMOS field.  Ellipticity is defined as:
\begin{equation}
E = 1 - b/a
\label{eq:e}
\end{equation}
with $a$ and $b$, the major and minor axes of the galaxy, computed from the spatial second order 
moments of the light along the x and y axes of the image in the same manner as Source Extractor 
\citep{se,seman}. We include this definition for completeness.

As input for all these parameters we need an estimate of the center of the object and a definition of 
the area over which they need to be computed. For the object centers, we use the right ascension 
and declination from \cite{s4g} and we use a limiting surface brightness of 25 mag/arcsec$^2$ to 
include pixels, excluding foreground and background objects masked by the S$^4$G pipeline.

\subsection{Computation of Morphological Parameters}
\label{s:prep}

To compute the morphological parameters, one needs the image, the center of the object 
and a criterion for which pixels to include. The object center is taken from the \s4g catalog 
(Mu\~{n}oz-Mateos et al. {\em in preparation}) and pixels are included if they exceed our 
surface brightness criterion (25 mag/arcsec$^2$) and are not excluded by the P2 masks.
We chose a practical limit of 25 mag/arcsec$^2$ (AB magnitude) in both bands 
\citep[][Mu\~{n}oz-Mateos et al. {\em in preparation}]{s4g}. \s4g is sensitive down to 
$\sim27$ mag$_{AB}$/arcsec$^2$ in both bands, in the case of smoothed isophotes but not individual pixels.
We cut out a section of the mosaic corresponding to $5 ~ R_{25} \times5 ~ R_{25}$, the radius from \cite{RC3} 
around the central position to speed computation.

The use of an isophotal criterion is uncommon in Hubbel Space Telescope ({\em HST}) surveys of distant galaxies which 
span large redshifts, since the selected area will be affected by cosmological surface brightness 
dimming, k-correction, evolution, and zeropoint offsets at different redshifts. 
For these applications, an aperture with the Petrosian radius is often employed \citep[see][]{Bershady00}. 
\cite{Munoz-Mateos09a} find that an elliptical aperture based on the Petrosian radius misses significant 
emission in the {\em Spitzer} Infrared Nearby Galaxy Survey \citep[SINGS,][]{SINGS} IRAC images, 
depending on the concentration of the galaxy (i.e., depending on Hubble Type). 
Since this is a local volume sample which suffers little from the above-mentioned redshift issues, we opt for an 
isophote-defined area (all pixels exceeding 25 mag/arcsec$^2$) to fully include all information, while excluding 
as much sky-noise as practical.

\subsection{Uncertainties}
\label{s:unc}

Uncertainties in the morphological parameters come from the uncertainty in the position of the center, 
the image segmentation, and shot noise in the pixel flux values. Potential biases are if the parameter values change 
also with viewing angle (i.e., disk inclination) and distance to the object.
We explore these issues in sections \ref{s:data} and \ref{s:morph} and Figures \ref{f:incl:6par} and \ref{f:dist:6par} in the the Appendix.

Some authors minimize the parameters --most often Asymmetry-- by varying the central position 
\citep{Bendo07, Munoz-Mateos09a}. However, \cite{de-Blok08} find that the dynamical center and 
the brightest point in the 3.6 \mum\ light distribution nearly always coincide. Instead of minimizing, we 
take the central position from the \s4g catalog as given but then vary this input center with a random 
Gaussian distribution with FWHM=3 pixels to define the variance in each morphological parameter. 
This variance then defines a measure of our uncertainty in these parameters.

The second uncertainty relates to the segmentation of the image, i.e., which parts of the image are assigned 
to the target object, which to other objects or masked because of image artifacts. 
Depending on crowding of objects in the field, a substantial fraction of the information on an extended object 
may be lost. DR1 of \s4g\ applies uniform criteria to mask objects not belonging to the target galaxy using a 
combination of a sextractor segmentation image \citep[see also][]{seman} and visual masking by the data team. 
While a different fraction of the image will be masked for each target galaxy, we can be confident that the masking 
is self-consistent across the sample.

Our remaining choice is which parts of the image to include as information on the target galaxy, i.e., which pixels 
contain enough flux from this galaxy to be included and which pixels are mostly background noise? Different 
authors have solved this in the literature. For example, both \cite{Bendo07} and \cite{Munoz-Mateos09a} use 
an elliptical aperture to define the boundaries on the image over which the morphological parameters are to 
be computed. The high-redshift studies, however, tend to use an isophotal cutoff, a minimum value or signal-to-noise 
for pixels to be included. The latter reasoning is that an elliptical aperture may both cut off outlying flux belonging 
to the target galaxy as well as include areas of near pure noise. Because our goals include to serve as a 
benchmark for higher redshift surveys, we opt for an isophotal approach.
But the choice of both masking and the threshold or aperture will influence the level of noise included in the pixel 
collection over which the morphological parameters are computed. One solution would be to take a random sub-set of the 
pixel collection that is the target galaxy and compute the parameters over these. The variance in the parameter values 
would be an indication how critical the parameters depend on the inclusion of certain pixels. In an extreme case for 
example, a single saturated pixel could throw all the morphological parameters and the variance would reflect that. 
However, taking sub-samples would change the signal-to-noise in each sub-set of pixels. 

For the majority of parameters the uncertainty is dominated by variance in the input central position and shot noise in 
the pixels. Computing the variance from sub-samples of pixels would count the pixel shot noise twice.
The exception is the Gini parameter, which does not depend on the input central position but critically on the size of 
the sample\footnote{The issue of the Gini parameter's dependence on signal-to-noise noted by \cite{Lisker08} is a 
direct result of the use of an aperture rather than an isophote. However, the \s4g galaxies are well above 
the S/N levels discussed by Lisker et al.}. Therefore, we use jackknifed (sub-sampled) Gini values to 
compute its uncertainty, using a set of ten random subsamples.

The third uncertainty is the Poisson noise in the pixel flux values. We estimate this by randomizing the 
pixel values around the mean with the same rms as the real pixel collection but keeping the general 
shape of the pixel collection. 
This has the advantage of keeping the total information going into the 
morphological parameter the same but it quantifies the effect of pixel value outliers on the overall 
parameter value, i.e., a single bright spot skewing the computed value.

The reported uncertainties in Tables \ref{t:ch1} and \ref{t:ch2} are the quadratic combination of the 
uncertainty due to variance in the central position and the uncertainty due to shot noise in the pixels. 
In the case of the Gini parameter, it is the quadratic combination of the uncertainties estimated from 
subsampling and pixel shot noise. These values are formal uncertainties of these parameters as the viewing 
angle and distance may still influence the perception of morphology, and affect the morphological parameter 
values. However, since the viewing angle is arguably part and parcel of a galaxy morphology 
(e.g., \cite{Scarlata07} treat it as such) and \s4g  is a sample of local galaxies 
(as described in the previous section), making distance less of an issue, we leave these effects out of the formal error.

\section{Morphology Catalogs}
\label{s:result}

Tables \ref{t:ch1} and \ref{t:ch2} list the morphological parameters for both IRAC wavelengths (3.6 and 4.5 $\mu$m) 
and their uncertainties for all \nsamp \ galaxies (full tables are available in the {\em electronic edition} of the paper).
We compute the $\rm C_{82}$ \ concentration index, Asymmetry, 
Smoothness (after a 3 pixel FWHM Gaussian smooth), the Gini, \m20, the Gini of the second order moment ($G_M$), 
and the Ellipticity of the images. Uncertainties are based on randomly changing the central position (with the 
exception of Gini) and a random reshuffle of the pixel values to simulate shot noise. We note that these uncertainties 
should be viewed as formal errors and do not include the effects of, for example, disk inclination which have a 
pronounced effect \citep[see also][]{Bendo07, Holwerda11a}. We explore the possibility of an ordered morphological 
list based on these parameters, and their ability to select out-of-the ordinary or merging morphology in the near-infrared.

\section{Galaxy Classification -- where normal galaxies lie}
\label{s:type}

As already noted, these parameters do not constitute an orthogonal parameter space, and most often some 
combination is used to define a sub-space populated by unperturbed galaxies on the Hubble tuning fork, i.e., ``normal'' galaxies. 
First we explore each parameter with Hubble type, and subsequently the two-parameter combinations from \cite{Lotz04} and \cite{CAS}. 

Normal spaces have been defined for local samples by \cite{CAS} and \cite{Lotz04} from the visible light 
image collection originally presented in \cite{Frei96}. 
Morphological parameters from {\em Spitzer}-IRAC images for the SINGS sample have been reported by 
\cite{Bendo07}, \cite{Munoz-Mateos09a} and \cite{Holwerda11a}. We will compare to each of these studies 
to explore where normal galaxies reside in the morphological parameter space measured at 3.6 and 4.5 
$\mu$m. In Figures \ref{f:AS} --\ref{f:CM20res2}, we use the RC3 numerical Hubble type (Table \ref{t:type}) from HyperLEDA \citep{hyperleda}
to color-code the data points. These are visual classifications in bluer wavelength images but their uniformity and numerical scale allow for a quick comparison.

\begin{table}
\caption{Legend for the Numerical Hubble types from HyperLEDA \citep{hyperleda}.}
\begin{center}
\begin{tabular}{l l }
Hubble	 & Type \\
\hline
\hline
 E      	&  -5  \\
 E-S0   	&  -3 \\
 S0     	&  -2  \\
 S0/a   	&  0  \\
 Sa     	&  1  \\
 Sab   	&  2  \\
 Sb     	&  3  \\
 Sbc   	&  4  \\
 Sc     	&  6 \\
 Sc-Irr 	&  8 \\
 Irr   		& 10  \\
\hline
\hline
\end{tabular}
\end{center}
\label{t:type}
\end{table}%

\begin{table*}
\caption{The Spearman ranking of the relation between Hubble Type or stellar mass with the morphological parameters in either 3.6 \mum\ or 4.5 \mum\ images for our full sample. The absolute z-values of significance for each of Spearman rankings are noted between parentheses. }
\begin{center}
\begin{tabular}{l l l l l l l l}
			   & C	 & A & S & $M_{20}$ & G & $G_M$ \\
\hline
\hline
3.6 $\mu$m			& & & & & & \\
\hline
$\rm log_{10}(M_*)$      & 0.11 (3.74)   & -0.06 (1.90)         & -0.15 (5.25)         & -0.08 (2.95)         & 0.06 (1.94)   & -0.09 (3.17) \\ 
Hubble Type      & -0.55 (29.20)        & 0.02 (0.88)   & 0.26 (12.67)          & 0.62 (33.86)          & -0.52 (27.21)        & 0.05 (2.36) \\ 
$[3.6-4.5]$        & 0.00 (0.08)   & 0.02 (0.55)   & -0.01 (0.39)         & 0.00 (0.11)   & 0.00 (0.03)   & -0.01 (0.23) \\ 
\hline
4.5 $\mu$m			& & & & & & \\
\hline
$\rm log_{10}(M_*)$      & 0.10 (3.57)   & -0.03 (1.18)         & -0.10 (3.45)         & -0.06 (2.01)         & 0.02 (0.68)   & -0.04 (1.37) \\ 
Hubble Type      & -0.43 (21.69)        & -0.05 (2.28)         & 0.09 (4.04)   & 0.50 (25.91)          & -0.38 (18.81)        & -0.06 (2.80) \\ 
$[3.6-4.5]$        & 0.00 (0.01)   & 0.04 (1.23)   & -0.00 (0.13)         & -0.02 (0.52)         & 0.02 (0.53)   & 0.02 (0.64) \\ 

\hline
\hline
\end{tabular}
\end{center}
\label{t:spear}
\end{table*}%

\subsection{Single Parameters}
\label{s:single}

\begin{figure*}
\begin{center}
\includegraphics[width=0.49\textwidth]{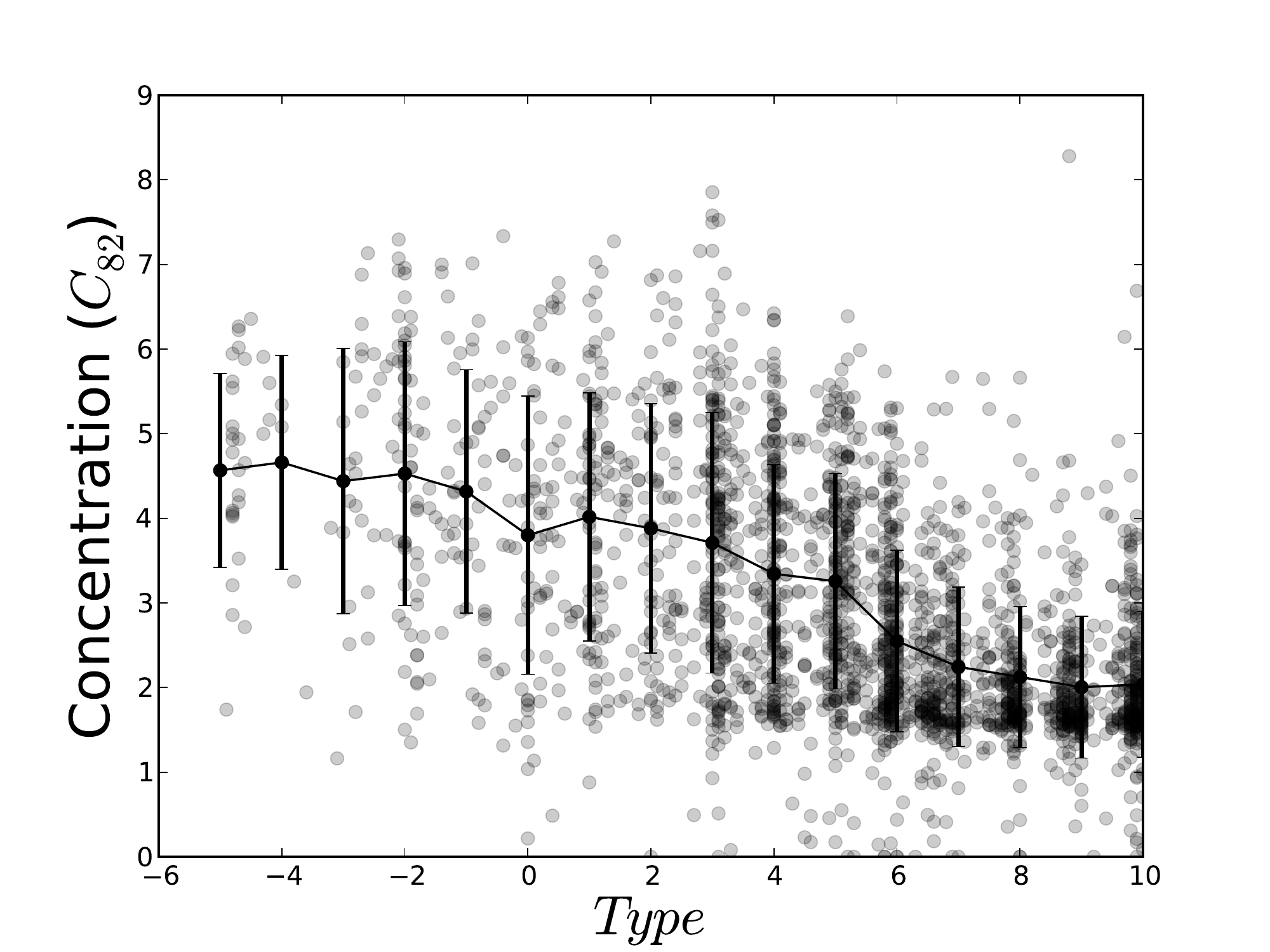}
\includegraphics[width=0.49\textwidth]{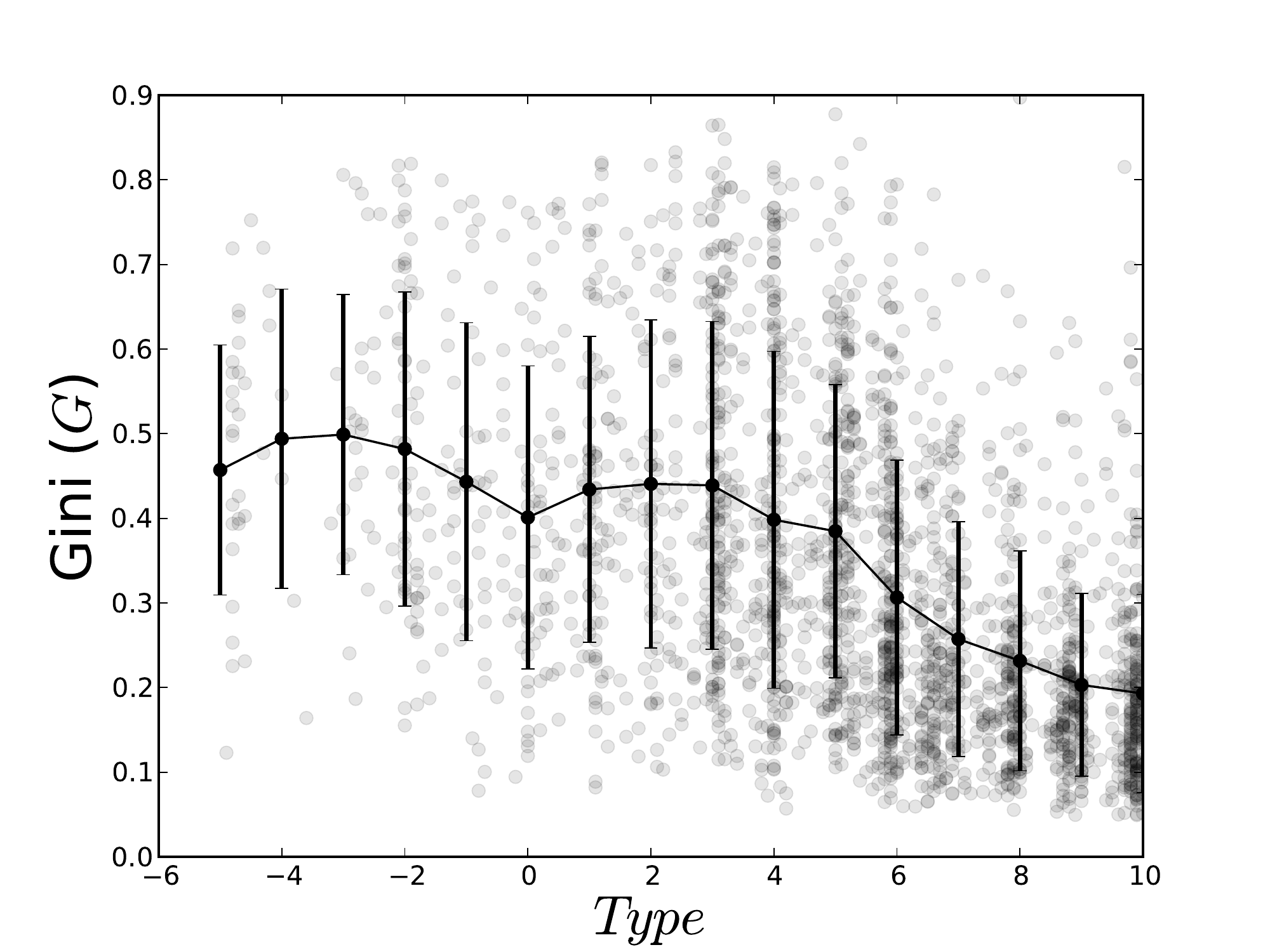}\\
\includegraphics[width=0.49\textwidth]{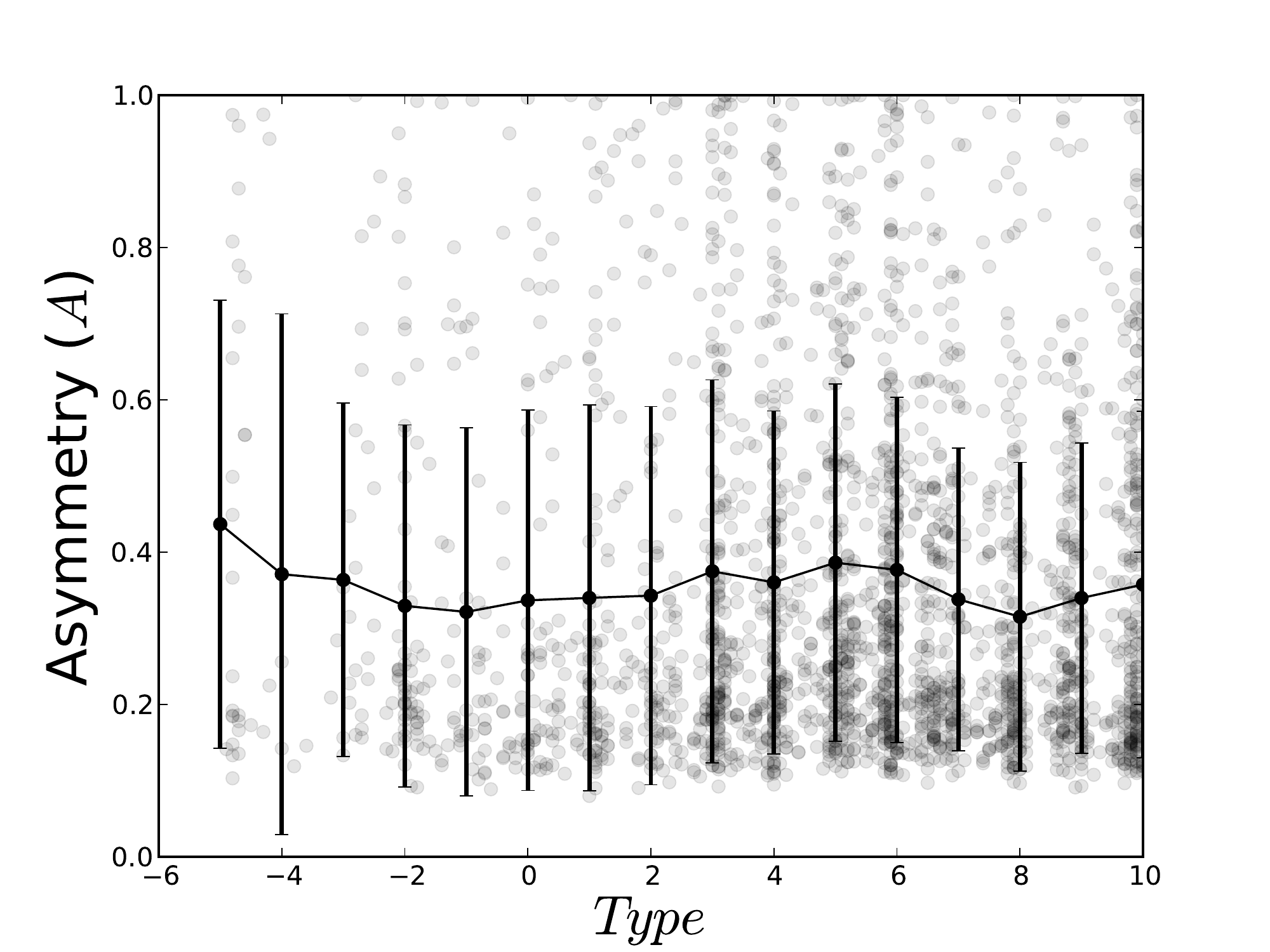}
\includegraphics[width=0.49\textwidth]{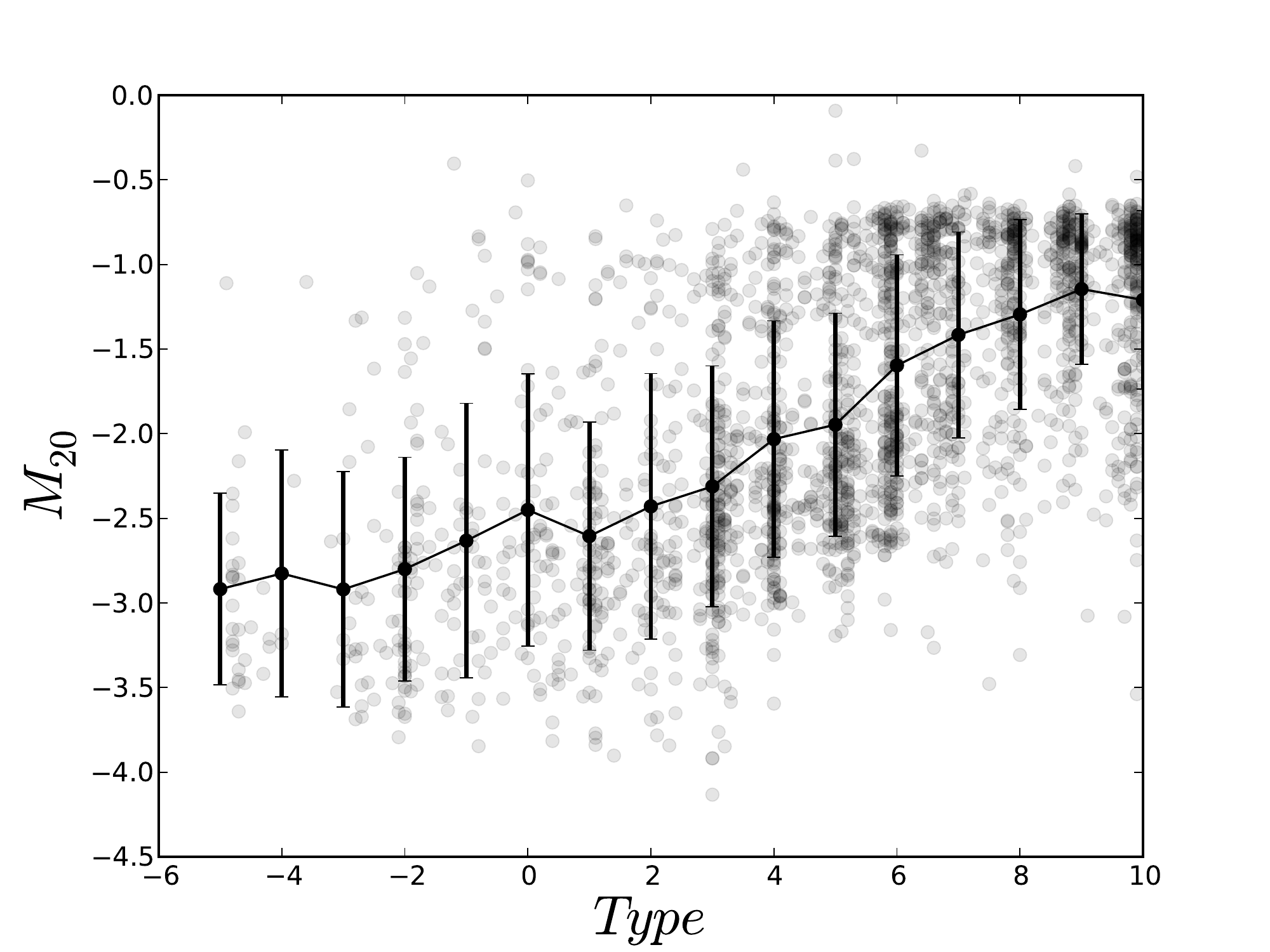}\\
\includegraphics[width=0.49\textwidth]{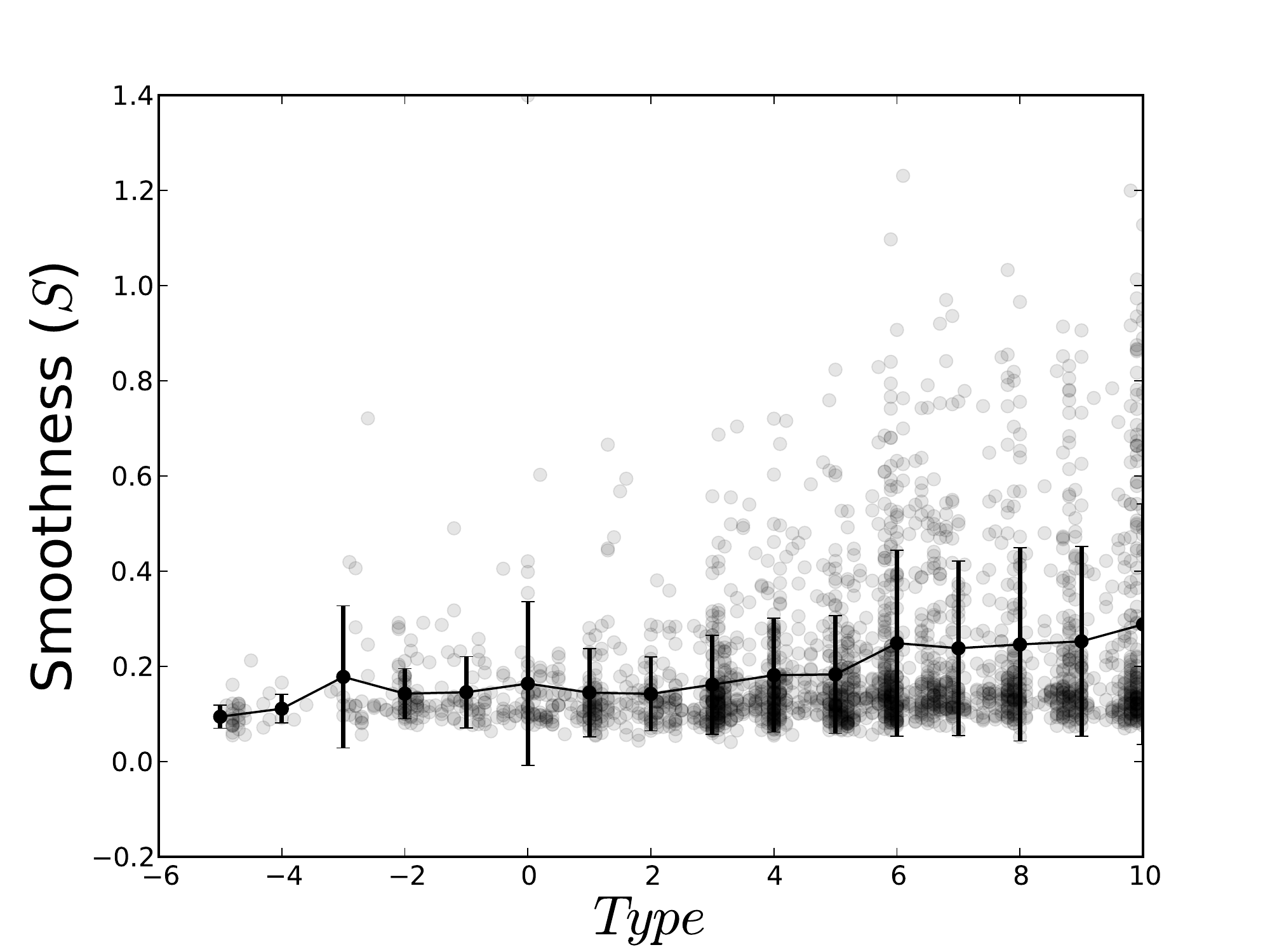}
\includegraphics[width=0.49\textwidth]{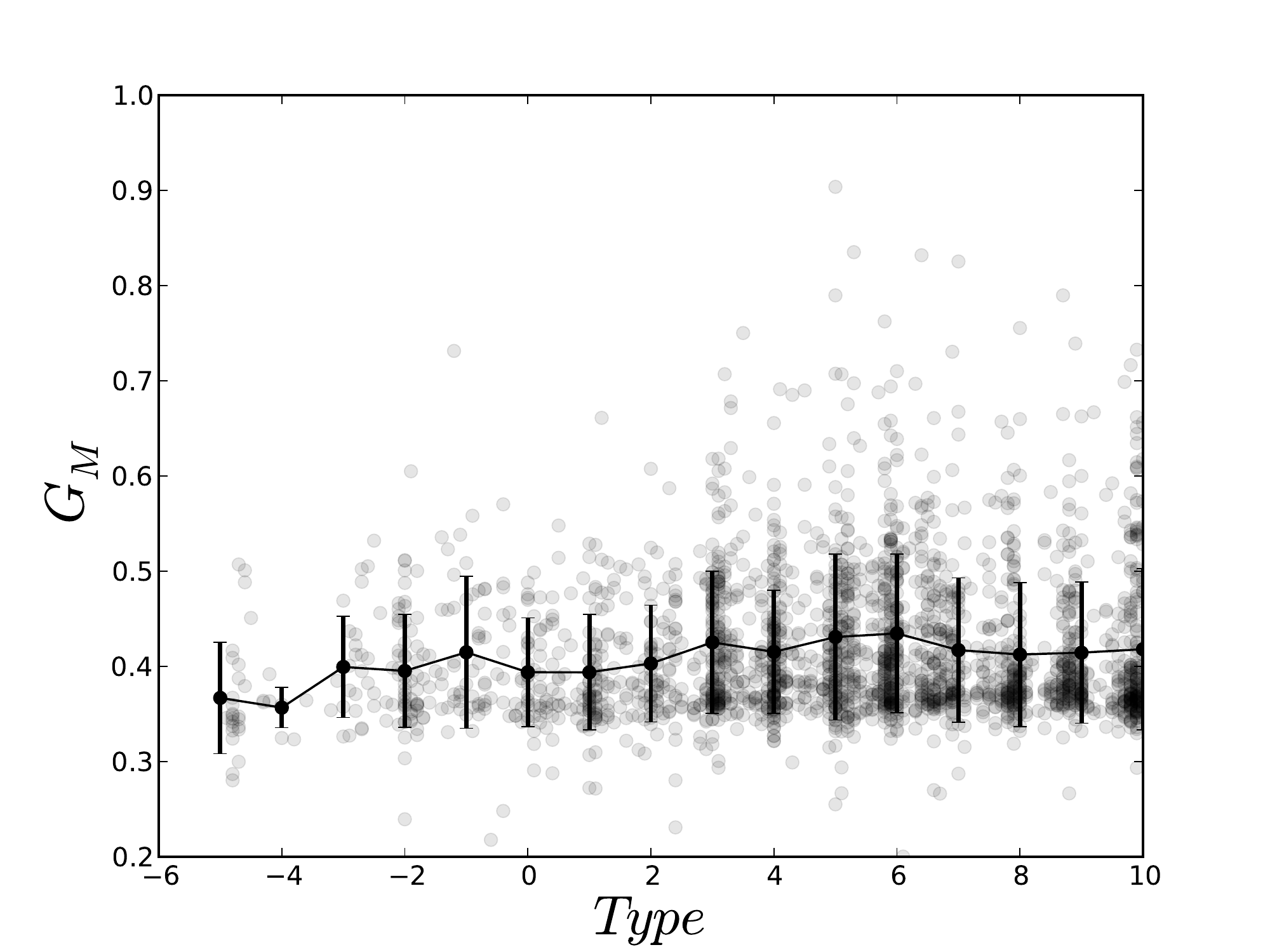}\\
\caption{The relation between the Hubble type and each of the morphological parameters in 3.6 $\mu$m. 
Solid points are the mean value for each Hubble type, the error bars the rms in each type.}
\label{f:single}
\end{center}
\end{figure*}

Figure \ref{f:single} shows the relation between Hubble Type (RC, Table \ref{t:type}) and each of the 
morphological parameters. Concentration, Gini and \m20 show the most promise for differentiating between 
Hubble types. No single parameter alone appears discerning enough to quantify Hubble type completely. 
This has been found previously for visible light morphologies by \cite{Lotz04, CAS}, and \cite{Scarlata07}.
 \m20 appears to have the most differentiating power for Hubble type classification, i.e., this 
parameter has the steepest dependence on Hubble Type in Figure \ref{f:single} (see \S \ref{s:CM20}).
The Spearman ranking with Hubble Type (Table \ref{t:spear}), ranks Concentration, \m20, and Gini as 
reasonably closely linked to Hubble Type (a ranking of 0 is unrelated, -1 anti-correlated and 1 linearly related). 
The link is stronger with 3.6 \mum\ parameters than 4.5 \mum. For comparison, the ranking with stellar mass 
is also listed in Table \ref{t:spear}, showing that the morphological relation is related to total stellar mass as well (from low-mass irregulars to massive ellipticals).

In the \m20\ panel, a clustering is visible near \m20=-1 for $T>2$. We inspected the \s4g\ images of 
some examples of these objects. They include many examples of edge-on and barred galaxies. 
In the case of edge-on galaxies, the line-of-sight integration of stellar light (with little extinction) results in
relatively more light at higher galactocentric radii; thus the same Hubble Type has a greater contribution 
from the top 20\% of pixels at greater radii. A similar effect happens if stars are dynamically concentrated 
in a bar: some of the brightest pixels will occur at higher radii, increasing the value of \m20.

\subsection{Parameter Pairs}
\label{s:pairs}
 
In this section we discuss a few of the parameter pairs, noted in the literature 
\citep{CAS,Lotz04,Munoz-Mateos09a} as useful to separate ``normal'' galaxies from ``disturbed'' ones 
and morphologically classify these ``normal'' galaxies. For example, figure \ref{f:A12} in the appendix illustrates the 
distribution of the \s4g\ sample over the parameter space.
\cite{Buta10} note that late-types (S0/a to Sc) galaxies appear ÒearlierÓ in type at 3.6 \mum, 
due to the slightly increased prominence of the bulge and the reduced effects of extinction.

\subsubsection{Asymmetry and Smoothness}
\label{s:AS}

\begin{figure*}
\begin{center}
\includegraphics[width=0.49\textwidth]{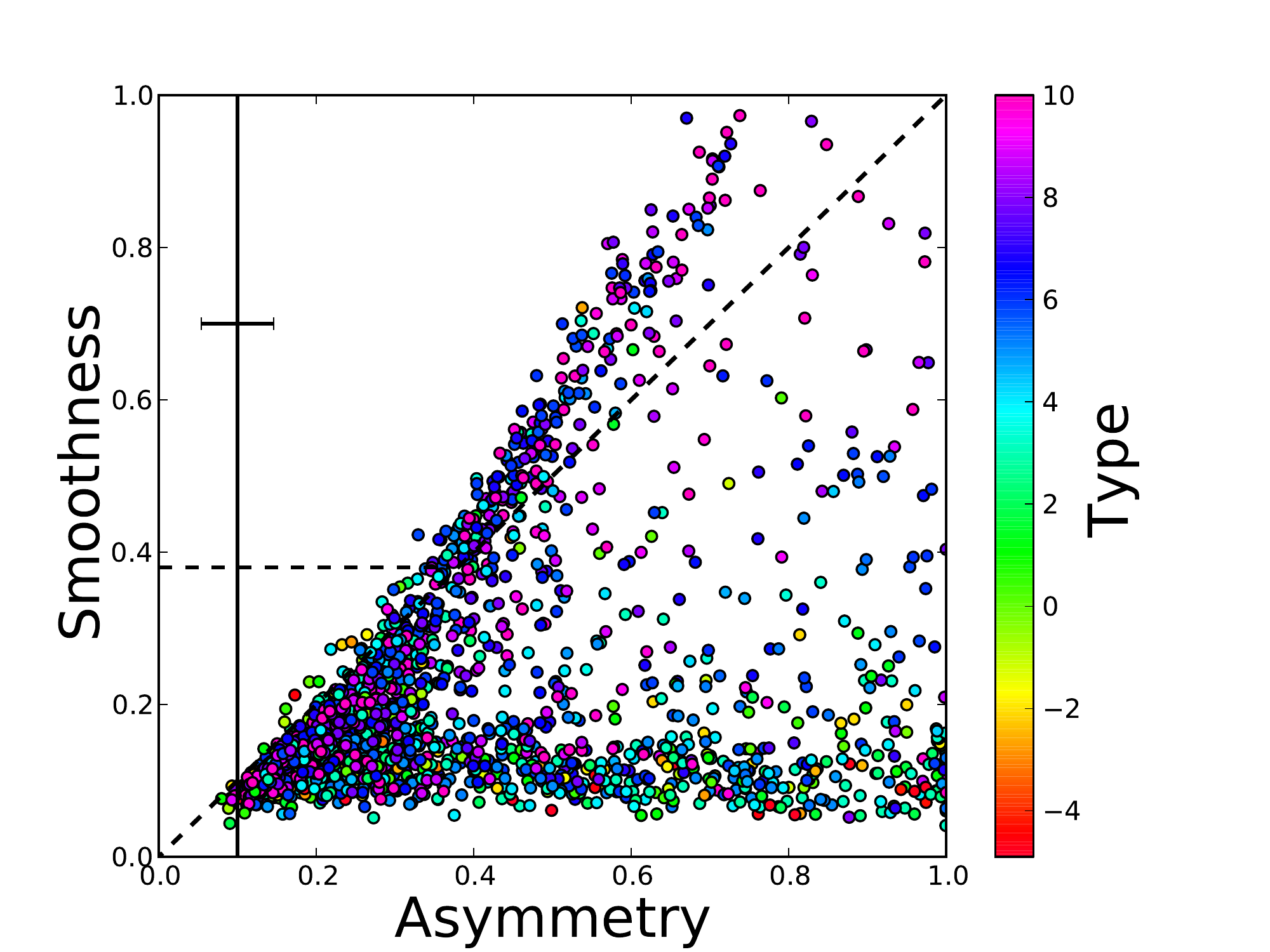}
\includegraphics[width=0.49\textwidth]{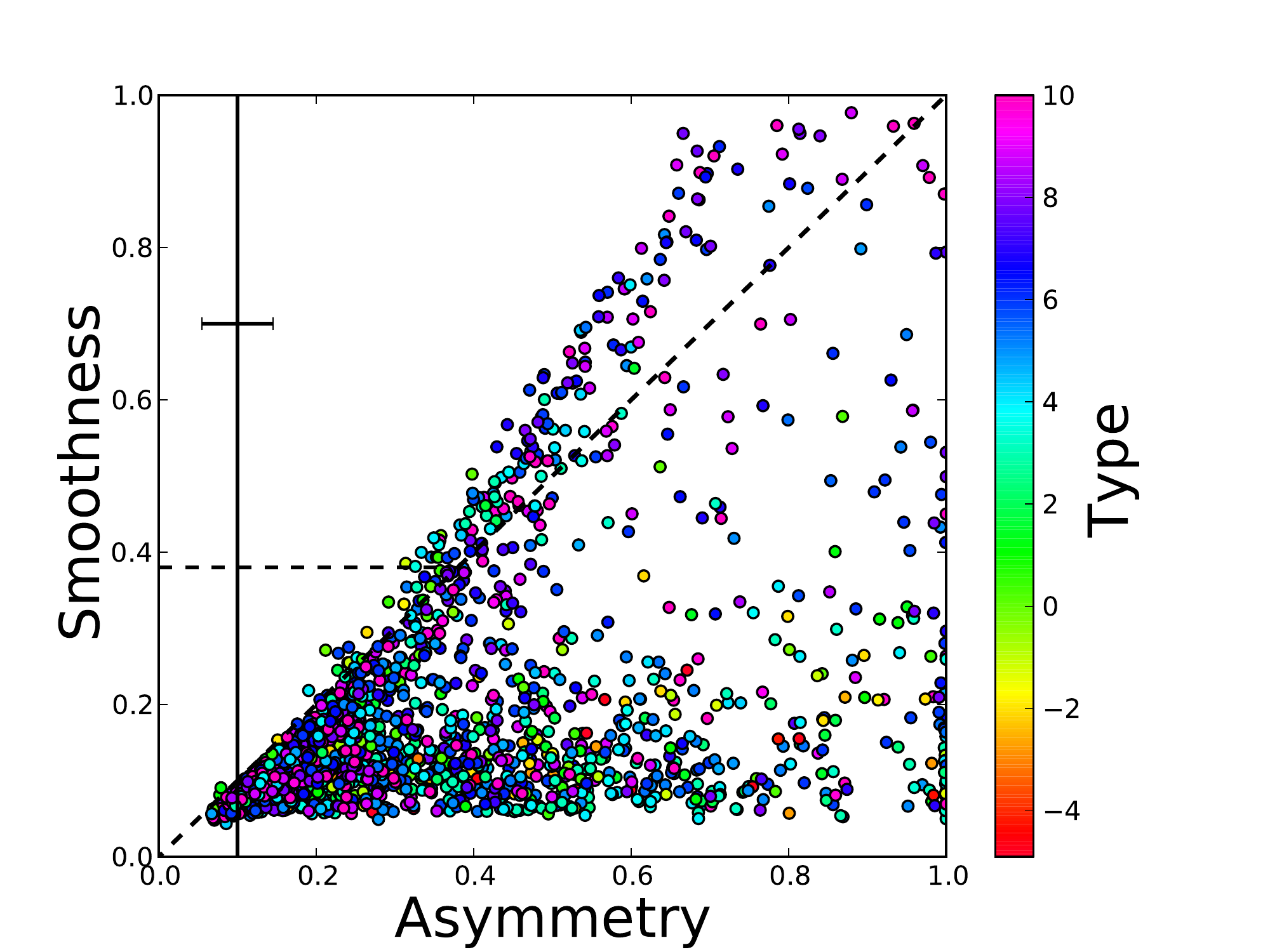}
\caption{The relation between Asymmetry and Smoothness for 3.6 (left) and 4.5 \mum\ (right) for \s4g\ galaxies. The dashed line is Asymmetry-Smoothness equality, a prerequisite for interaction from \protect\cite{CAS} for interacting systems (equation \ref{eq:AS}). Galaxies above this dashed line and with Asymmetry greater than A=0.4 are candidates for ongoing or recent interaction.}
\label{f:AS}
\end{center}
\end{figure*}

\cite{CAS} define an Asymmetry--Smoothness relation ($A = 0.35\times S + 0.02$) for R-band images where normal galaxies reside.
Figure \ref{f:AS} shows the relation between Asymmetry and Smoothness. The population is split between two 
sequences: one where Smoothness follows Asymmetry, mostly populated by irregulars and spirals, and one 
where these parameters are completely unrelated. Neither case presents a clear separation between early 
and late types.

In these near-infrared images (and perhaps with our implementation of the parameters) there is little use for this pair as a classifier.
One obvious difference between this study and any previous one is the wavelength: in their comparisons between 
wavelengths in the SINGS/THINGS sample both \cite{Munoz-Mateos09a} and \cite{Holwerda11a} find much 
lower values of Asymmetry for 3.6 \mum\ compared to other, especially optical, wavelengths. 
The typically lower values of Asymmetry are the cause of the poor separation of early and late types.

We note that our simple implementation of Smoothness, i.e., a fixed-sized smoothing kernel, could be affected 
by distance effects, but the mean of the Smoothness parameter for a given Hubble Type does not change between 
nearby and distant subsamples. A more likely reason is that \cite{CAS} use R-band optical images and \s4g\ is in the near-infrared with the resulting different dependencies on star-formation and dust extinction.

\subsubsection{Gini and \m20}
\label{s:GM20}

\begin{figure*}
\begin{center}
\includegraphics[width=0.49\textwidth]{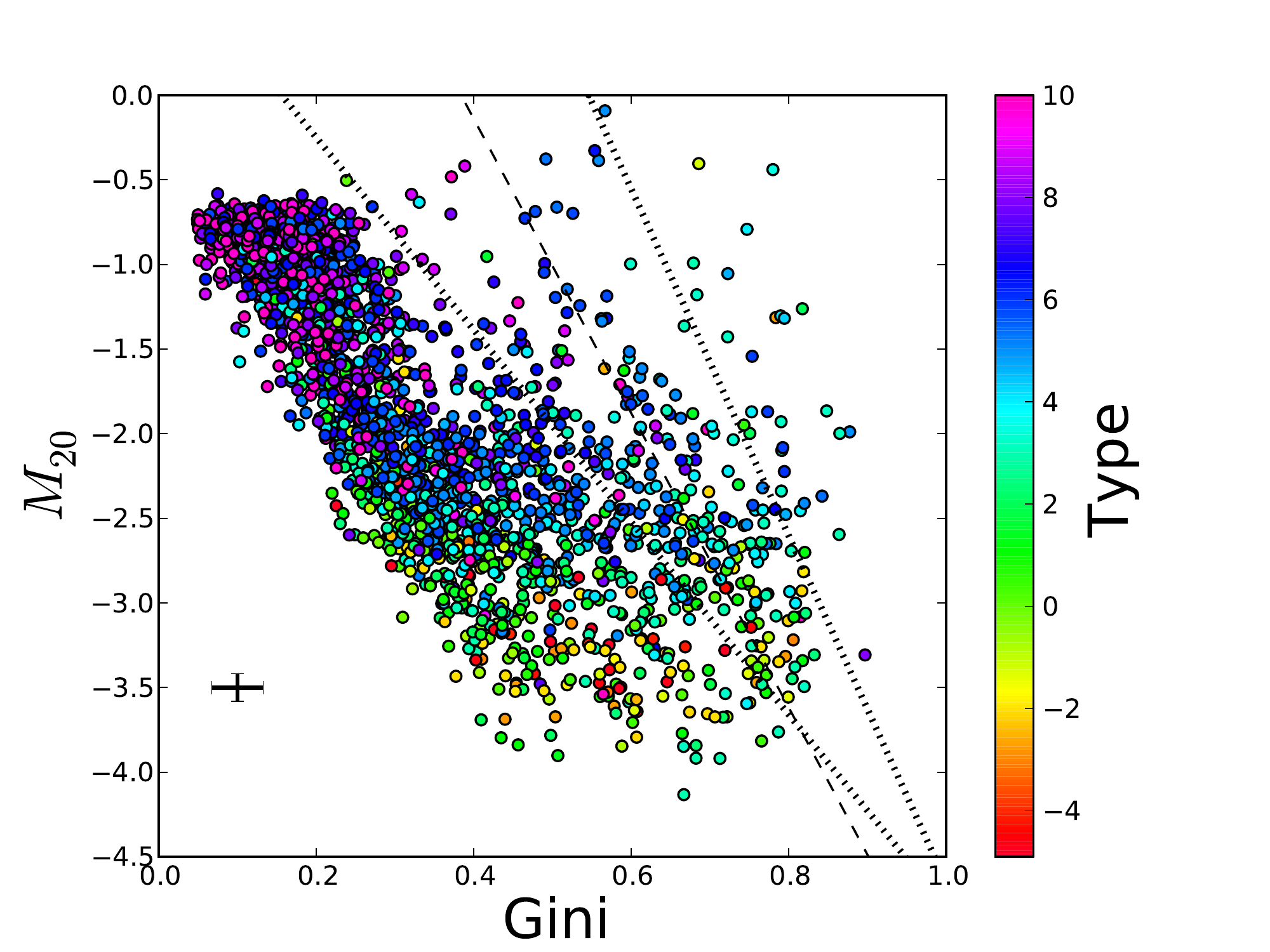}
\includegraphics[width=0.49\textwidth]{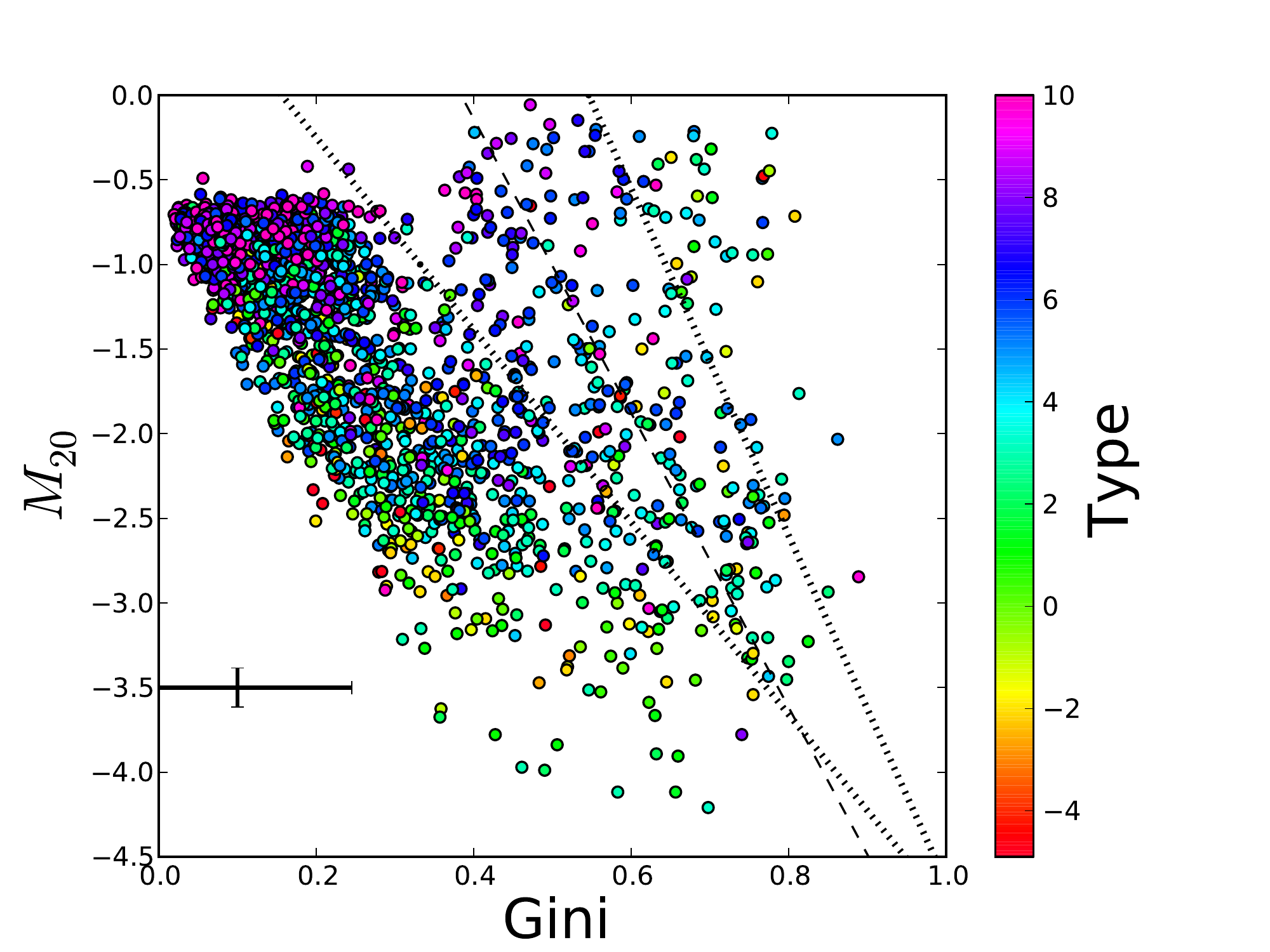}
\caption{The relation between Gini and \m20\ for 3.6 (left) and 4.5 \mum\ (right) for \s4g\ galaxies. 
The dashed line is the interaction criterion in equation \ref{eq:GM20} from \protect\cite{Lotz04}
The dotted lines in the 3.6 \mum\ plot are the limits of the envelope from \protect\cite{Munoz-Mateos09a}. }
\label{f:GM20}
\end{center}
\end{figure*}

\begin{figure*}
\begin{center}
\includegraphics[width=0.49\textwidth]{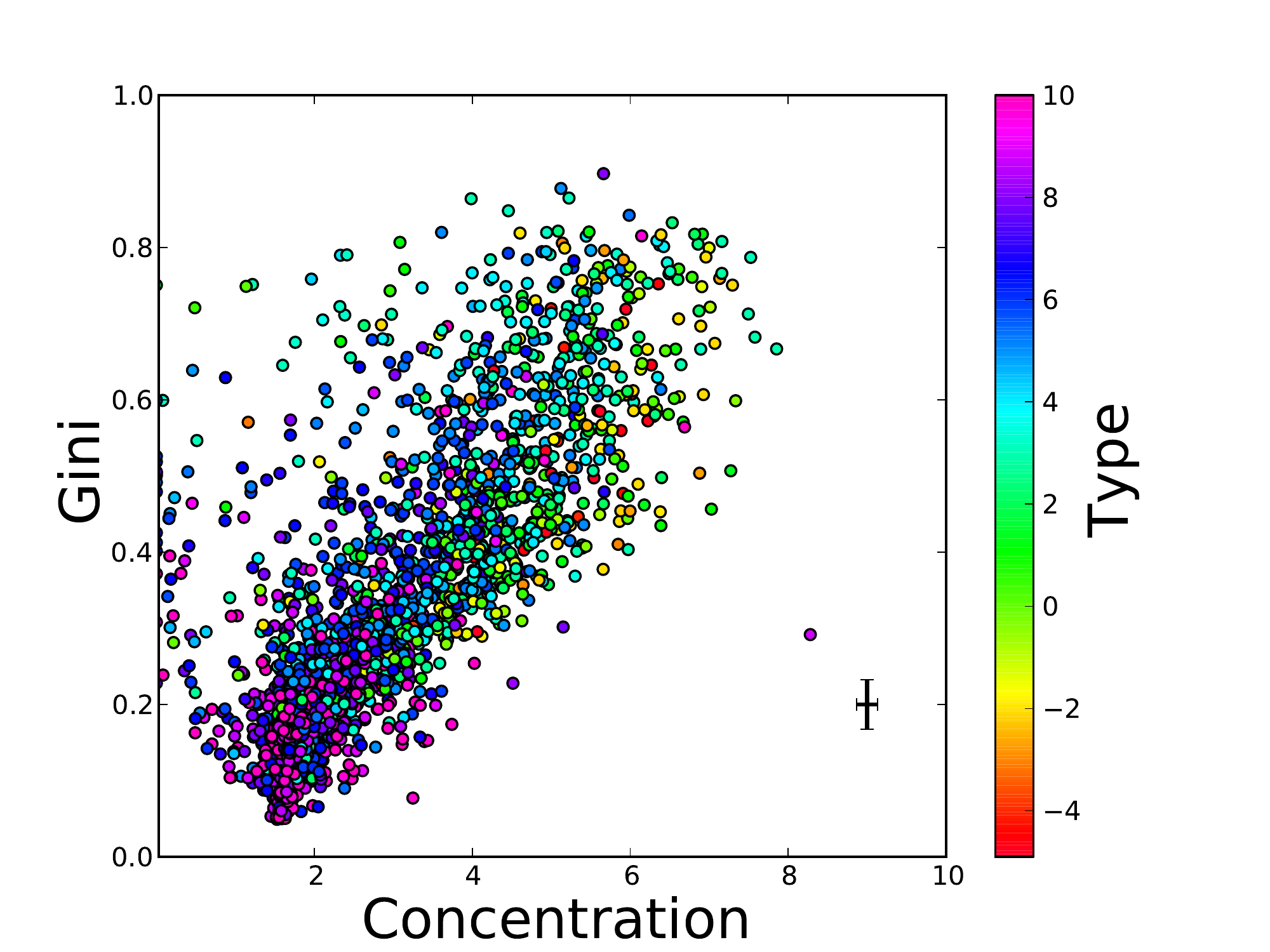}
\includegraphics[width=0.49\textwidth]{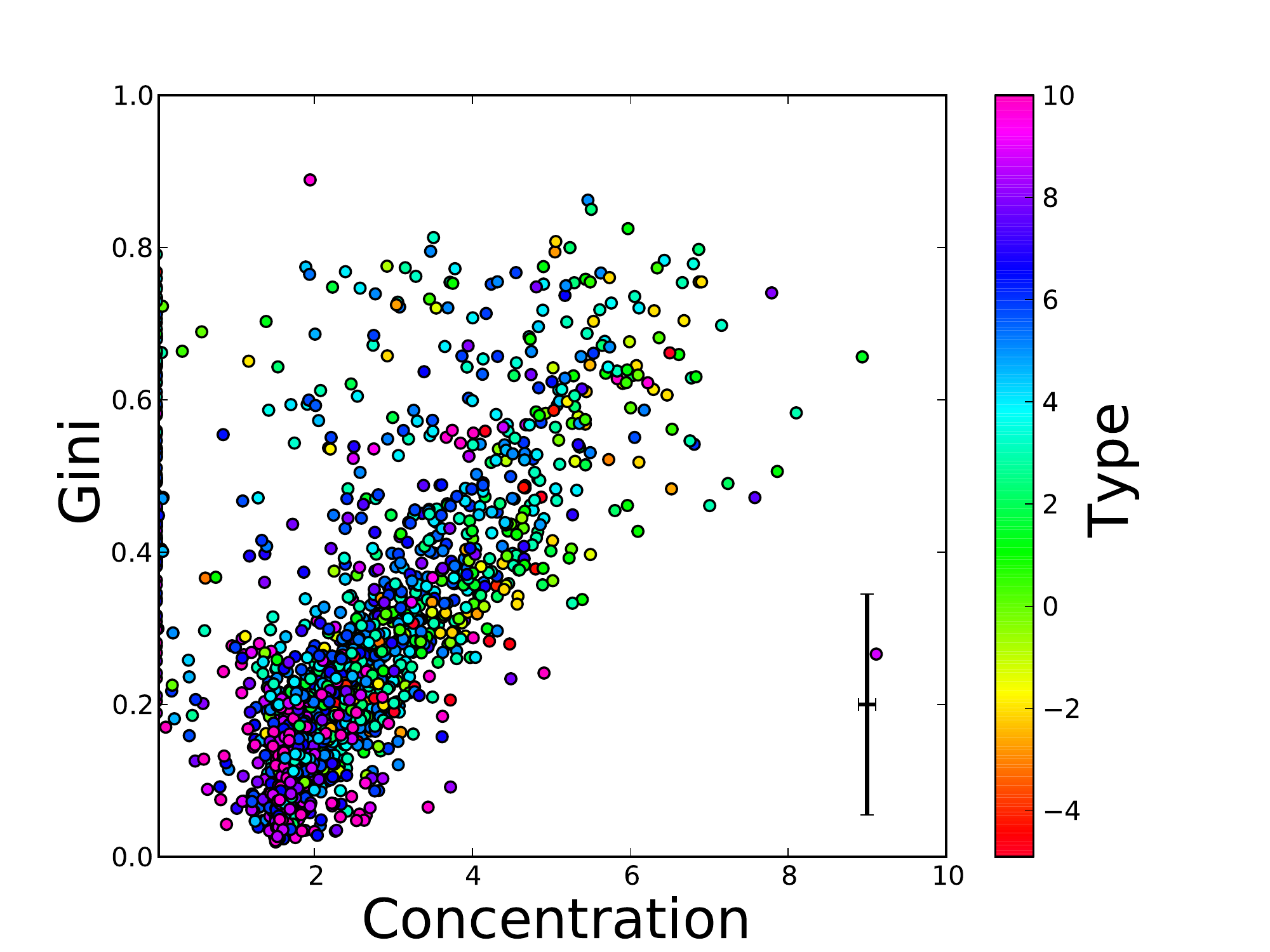}
\caption{The relation between Concentration and Gini for 3.6 (left) and 4.5 \mum\ (right) for \s4g\ galaxies. Points are color-coded according to Hubble Type. Both parameters are closely related for early-types but diverge for late-types as additional structure influences both differently.}
\label{f:CG}
\end{center}
\end{figure*}

\cite{Lotz04} showed that Gini and \m20 together separate early from late types based on visible light images. 
They define a criterion between normal and disturbed galaxies (see equation \ref{eq:GM20}).  Figure \ref{f:GM20} 
shows the Gini--\m20 space for \s4g. There seems to be a (noisy) sequence between Gini  and \m20\ with Hubble 
type. This correlation reflects the well known trend of an increase in central/bulge prominence from late to early type.
Surprisingly, early-types (elliptical and S0) galaxies appear to display a range of Gini values. This is somewhat 
unusual as these are the smoothest galaxies, with the smallest contribution to the second order moment by the 
brightest 20\% of the flux (\m20) because these are all in the center. Several of these are selected as ``disturbed'' 
galaxies (see also the discussion of Figure \ref{f:disthis} below). However, their pixel values are not homogeneous --each pixel contributing the same 
fraction of the flux-- and thus the Gini parameter becomes akin to a concentration index (Figure \ref{f:CG}) as ranking by flux becomes similar to ranking by radius.

\cite{Munoz-Mateos09a} define an envelope based on the morphology of the SINGS galaxies for the Gini-\m20\ 
space (dotted lines in the left panel of Figure \ref{f:GM20}). The \s4g\ parameters do not appear to adhere to this 
envelope. Our implementation is different on two points: first we use an isophotal definition of the pixels to be 
included and secondly, we include any bright central source in our computation.
%
In \cite{Holwerda11a}, we compared our results to those from \cite{Bendo07} for the SINGS sample, computed over a similar elliptical aperture. There is 
an offset in the Gini parameter --our Gini values are 0.15 higher-- which can be attributed to the convolution of the 3.6 \mum\ images to 24 \mum resolution by \cite{Bendo07}.
%
Similarly, \cite{Munoz-Mateos09a} find that for different apertures, the values of Gini change between the Gini values computed over the $R_{25}$ elliptical aperture and the Petrosian radius elliptical aperture. The difference is $\sim0.1$. Thus, the choice to include central sources, the choice of aperture and thirdly, any convolution, all add a shift to the Gini parameter values for the whole sample.

In Figure \ref{f:GM20}, we find that the offset in \m20\ is 0.5 lower than those typically found by \cite{Munoz-Mateos09a}, which would be the result of 
our choice of an isophotal area over an elliptical aperture: $M_{tot}$ is higher as low-flux pixels are excluded, and therefore 
the relative contribution by the brightest 20\% is smaller.  A larger number of pixels contributing a small fraction of the total 
flux would {\em increase} the value of the Gini parameter. However, the isophotal criterion does away with low 
contribution pixels and this may explain our lower values for Gini.

\subsubsection{Concentration -- \m20}
\label{s:CM20}

\begin{figure*}
\begin{center}
\includegraphics[width=0.49\textwidth]{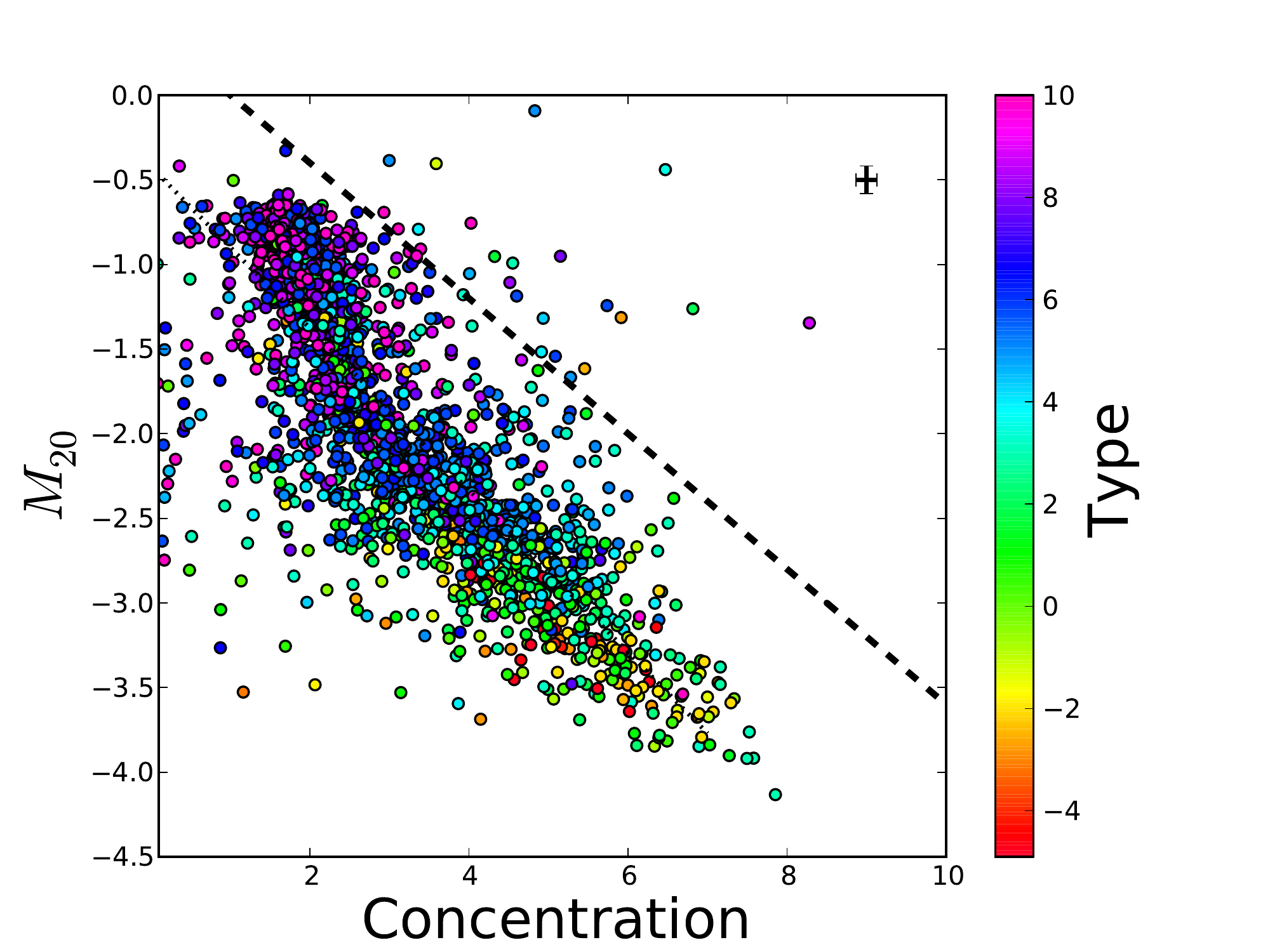}
\includegraphics[width=0.49\textwidth]{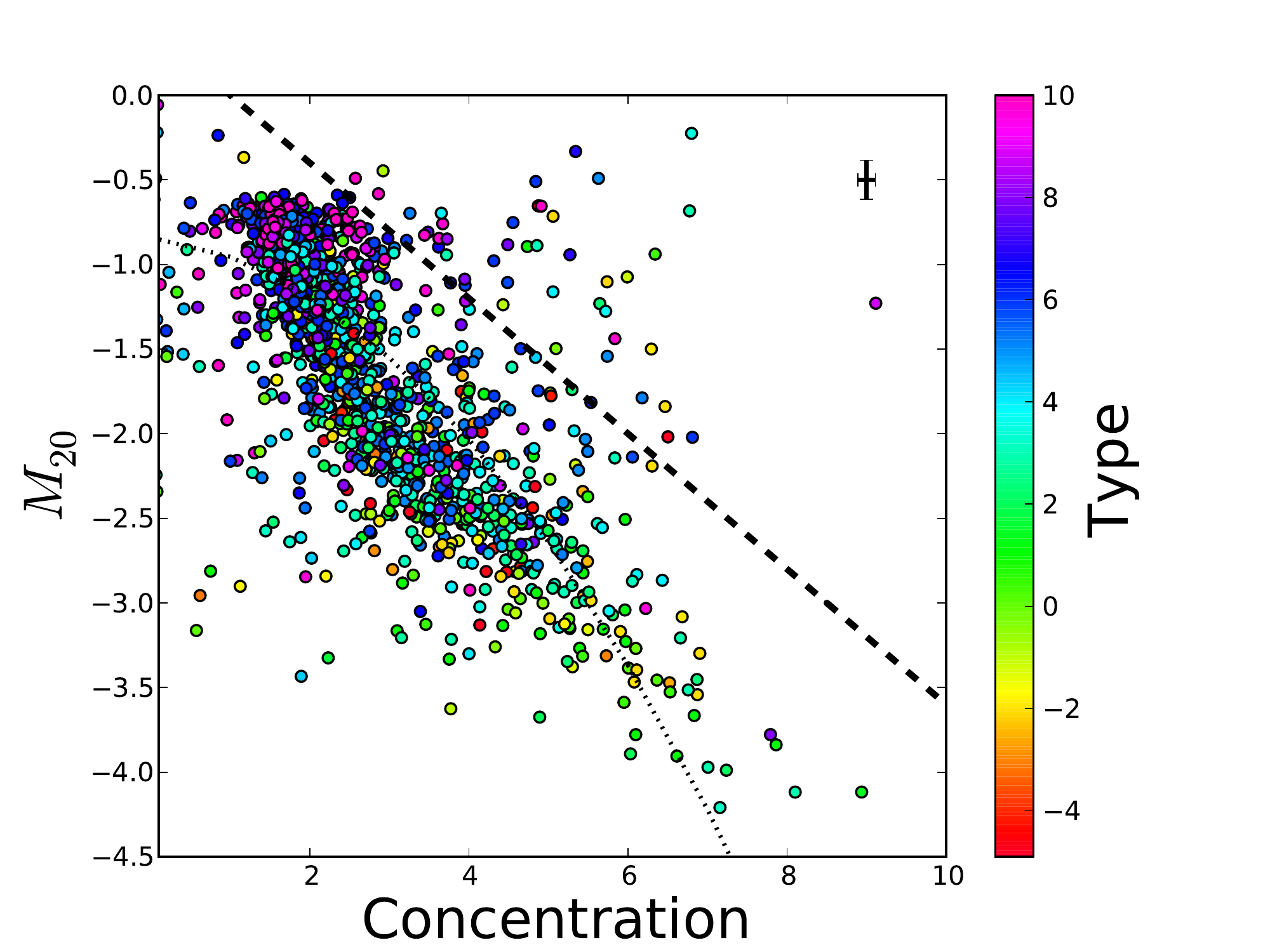}
\caption{The relation between Concentration and \m20\ for 3.6 (left) and 4.5 \mum\ (right) for \s4g\ galaxies. The dashed line is the interaction criterion for optical classification from \protect\cite{Lotz04} (equation \ref{eq:CM20}) and the dotted line the best fit to the Concentration and \m20\ relation (equations \ref{eq:cm20:ch1} and \ref{eq:cm20:ch2}).}
\label{f:CM20}
\end{center}
\end{figure*}

Originally, \cite{Lotz04} introduced the \m20\ parameter as a possible alternative to the concentration parameter from \cite{CAS}. 
The definition of \m20\ does not hinge on the placement of circular or elliptical apertures and is thus more sensitive to ``any bright nuclei, bars,
spiral arms, and off-center star clusters." 
\cite{Munoz-Mateos09a} find a clean relation between  $\rm C_{82}$ \ and \m20 at 3.6 \mum\ for galaxies in 
the SINGS sample that represents a clear sequence of Hubble morphologies. \cite{Scarlata07} also single 
out the concentration ($\rm C_{82}$) and \m20, as well as Gini-\m20 combinations for their discriminatory power.

\begin{figure*}
\begin{center}
\includegraphics[width=0.49\textwidth]{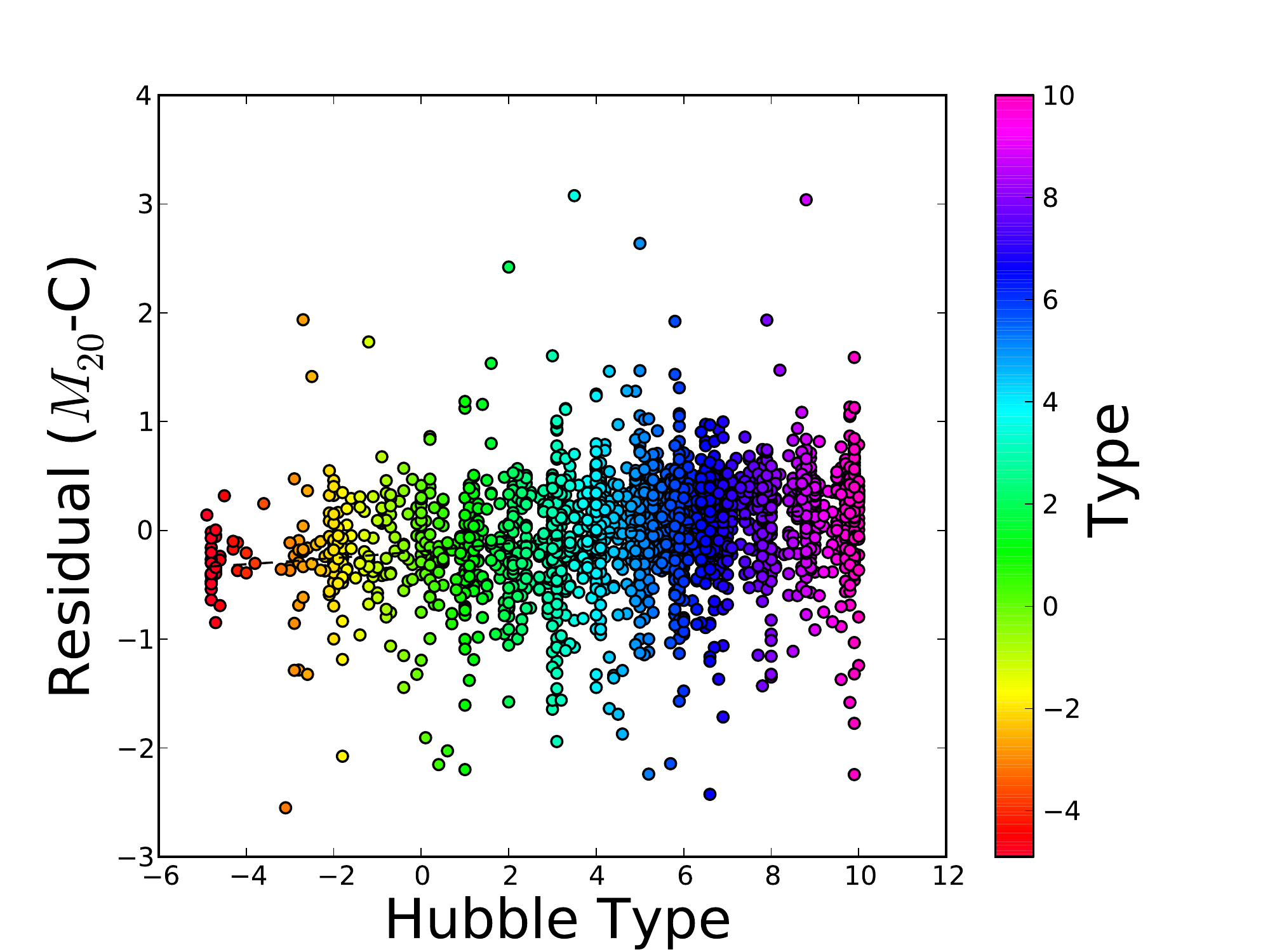}
\includegraphics[width=0.49\textwidth]{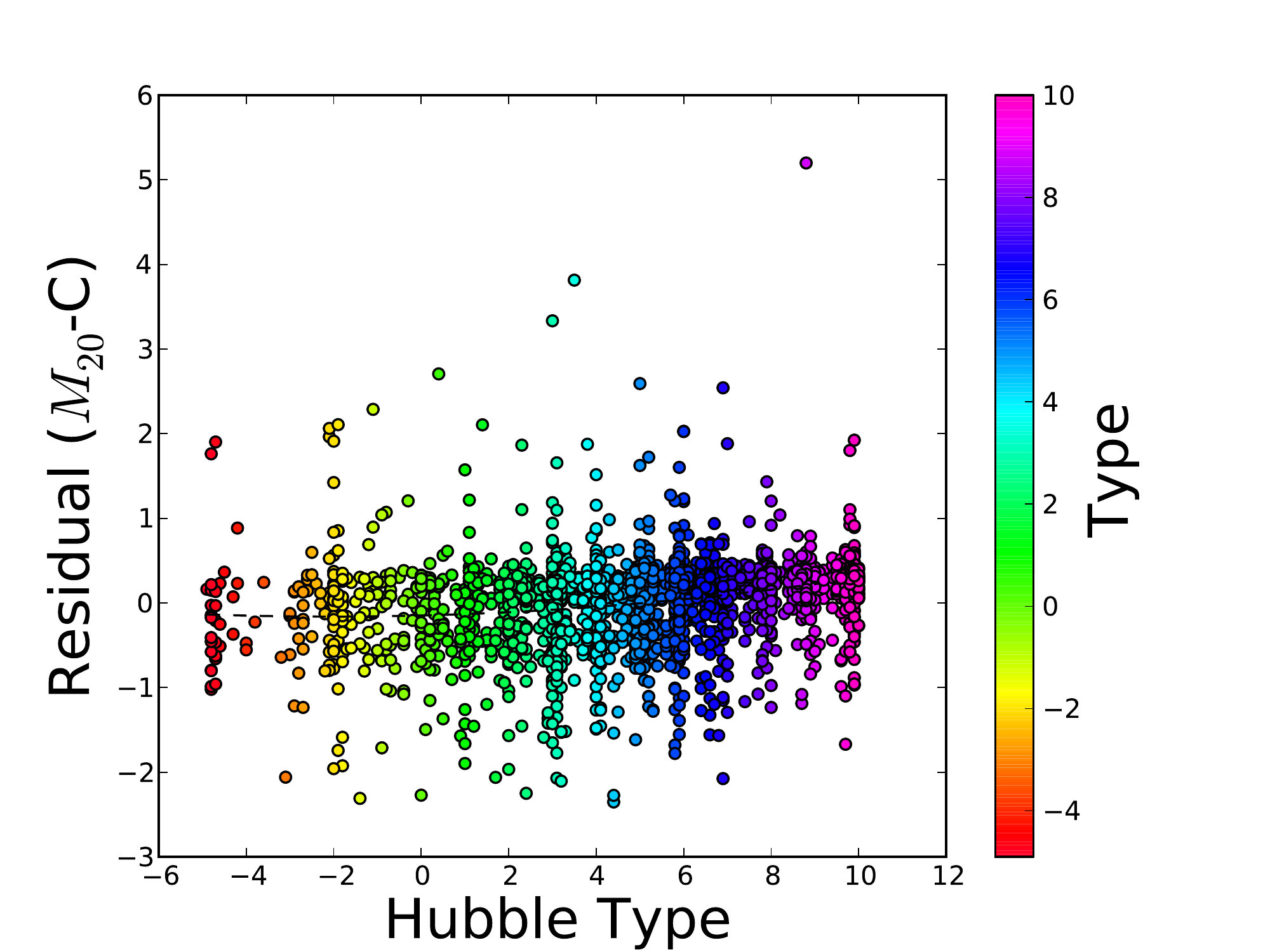}
\caption{The residual at 3.6 (left) and 4.5 \mum\ (right) for \s4g\ galaxies after subtracting the Concentration-\m20 relations in equations \ref{eq:cm20:ch1} (left) and \ref{eq:cm20:ch2} (right panel).}
\label{f:CM20res1}
\end{center}
\end{figure*}

Figure \ref{f:CM20} shows the relation between our $\rm C_{82}$ \ and \m20. Obviously the relation is different from the 
one in \cite{Munoz-Mateos09a} as we use the same definition of concentration but include any central source. 
In particular, the 3.6 $\mu$m relation is much tighter between these two parameters than any other pair, with only 
a few objects that are possibly disturbed galaxies (see below) away from the correlation.
A second order polynomial fit between $\rm C_{82}$ \ and \m20 at 3.6 $\mu$m yields a relation of: 
\begin{equation}
\label{eq:cm20:ch1}
M_{20} = -0.0017 \times  (C_{82})^2 - 0.47 \times C_{82} - 0.43
\end{equation}
\noindent and for 4.5 $\mu$m:
\begin{equation}
\label{eq:cm20:ch2}
M_{20}  = -0.064 \times (C_{82})^2 -0.04 \times  C_{82} - 0.85,
\end{equation}
\noindent after excluding the points above the ``disturbed'' line. Thus the concentration and \m20 at both wavelengths 
are closely related for the majority of the \s4g galaxies. Figure \ref{f:CM20res1} shows the residual as a function of Hubble Type. 

It appears that in the case of the 3.6 and 4.5 $\mu$m images, one can define the normal galaxy sequence of $C_{82}$ 
and \m20, and any galaxy with morphology that deviates from this relation by more than 0.5 can be marked as ``peculiar''. 
In the case of the 4.5 $\mu$m images, there are many more galaxies that would be marked as peculiar by this selection. 

Moreover, we can use the C--\m20 selection to identify ``normal''/unperturbed galaxies, and subsequently classify these using the \m20 parameter.
One potential use of this relation is a check of galaxy models. Typical stellar mass maps should lie on this $C_{82}$--\m20 sequence.

We now fit the relation between the Hubble type from HyperLEDA and the \m20 parameter, after excluding outliers from the Concentration--\m20 relation.
The numerical Hubble type, derived from \m20 only, either at 3.6 or 4.5 $\mu$m, can be expressed as:
\begin{equation}
\label{eq:t-M20:ch1}
T (3.6 \mu m) = -0.57 \times (M_{20})^2 - 0.31 \times M_{20} + 7.91,
\end{equation}
\noindent or 
\begin{equation}
\label{eq:t-M20:ch2}
T (4.5 \mu m)= 0.86 \times (M_{20})^2 - 5.3 \times M_{20} + 10.2,
\end{equation}
\noindent respectively.

\begin{figure}
\begin{center}
\includegraphics[width=0.49\textwidth]{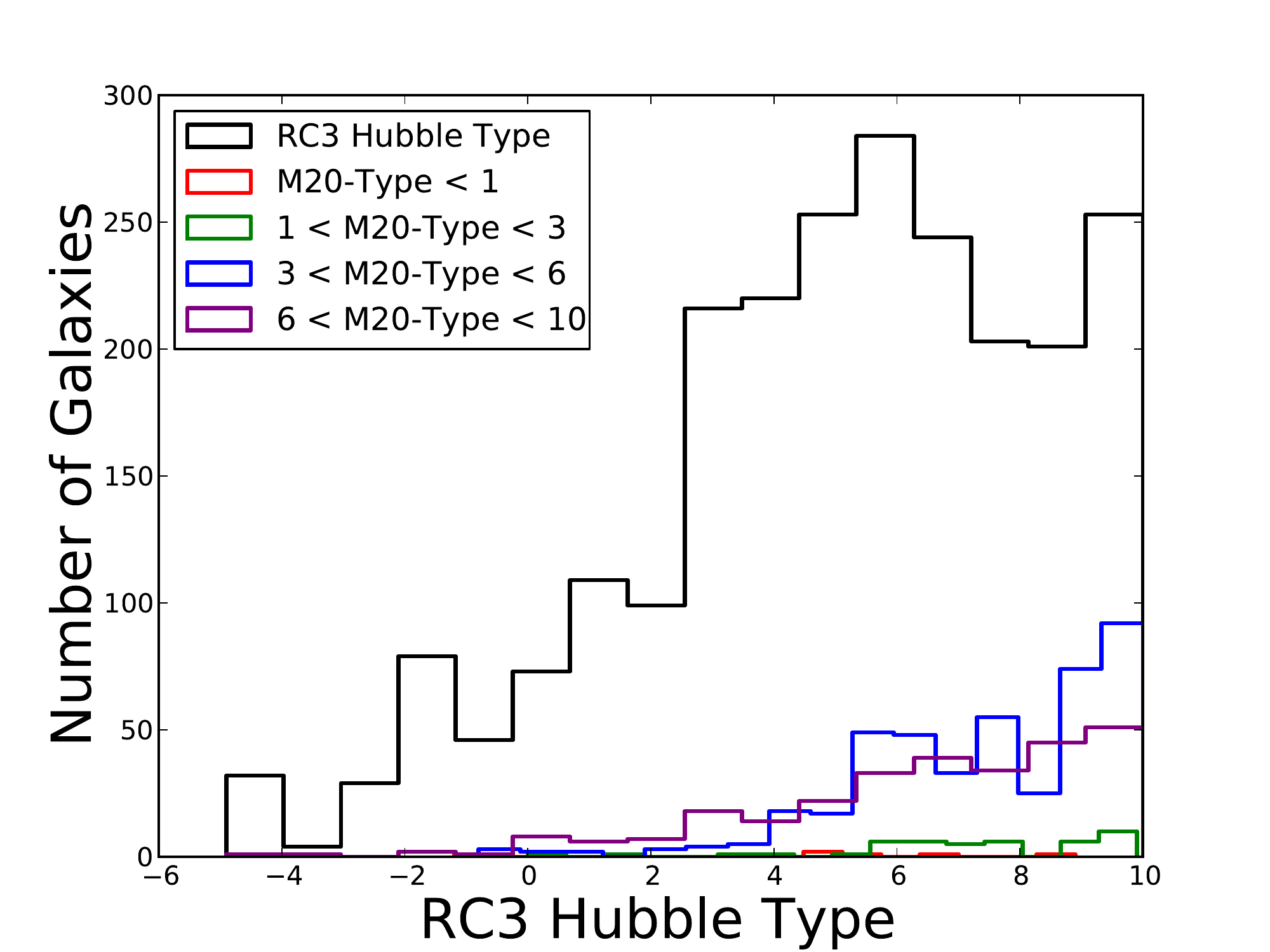}
\caption{The histogram of Hubble types from the RC3 \citep{RC3} for the complete S$^4$G sample and the
distribution of RC3 types for different selections of \m20-derived types, following equation \ref{eq:t-M20:ch1}.
Those galaxies classified as late-type by equation \ref{eq:t-M20:ch1} are in fact late-type according the the RC3 
but a finer distinction cannot be made, i.e., Sa from Sc type spirals.}
\label{f:m20type}
\end{center}
\end{figure}

Figure \ref{f:m20type} shows the Hubble type distribution of \m20-selected sub-samples.
One can retrieve the broad classifications from {\em HyperLEDA}, i.e., late- versus early-type galaxies or irregulars from spirals
but a more detailed distinction cannot be made using this approach alone.

Given the subjective nature of visual galaxy classifications, one could use this automatically-derived Hubble type as an 
alternative to catalogs such as RC3 or {\em HyperLEDA} in future uses of \s4g or in future near-infrared imaging surveys. 
As the 3.6 \mum\ relation has the least scatter, we recommend this band for this broad typing in particular.

\section{Galaxy Classification -- Disturbed Systems}
\label{s:disturb}

There are a few established criteria to select morphologically disturbed galaxies based on these parameters in the literature. 
These are mostly based on visible light data and select galaxies during the first and second passes of a merger and ofttimes 
recent merger remnants as well. Here we compare how well such parameters could be applied to the \s4g near-infrared imaging.

For visible light data, \cite{CAS} define the following criterion:
\begin{equation} 
A > 0.38 ~ {\rm and} ~S>A
\label{eq:AS}
\end{equation}
\noindent In general, they consider any highly asymmetric galaxy as a candidate merging system.
The vast majority of our galaxies is not disturbed according to this criterion and those selected are classified as late-types 
(Figure \ref{f:AS}). The definition of Smoothness fluctuates somewhat but this may be a viable way to select disturbed or 
irregular galaxies. 


\cite{Lotz04} added two different criteria using Gini and $M_{20}$:
\begin{equation} 
G > -0.115 \times M_{20} + 0.384
\label{eq:GM20}
\end{equation}
\noindent and Gini and Asymmetry: 
\begin{equation} 
G > -0.4 \times A + 0.66 ~ \rm or ~ A > 0.4.
\label{eq:GA}
\end{equation}
\noindent The latter is a refinement of the Conselice et al. criterion in equation \ref{eq:AS}.

\begin{figure*}
\begin{center}
\includegraphics[width=0.49\textwidth]{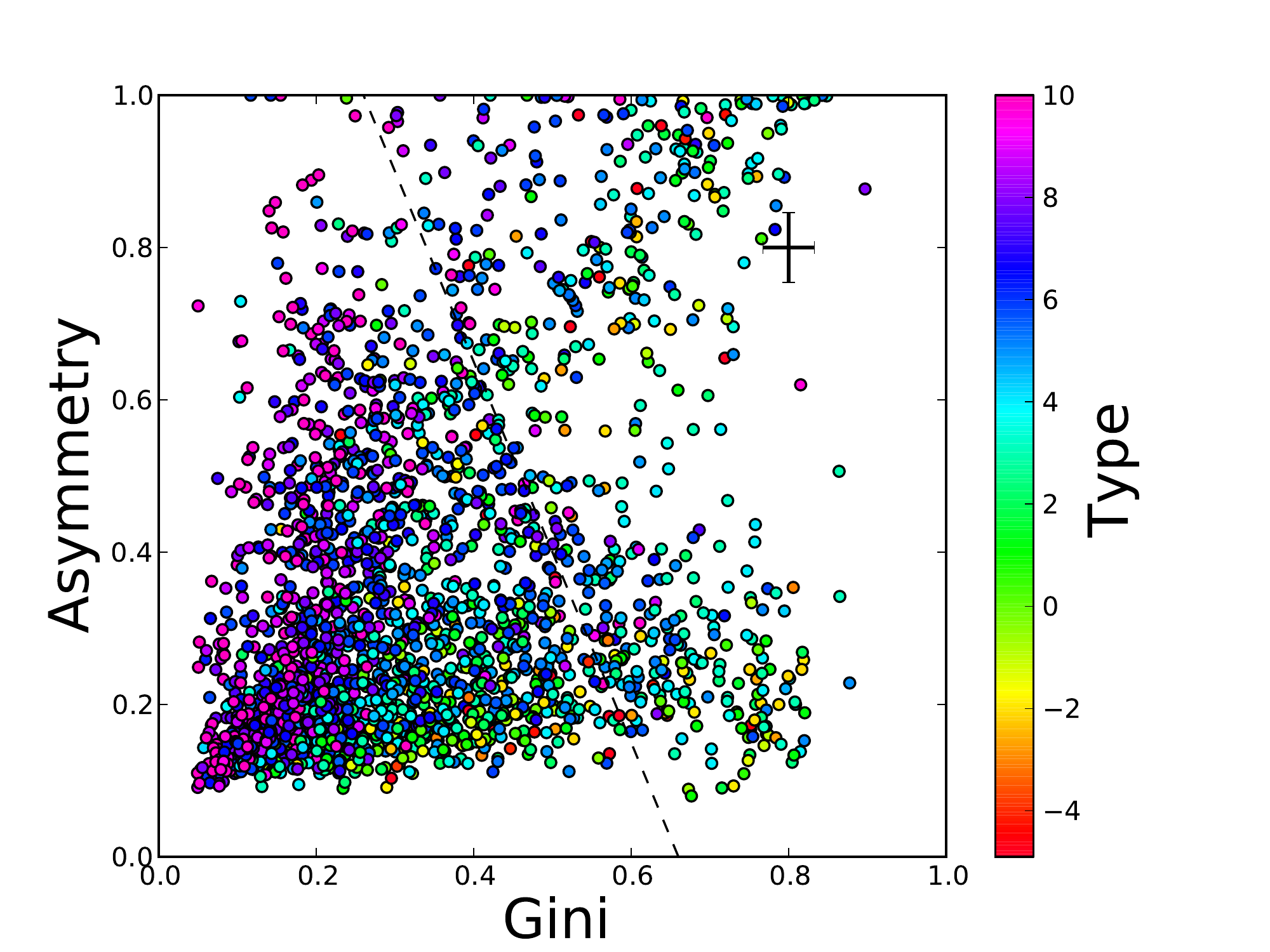}
\includegraphics[width=0.49\textwidth]{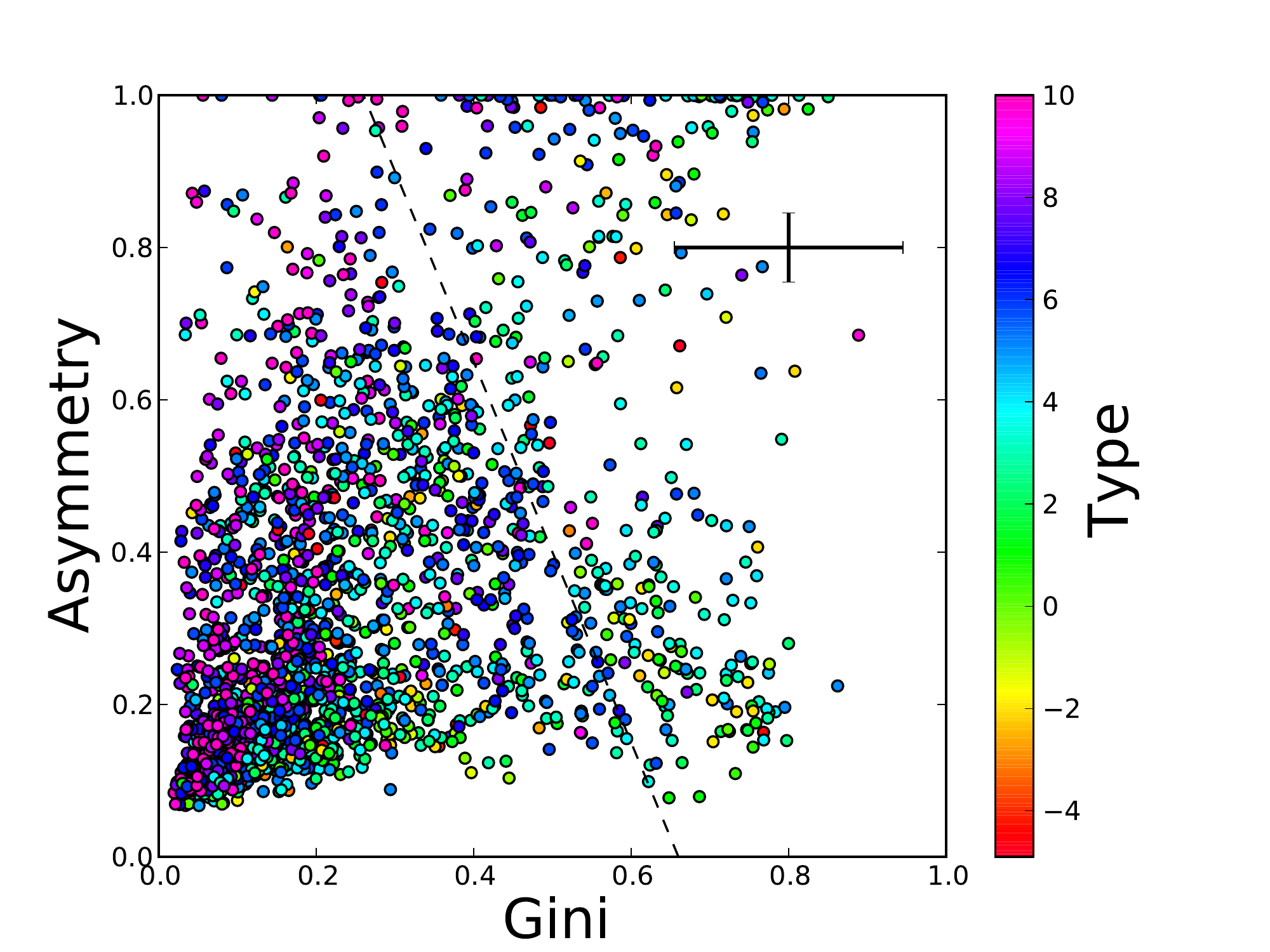}
\caption{The relation between Gini and Asymmetry  for 3.6 (left) and 4.5 \mum\ (right) for \s4g\ galaxies. The dashed line is the interaction criterion from \protect\cite{Lotz04} (equation \ref{eq:GA}). Galaxies to the right and above this line are candidate interactions.}
\label{f:GA}
\end{center}
\end{figure*}

The G-\m20 criterion selects the scatter away from the locus of galaxies which includes a variety of Hubble types 
(Figure \ref{f:GM20}). Slope and normalization will have to be adjusted to select all the disturbed systems. 
The bigger spread in Gini parameters for early types (E and S0) for their given \m20 value is the cause for this. 
These early-type ``disturbed'' systems do indeed include some galaxies with close companions 
(notably NGC 5195, M51's companion) but also some S0s with faint spiral structure and especially 
many S0 galaxies with rings visible in the 3.6 and 4.5 $\mu$m images.
The second criterion, using Gini and Asymmetry, does not seem to be applicable to the \s4g data (Figure \ref{f:GA}).

\cite{Holwerda11b} defined some criteria for 21 cm radio data (H{\sc i}), which has a much lower spatial resolution than \s4g and 
show the atomic hydrogen gas, not the stellar content. They define ``morphologically disturbed'', based on their $G_M$ parameter, as:
\begin{equation} 
G_M > 0.6,
\label{eq:GM}
\end{equation}
\noindent or based on Asymmetry and \m20:
\begin{equation} 
A > -0.2 \times M_{20} + 0.25,
\label{eq:AM20}
\end{equation}
\noindent or concentration and \m20, similar to the criteria from \cite{Lotz04} (equations \ref{eq:GM20} and \ref{eq:GA}). Based on Figure \ref{f:CM20}, we define a C-\m20\ criterion for Spitzer imaging, similar to equation 11 in \cite{Holwerda11c}, as:
\begin{equation} 
C_{82} > -2.5 \times M_{20} + 1 
\label{eq:CM20}
\end{equation}

\begin{figure*}
\begin{center}
\includegraphics[width=0.49\textwidth]{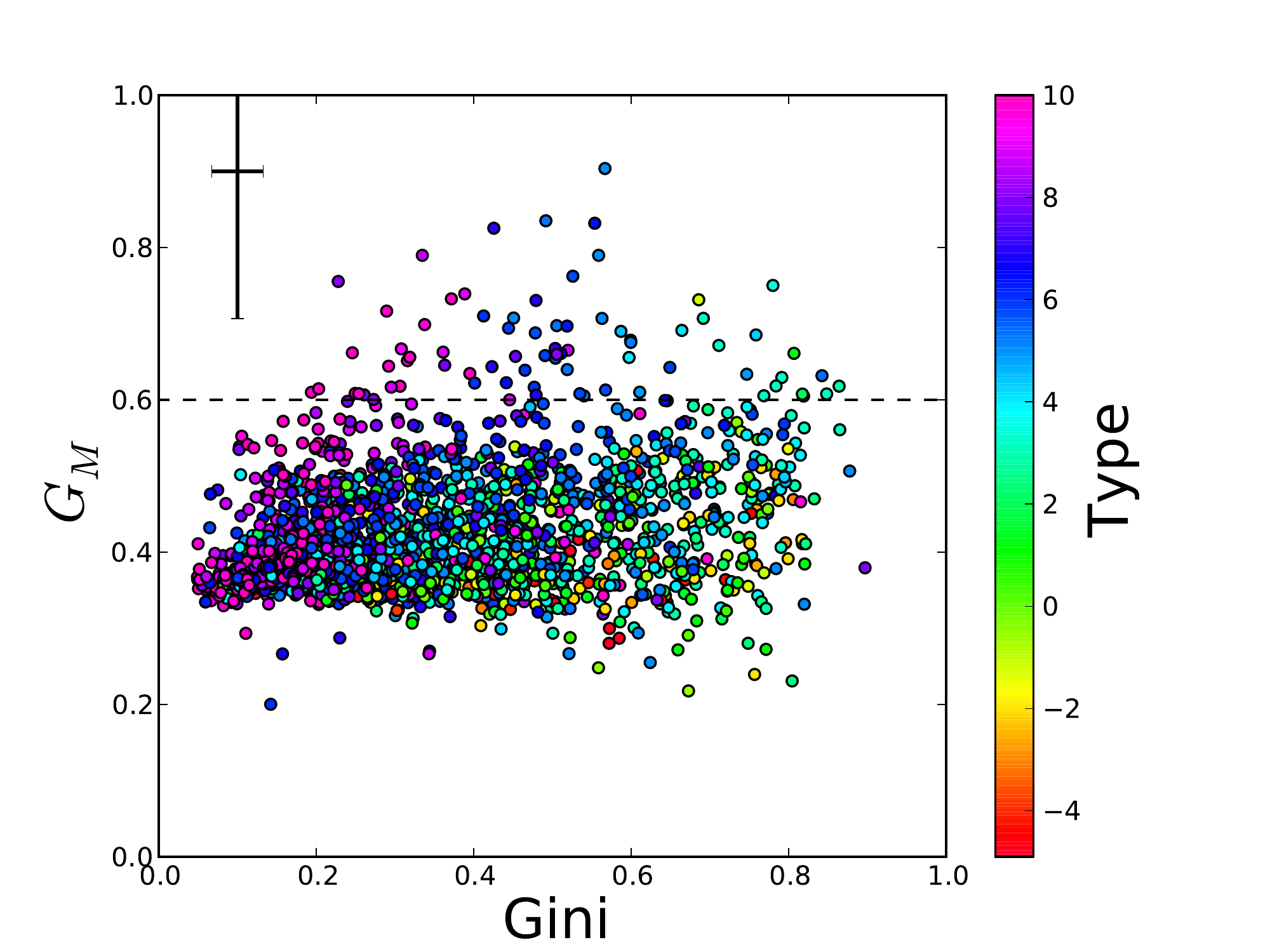}
\includegraphics[width=0.49\textwidth]{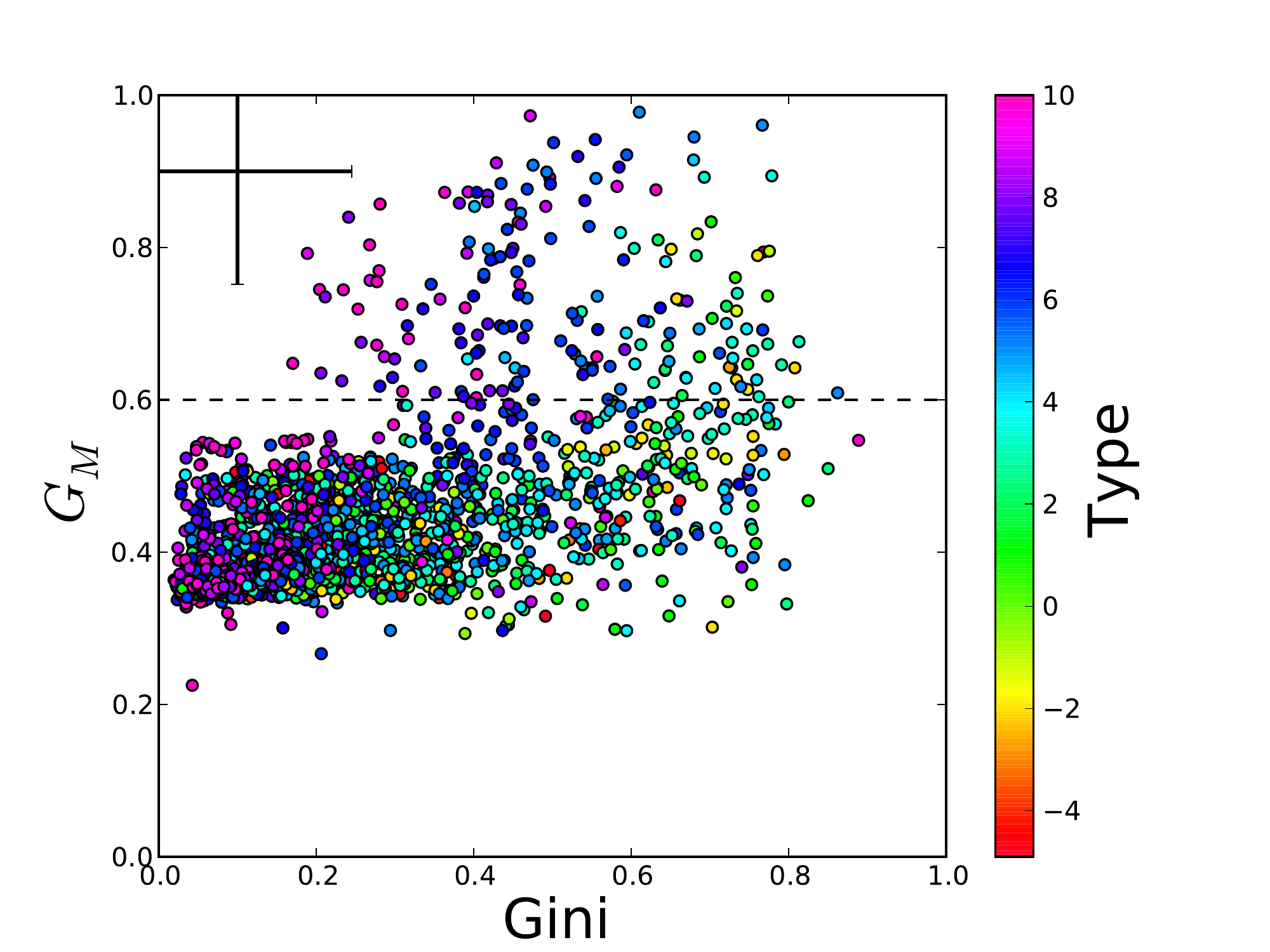}

\caption{The relation between Gini and $G_M$ for 3.6 (left) and 4.5 \mum\ (right) for \s4g\ galaxies. The horizontal dashed line is the interaction criterion for H\,I maps from \protect\cite{Holwerda11c} (equation \ref{eq:GM}). Galaxies above the line are candidate interactions.}
\label{f:GGM}
\end{center}
\end{figure*}

Of these, the $G_M$ and the $C_{82}$--\m20 criteria seem to be applicable to the \s4g data (Figures \ref{f:CM20} and \ref{f:GGM}), in the latter case with a slight renormalization. In the latter criterion's case, 4.5 $\mu$m morphology is more often disturbed than the 3.6 $\mu$m. 
One possibility is in our view that in these galaxies there is a hot dust contribution from HII regions to the global morphology of these disks (see \S \ref{s:clr}). 
This leaves us with four criteria that may well select the outlying ``disturbed'' galaxies. 

\begin{figure}
\begin{center}%
\includegraphics[width=0.5\textwidth]{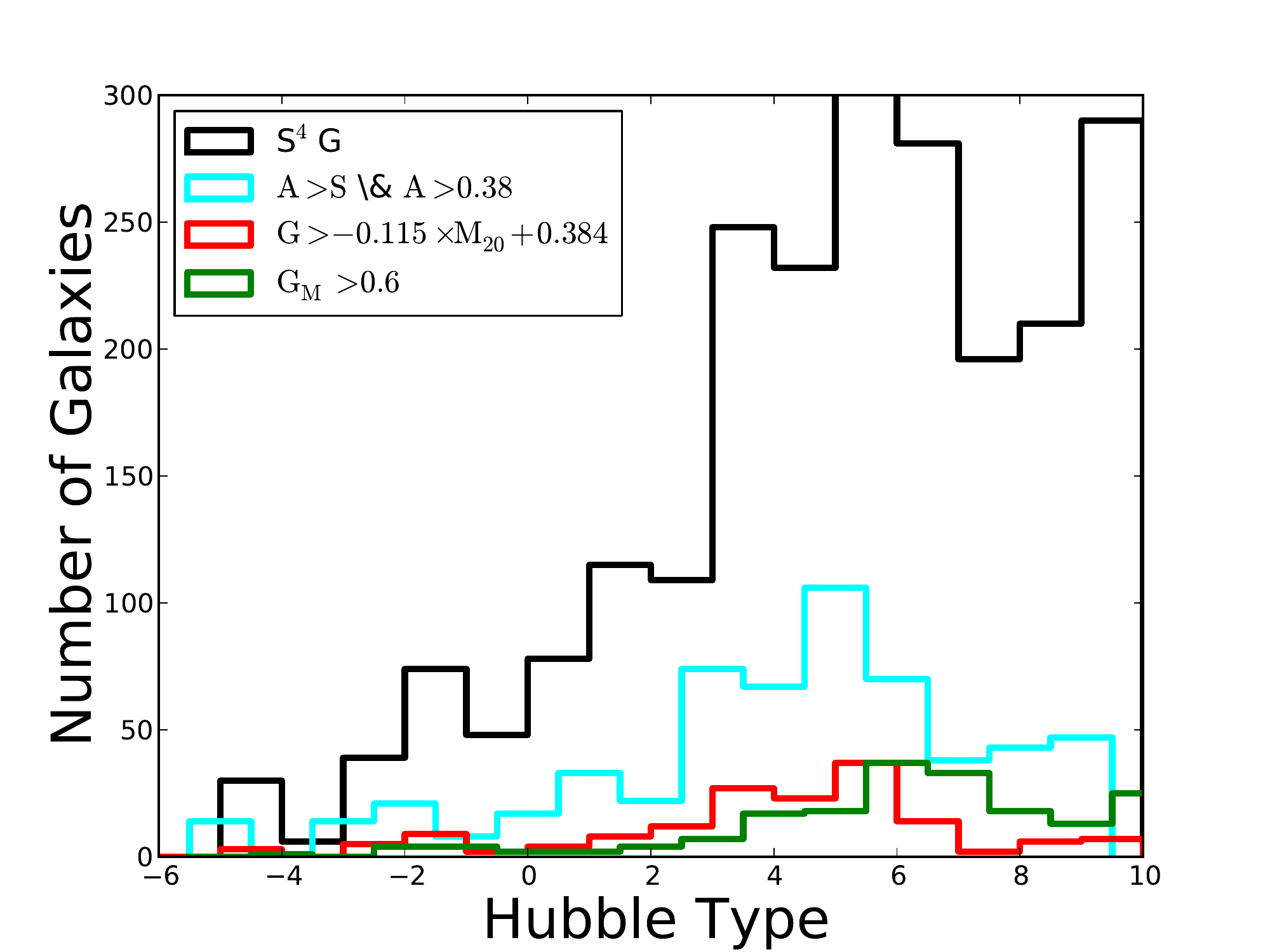}
\caption{The histogram of Hubble types for the full \s4g sample and those galaxies selected by the four criteria as disturbed.
The Asymmetry-Smoothness criterion (equation \ref{eq:AS}) from \protect\cite{CAS}, the Gini-\m20\ criterion (equation \ref{eq:GM20}) from \protect\cite{Lotz04} and the \gm\ (equation \ref{eq:GM}) and Concentration-\m20\ (equation \ref{eq:CM20}) criteria from \protect\cite{Holwerda11c}.}
\label{f:disthis}
\end{center}
\end{figure}

\begin{table}
\caption{Morphological Selection Parameters }
\begin{center}
\begin{tabular}{l l l l l}
Criterion							& 3.6 \mum		& 		& 4.5 \mum  		&	\\
 								& No. 		& \%		& No. 		& \% \\
\hline
\hline
$S > A$ \& $A> 0.38$ 				& 600			& 25.5	& 574			& 24.4 \\ 
$G > -0.115 \times M_{20} + 0.384$		& 166			& 7.1		& 159			& 6.7 \\ 
$G_M > 0.6$						& 76				& 3.2		& 185			& 7.9 \\ 
$C > -2.5 \times M_{20} + 1$  			& 24				& 1.0		& 46 				& 2.0 \\
\hline
\end{tabular}
\end{center}
\label{t:crit}
\end{table}%

\subsection{What kind of galaxies are selected as disturbed?}
\label{s:dgals}

The four different criteria in Table \ref{t:crit} select different Hubble types as ``disturbed''. 
Figure \ref{f:disthis} shows the distribution of Hubble types for the four criteria that seem promising for use on {\em Spitzer} data to identify disturbed galaxies.
The Asymmetry-Smoothness criterion (equation \ref{eq:AS}) selects many more galaxies than the other criteria, with a preference for spirals. The fact that so many galaxies are selected makes this criterion suspect to use for the selection of unusual or interacting systems.
The Gini-\m20 criterion (equation \ref{eq:GM20}) also selects a mix of Hubble types, mostly earlier type spirals (Sb or Sc). As noted, the early-types that are selected appear to be a mix of actually interacting galaxies and S0 galaxies with rings or spiral structure.
The $G_M$ criterion selects predominantly the latest types (Sc and Irr). 
The C-\m20 criterion (equation \ref{eq:CM20}) selects later types as well. A large fraction of these are edge-on spirals or very faint irregular galaxies, as well as interactions, e.g., NGC 5194 (M51A).
We note that \s4g does not include many ellipticals in its selection and that --- as with all morphological selection --- some objects are selected because of image artifacts. In the case of IRAC images, pulldown columns due to a bright nearby star seem to do so from time to time, even though most of these are masked in the pipeline products. A quick visual check of the selected objects reveals some are indeed selected for that reason.

Our sample of \nsamp~ galaxies is the full, volume- and mass-limited, sample of \s4g but the values in Table \ref{t:crit} show that not every criterion translates well to Spitzer IRAC morphologies for  tidally disturbed galaxies defined for other wavelengths. Only the $G_M$ and G-\m20 criteria select a similar fraction of galaxies to be interacting in the local Universe as previous studies \citep[$\sim5$\%][Laine et al. ({\em in preparation})]{Darg09, Knapen09, Holwerda11d}, using techniques such as visual inspection of the SDSS images and morphological selection of disturbed H\,I maps.

These two criteria ($G_M$ and G-\m20) predominantly select later types as ``unusual" so these selection criteria may not work as well for the early-type galaxies observed in the near-infrared that are interacting. Combined with the relative lack of early-types in the \s4g\ sample, we note that these two morphological selections to estimate a merger rate for the later-types from \s4g.
%



\begin{figure*}
\begin{center}
\includegraphics[width=0.49\textwidth]{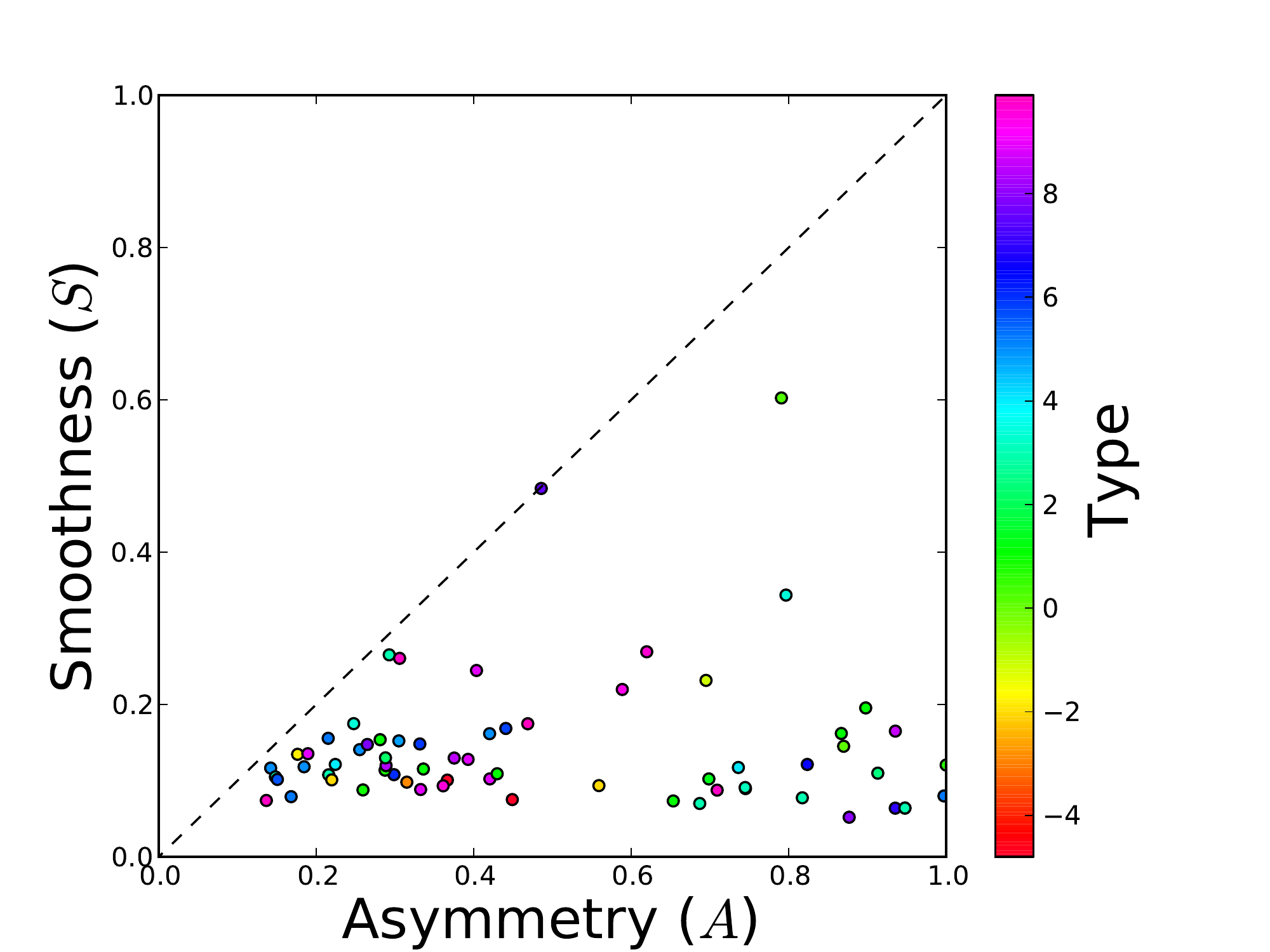}
\includegraphics[width=0.49\textwidth]{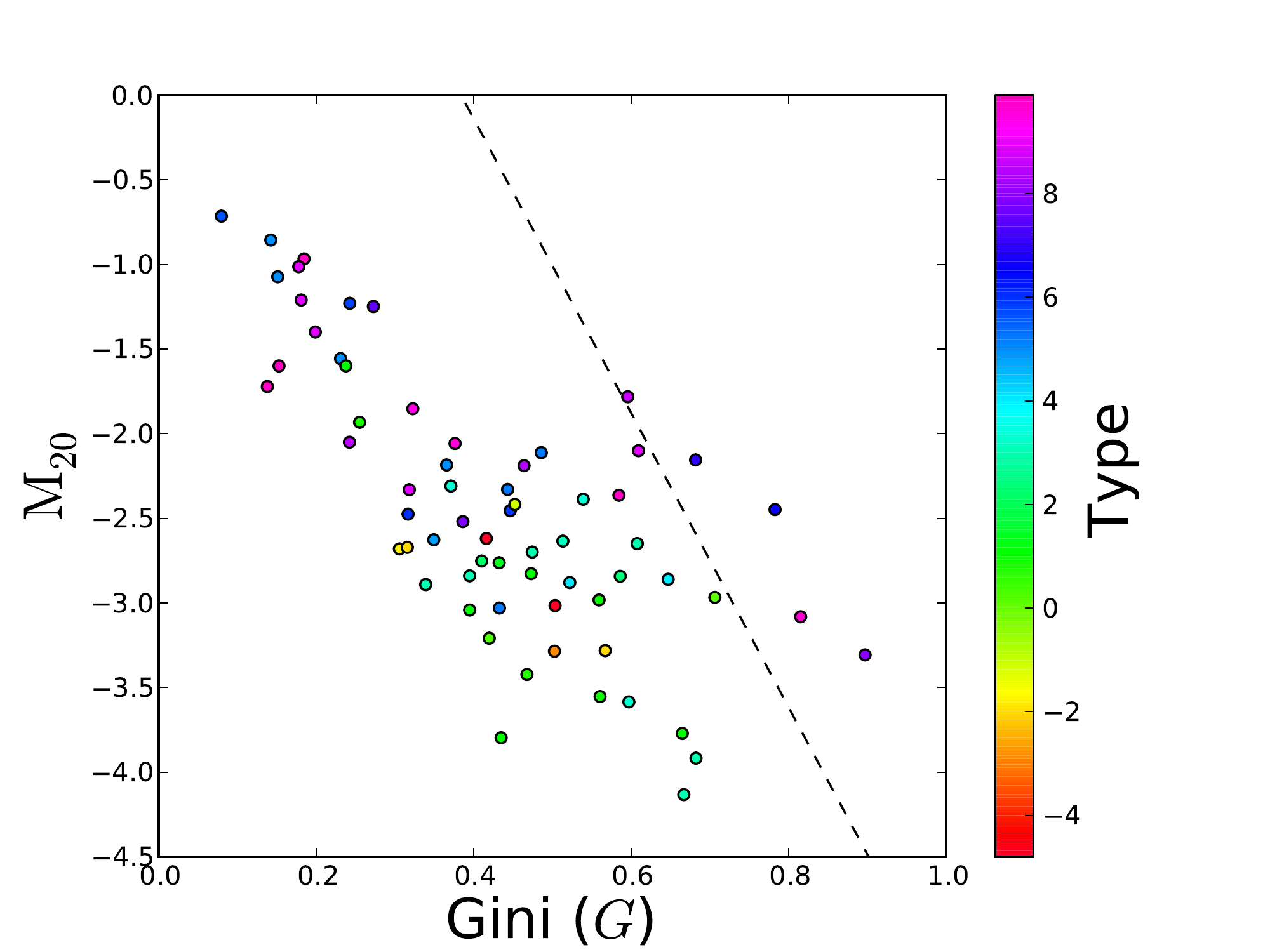}
\includegraphics[width=0.49\textwidth]{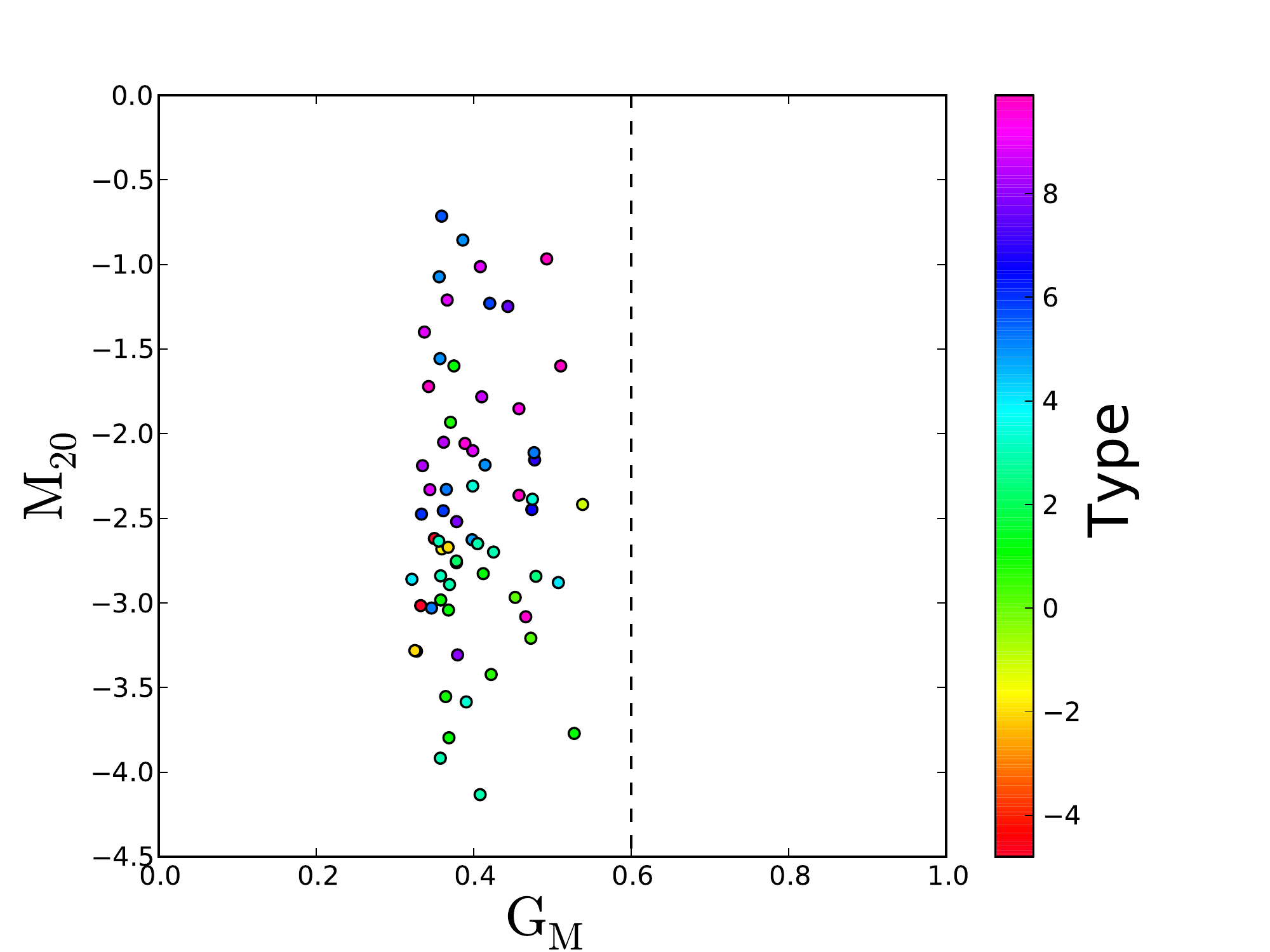}
\includegraphics[width=0.49\textwidth]{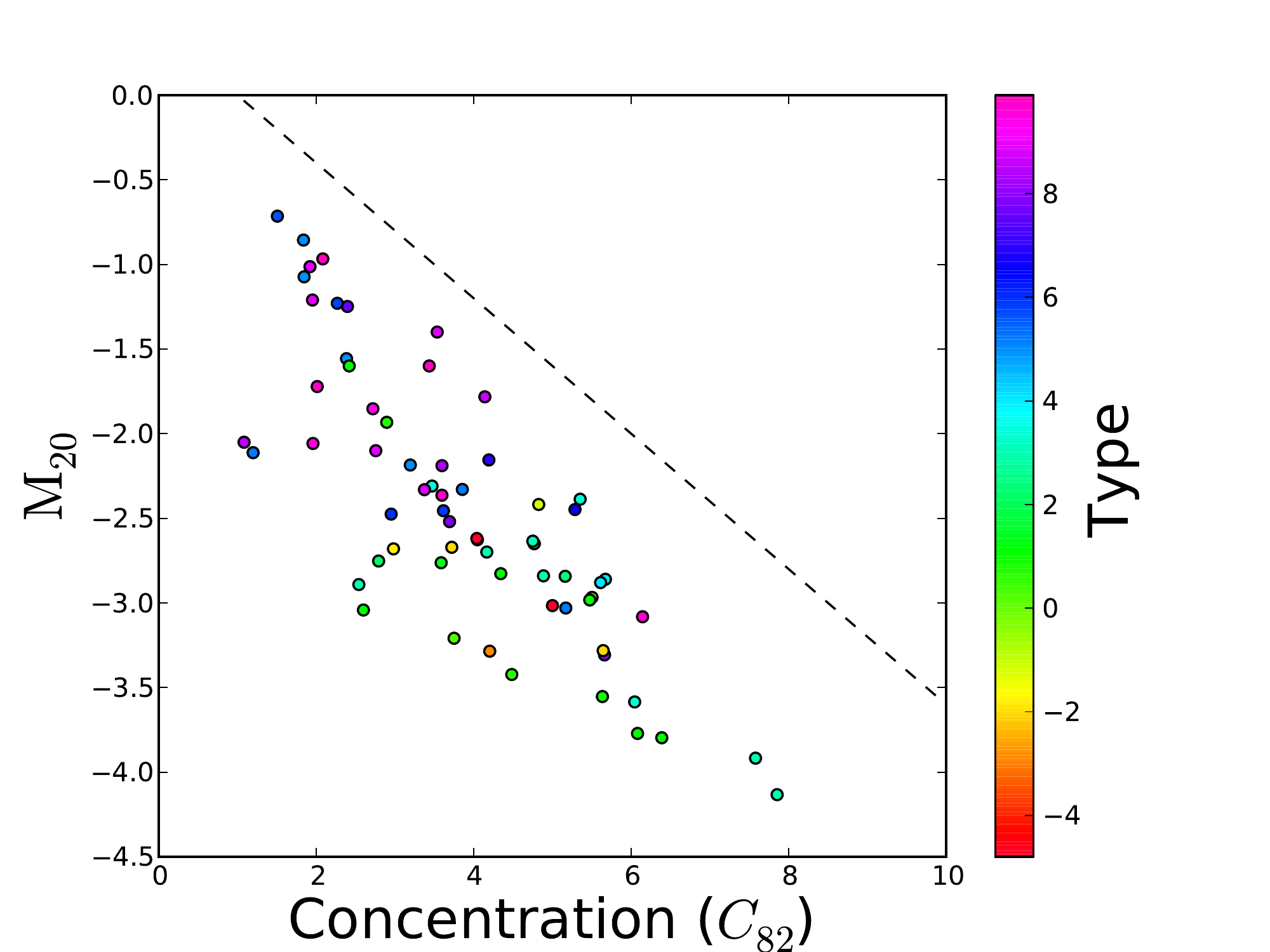}
\caption{The distribution of morphological parameters in 3.6 \mum\ of the 104 Arp galaxies in our sample. 
Dashed lines are the selection criteria (equation \ref{eq:AS}--\ref{eq:GM20}) for disturbed galaxies:
the Asymmetry-Smoothness criterion from \protect\cite{CAS} (equation \ref{eq:AS}).
the Gini-\m20\ criterion from \protect\cite{Lotz04} (equation \ref{eq:GM20}),
the \gm\ and Concentration-\m20\ criteria from \protect\cite{Holwerda11c} (equations \ref{eq:GM} and \ref{eq:CM20} respectively).
Only the Gini-\m20\ criterion selects a sizable number of Arp atlas galaxies based on their \s4g\ morphology. }
\label{f:arp}
\end{center}
\end{figure*}

\subsection{Arp Atlas}
\label{s:arp}

In our \nsamp \ galaxies, there are 104 galaxies out of the 338 in the Arp catalog of peculiar galaxies \citep{Arp66, Arp95}. Figure \ref{f:arp} shows the distributions of these 104 galaxies. The first thing to note is that hardly any are selected by the morphological criteria for disturbed systems, i.e., ``peculiar'' does not equate ``disturbed'' in the quantified morphology sense. This is to caution the use of a selection of galaxies with outlying morphologies. Galaxies with a peculiar appearance in the visible light are not those identified in outlying morphological parameters in the near-infrared. 

And the second thing to note is that while these morphological parameters may contain enough information to approximately morphologically classify and single out tidally disturbed systems, they do not contain enough power to single out galaxies with peculiar properties. That remains in the scope of visual classification (Laine et al. {\em in preparation}).


\begin{figure*}
\begin{center}
\includegraphics[width=0.49\textwidth]{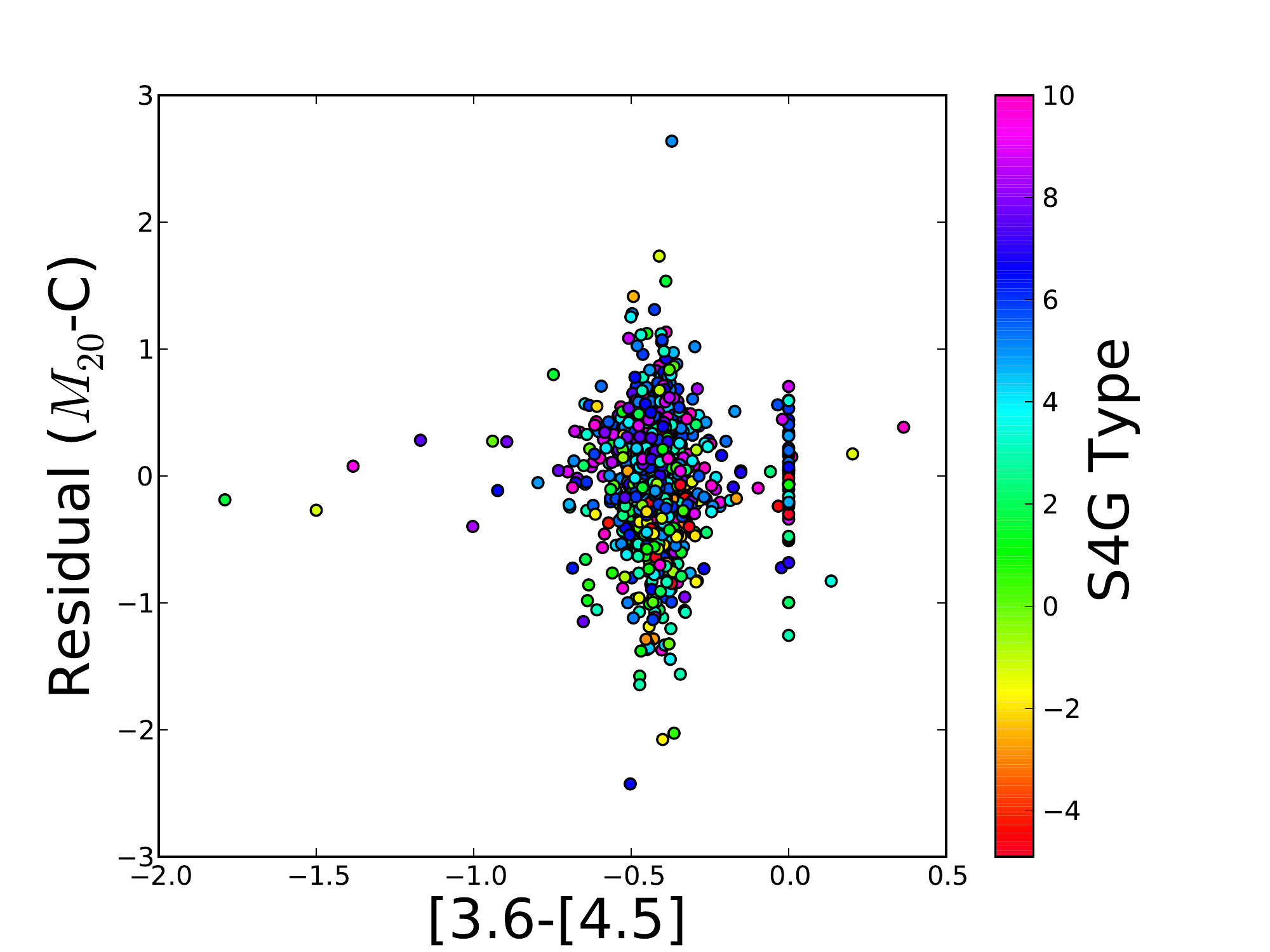}
\includegraphics[width=0.49\textwidth]{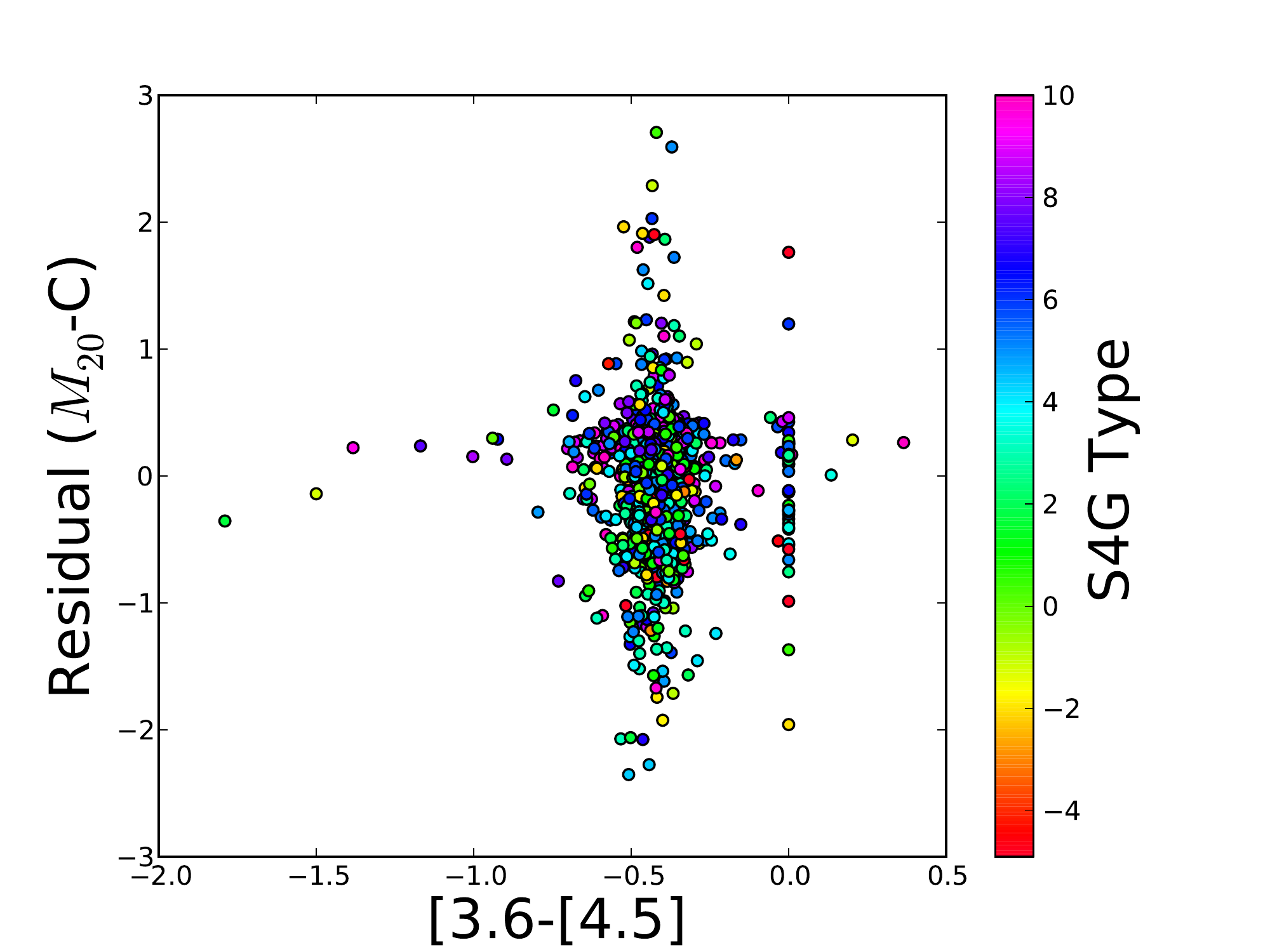}

\caption{The residual after subtracting the Concentration-\m20\ relations in equations \ref{eq:cm20:ch1} at 3.6 \mum\ 1 (left) and \ref{eq:cm20:ch2} for 4.5 \mum\ (right panel) as a function of global galaxy color from Munoz-Mateos et al. ({\em in preparation}). Points are color-coded by HyperLEDA type \citep{hyperleda}. The color range for the \s4g\ is galaxies is very narrow. Those with extreme colors have few outliers and those with great residuals, a typical color.}
\label{f:CM20res2}
\end{center}
\end{figure*}

\subsection{[3.6--4.5] Color}
\label{s:clr}

One option that may influence the difference in the Concentration-\m20\ relations between the 3.6 and 4.5 \mum\ images is a contribution from the PAH feature at 3.1 \mum\ to the 3.6 \mum\ images or relatively brighter hot dust emission in the 4.5 \mum\ channel. Figure \ref{f:CM20res2} shows the residual from the C-\m20\ fit as a function of global [3.6--4.5] color from Munoz-Mateos et al. ({\em in preparation}). The majority of \s4g\ galaxies lie in a narrow range of color [3.6--4.5] = -0.7 to -0.3 with a mean color of -0.427.  The most extreme colors paradoxically have the smallest residuals and the morphological outliers (residual $>$1.5) have typical colors. 
Table \ref{t:spear} also lists the correlation between the morphological parameters and the [3.6--4.5] color. The Spearman values are as close to unrelated as one can expect.

In our opinion, this points to a scenario where one or more bright HII regions or other features, likely at higher radii, displace the galaxy from the C-\m20\ morphological relation but not from the typical [3.6--4.5] color, i.e., there is not enough hot dust or PAH emission in the bright HII regions to change the galaxy-wide color but enough flux (at a greater distance to the center) to change the appearance.



\begin{table*}
\begin{center}
\caption{The Spearman ranking of the relations between the Lopsidedness parameterizations from \cite{Zaritsky13b}) and the morphological parameters in 3.6 \mum\ and 4.5 \mum\ for the available subsample of 187 galaxies. The absolute z-values of significance for each of Spearman rankings are noted between parentheses.}
\begin{tabular}{l l l l l l l l}
Lopsidedness	   & C	 & A & S & $M_{20}$ & G & $G_M$ \\
\hline
\hline
3.6 $\mu$m			& & & & & & \\
\hline
$\langle A_{1}\rangle_{i}$       & -0.30 (3.86)         & -0.03 (0.39)         & 0.12 (1.50)   & 0.35 (4.45)   & -0.22 (2.81)         & 0.03 (0.34) \\ 
$\langle A_{2}\rangle_{i}$       & -0.31 (3.99)         & -0.20 (2.48)         & 0.09 (1.15)   & 0.29 (3.61)   & -0.27 (3.40)         & -0.02 (0.31) \\ 
$\langle A_{1}\rangle_{o}$       & -0.21 (2.61)         & -0.11 (1.37)         & 0.19 (2.35)   & 0.18 (2.30)   & -0.19 (2.31)         & 0.03 (0.41) \\ 
$\langle A_{2}\rangle_{o}$       & -0.48 (6.38)         & -0.16 (1.93)         & 0.16 (2.03)   & 0.39 (5.09)   & -0.40 (5.17)         & 0.00 (0.05) \\ 
\hline
\hline
4.5 $\mu$m			& & & & & & \\
\hline
$\langle A_{1}\rangle_{i}$       & -0.22 (2.76)         & -0.10 (1.26)         & 0.02 (0.22)   & 0.22 (2.72)   & -0.18 (2.18)         & -0.16 (1.96) \\ 
$\langle A_{2}\rangle_{i}$       & -0.24 (2.94)         & -0.20 (2.47)         & 0.01 (0.18)   & 0.13 (1.58)   & -0.17 (2.14)         & -0.19 (2.39) \\ 
$\langle A_{1}\rangle_{o}$       & -0.12 (1.52)         & -0.09 (1.16)         & 0.09 (1.16)   & 0.07 (0.86)   & -0.10 (1.21)         & -0.10 (1.26) \\ 
$\langle A_{2}\rangle_{o}$       & -0.38 (4.94)         & -0.25 (3.19)         & -0.05 (0.65)         & 0.32 (4.13)   & -0.32 (4.04)         & -0.18 (2.23) \\ 
\hline
\hline
\end{tabular}
\end{center}
\label{t:lop}
\end{table*}%


\section{Lopsidedness}
\label{s:lop}

Morphological lopsidedness is a distinct displacement of the disk with respect to its apparent center (photometric or kinematic). The effect was initially noticed in HI and then in stellar disks \citep[see for a comprehensive review,][]{Jog09}. The first comprehensive study on stellar disks was by \cite{Rix95} and a study on \s4g\ was just completed \citep{Zaritsky13b}. The motivation for this analysis is very similar to our own: the advent of large surveys, ample computing power, and the desire for more reproducible results in morphological studies.

\cite{Zaritsky13b} perform an azimuthal Fourier decomposition of the luminosity in circular annuli at two radial intervals on the \s4g data. 
Similar to \cite{Rix95}, they calculate the relative strength of the first and second Fourier component: $\langle A_1 \rangle$ is the average of $A_1/A_0$ and $\langle A_2 \rangle$ is the average of $A_2/A_0$. $A_m$ is the amplitude of the {\em m} mode (m=0-4) in the image. The m=0 mode corresponds to the central amplitude (concentration of flux in the center), the m=1 mode to a displacement of the flux in one direction with respect to the center, i.e., lopsidedness, and m=2 is a axisymmetric displacement of flux with respect to the center, e.g., a strong bar.

\cite{Zaritsky13b} report these values calculated between 1.5 to 2.5 disk scale-lengths (\Aonetwo\ \& \Atwotwo), and 2.5 to 3.5 scale lengths (\Aonethree\ \& \Atwothree) in the \s4g\ data. The deep \s4g\ data allows the additional measurement at larger radii compared to the earlier studies. The $m=1$ modes (\Aonetwo\ and \Aonethree) trace the lopsidedness of the disk. 

There is only a weak link in H\,I between the C, A, S, Gini, \m20\ and \gm\ parameters and the presence of lopsidedness \citep{Holwerda11c}. 
However, the new morphology catalog and the lopsidedness parameterization for an \s4g\ subsample, allows us to compare the relation between the strength of lopsidedness and the morphology parameters. Figures \ref{f:A12}, \ref{f:A13}, \ref{f:A22}, \ref{f:A23} show our catalog color coded with the m=1 and m=2 modes at both radii. They illustrate that the lopsidedness sample does not cover the full morphological parameter space, with only few points above the traditional interacting criteria, e.g., equation \ref{eq:GA} or \ref{eq:GM}.
Figures \ref{f:A12:6par} and \ref{f:A22:6par} show the direct relation between the m=1 and m=2 modes with the morphology parameters, for the inner and \ref{f:A13:6par} and \ref{f:A23:6par} for the outer ring respectively. Table \ref{t:lop} lists the Spearman ranking of all our parameters with the m=1 and m=2 modes at both radii for both wavelengths. 
Between lopsidedness (m=2) at either radii and most of our parameters, there is only a weak correlation or none at all.
The strongest anti-correlation is between \Aonetwo\ and Concentration and strongest correlation is between \Aonetwo\ and \m20. The concentration and \m20\ parameters are equally strongly correlated to the m=2 mode. The lack of a strong relation between the morphology parameters and the Fourier components shows that Asymmetry and lopsidedness are not related phenomena, i.e., an asymmetric galaxy need not be lopsided and a lopsided one not strongly asymmetric.

\section{Concluding Remarks}
\label{s:concl}

Based on 3.6 and 4.5 \mum\ images of \nsamp \ galaxies from the \s4g survey, we can conclude that:

\begin{enumerate}
\item There is a close relation for normal galaxies between their concentration ($C_{82}$) and \m20 (Figure \ref{f:CM20}), at 3.6 $\mu$m images.
\item To first order, a Hubble type can be found from \m20 (or $C_{82}$): samples of early and late types could be identified using \m20~ alone (Figure \ref{f:m20type}) but sub-type classification is impossible. 
\item Four morphological criteria work to identify ``disturbed'' or unique systems (Table \ref{t:crit}), but each selects a different subgroup of our sample. Only the \gm\ and G--\m20 criteria select close to the typical merger fraction of the local Universe. The C--\m20 criterion provides a lower limit for the interaction fraction.
\item General morphological type, i.e., early vs late, can be inferred from \m20 parameter in 3.6 $\mu$m but not finer than that. 
\item The lack of a relation between the Concentration--\m20\ residual and global galaxy color points to distinct substructures causing the residual, not global hot dust or PAH contributions.
\item There is only a weak link between Concentration and \m20\ and lopsidedness in a subsample (Table \ref{t:lop}) and these parameters are not suited to detection of this phenomenon.
\end{enumerate}

\section*{Acknowledgements}

The authors thank the entire \s4g team for their efforts in this project and the anonymous referee for his or her insightful comments for the improvement of the manuscript. 
The lead author thanks the European Space Agency for the support of the Research Fellowship program.
This work was co-funded under the Marie Curie Actions of the European Commission (FP7-COFUND).
We acknowledge financial support to the DAGAL network from the People Programme (Marie Curie Actions) of the European Union's Seventh Framework Programme FP7/2007-2013/ under REA grant agreement number PITN-GA-2011-289313.
%
E.A. and A.B. acknowledge the CNES (Centre National d'Etudes Spatiales - France) for financial support. They also acknowledge the support from the FP7 Marie Curie Actions of the European Commission, via the Initial Training Network DAGAL under REA grant agreement n¡ 289313.
This work is based on observations made with the {\em Spitzer Space Telescope}, which is operated by the Jet Propulsion Laboratory, California Institute of Technology under a contract with NASA.
This research has made use of the NASA/IPAC Extragalactic Database (NED) which is operated by the Jet Propulsion Laboratory, California Institute of Technology, under contract with the National Aeronautics and Space Administration. 
This research has made use of NASA's Astrophysics Data System.
We acknowledge the usage of the HyperLEDA database \citep[\protect\url{http://leda.univ-lyon1.fr}][]{hyperleda}.


\newpage

\appendix
\begin{table*}
\section{Tables of Morphological Parameters}
\caption{The morphological parameters at 3.6 $\mu$m for the \nsamp \ $S^4G$ galaxies. Full table is available in the {\em electronic edition}.}
\setcounter{table}{1}
\begin{center}
\begin{tabular}{llllllllllll}
Name	& Gini & $M_{20}$  & C82 &  A   & S & E & $G_M$  \\ 

\hline
\hline

ESO011-005 & 0.306 $\pm$ 0.009 	 & -2.359 $\pm$ 0.082 	 & 2.190 $\pm$ 0.108 	 & 1.206 $\pm$ 0.042 	 & 0.741 $\pm$ 0.017 	 & 0.068 $\pm$ 0.010 	 & 0.432 $\pm$ 0.004 \\ 
ESO012-010 & 0.372 $\pm$ 0.016 	 & -1.959 $\pm$ 0.091 	 & 2.879 $\pm$ 0.177 	 & 0.710 $\pm$ 0.136 	 & 0.332 $\pm$ 0.035 	 & 0.421 $\pm$ 0.014 	 & 0.438 $\pm$ 0.014 \\ 
ESO012-014 & 0.208 $\pm$ 0.009 	 & -0.906 $\pm$ 0.054 	 & 1.451 $\pm$ 0.082 	 & 1.347 $\pm$ 0.035 	 & 0.850 $\pm$ 0.033 	 & 0.041 $\pm$ 0.026 	 & 0.532 $\pm$ 0.005 \\ 
ESO013-016 & 0.476 $\pm$ 0.005 	 & -1.993 $\pm$ 0.105 	 & 3.734 $\pm$ 0.177 	 & 0.898 $\pm$ 0.095 	 & 0.477 $\pm$ 0.035 	 & 0.070 $\pm$ 0.028 	 & 0.530 $\pm$ 0.005 \\ 
ESO015-001 & 0.316 $\pm$ 0.006 	 & -2.281 $\pm$ 0.145 	 & 1.025 $\pm$ 0.223 	 & 1.111 $\pm$ 0.152 	 & 0.713 $\pm$ 0.046 	 & 0.084 $\pm$ 0.024 	 & 0.515 $\pm$ 0.004 \\ 
ESO026-001 & 0.447 $\pm$ 0.007 	 & -2.108 $\pm$ 0.140 	 & 3.096 $\pm$ 0.145 	 & 1.035 $\pm$ 0.067 	 & 0.456 $\pm$ 0.025 	 & 0.042 $\pm$ 0.017 	 & 0.436 $\pm$ 0.007 \\ 
ESO027-001 & 0.637 $\pm$ 0.006 	 & -2.550 $\pm$ 0.103 	 & 4.374 $\pm$ 0.289 	 & 1.783 $\pm$ 0.044 	 & 0.252 $\pm$ 0.045 	 & 0.250 $\pm$ 0.026 	 & 0.490 $\pm$ 0.006 \\ 
ESO027-008 & 0.631 $\pm$ 0.006 	 & -2.671 $\pm$ 0.080 	 & 3.768 $\pm$ 0.331 	 & 1.859 $\pm$ 0.011 	 & 0.296 $\pm$ 0.039 	 & 0.323 $\pm$ 0.018 	 & 0.507 $\pm$ 0.005 \\ 
ESO048-017 & 0.263 $\pm$ 0.008 	 & -1.741 $\pm$ 0.096 	 & 1.914 $\pm$ 0.122 	 & 1.235 $\pm$ 0.028 	 & 0.756 $\pm$ 0.024 	 & 0.037 $\pm$ 0.014 	 & 0.468 $\pm$ 0.003 \\ 
ESO054-021 & 0.440 $\pm$ 0.010 	 & -2.018 $\pm$ 0.083 	 & 3.614 $\pm$ 0.125 	 & 0.840 $\pm$ 0.070 	 & 0.330 $\pm$ 0.038 	 & 0.238 $\pm$ 0.026 	 & 0.491 $\pm$ 0.009 \\ 
ESO079-003 & 0.848 $\pm$ 0.004 	 & -1.866 $\pm$ 0.313 	 & 4.457 $\pm$ 0.449 	 & 1.998 $\pm$ 0.001 	 & 0.138 $\pm$ 0.035 	 & 0.128 $\pm$ 0.075 	 & 0.608 $\pm$ 0.003 \\ 
ESO079-005 & 0.504 $\pm$ 0.008 	 & -1.702 $\pm$ 0.166 	 & 1.564 $\pm$ 0.364 	 & 0.881 $\pm$ 0.178 	 & 0.368 $\pm$ 0.069 	 & 0.186 $\pm$ 0.036 	 & 0.667 $\pm$ 0.005 \\ 
ESO079-007 & 0.419 $\pm$ 0.006 	 & -2.501 $\pm$ 0.097 	 & 3.327 $\pm$ 0.190 	 & 0.926 $\pm$ 0.150 	 & 0.543 $\pm$ 0.038 	 & 0.066 $\pm$ 0.019 	 & 0.479 $\pm$ 0.003 \\ 
ESO085-014 & 0.446 $\pm$ 0.005 	 & -1.334 $\pm$ 0.093 	 & 1.107 $\pm$ 0.129 	 & 1.869 $\pm$ 0.017 	 & 0.538 $\pm$ 0.037 	 & 0.277 $\pm$ 0.021 	 & 0.600 $\pm$ 0.003 \\ 
ESO085-030 & 0.734 $\pm$ 0.008 	 & -2.731 $\pm$ 0.291 	 & 6.022 $\pm$ 0.360 	 & 0.380 $\pm$ 0.515 	 & 0.181 $\pm$ 0.058 	 & 0.120 $\pm$ 0.036 	 & 0.570 $\pm$ 0.006 \\ 
ESO085-047 & 0.199 $\pm$ 0.005 	 & -1.124 $\pm$ 0.028 	 & 1.804 $\pm$ 0.037 	 & 0.794 $\pm$ 0.030 	 & 0.404 $\pm$ 0.008 	 & 0.021 $\pm$ 0.005 	 & 0.394 $\pm$ 0.003 \\ 
ESO107-016 & 0.183 $\pm$ 0.006 	 & -0.693 $\pm$ 0.008 	 & 1.587 $\pm$ 0.011 	 & 0.588 $\pm$ 0.026 	 & 0.267 $\pm$ 0.005 	 & 0.013 $\pm$ 0.012 	 & 0.401 $\pm$ 0.007 \\ 
ESO114-007 & 0.361 $\pm$ 0.009 	 & -1.994 $\pm$ 0.289 	 & 3.453 $\pm$ 0.360 	 & 1.221 $\pm$ 0.167 	 & 0.626 $\pm$ 0.063 	 & 0.087 $\pm$ 0.038 	 & 0.663 $\pm$ 0.005 \\ 
ESO115-021 & 0.303 $\pm$ 0.003 	 & -1.673 $\pm$ 0.053 	 & 3.163 $\pm$ 0.090 	 & 1.955 $\pm$ 0.008 	 & 0.649 $\pm$ 0.021 	 & 0.528 $\pm$ 0.009 	 & 0.575 $\pm$ 0.002 \\ 
ESO116-012 & 0.408 $\pm$ 0.006 	 & -1.985 $\pm$ 0.078 	 & 0.410 $\pm$ 0.063 	 & 1.235 $\pm$ 0.018 	 & 0.252 $\pm$ 0.009 	 & 0.568 $\pm$ 0.007 	 & 0.476 $\pm$ 0.004 \\ 
ESO119-016 & 0.228 $\pm$ 0.005 	 & -1.356 $\pm$ 0.037 	 & 1.717 $\pm$ 0.052 	 & 1.258 $\pm$ 0.032 	 & 0.683 $\pm$ 0.015 	 & 0.025 $\pm$ 0.010 	 & 0.461 $\pm$ 0.003 \\ 
ESO120-012 & 0.252 $\pm$ 0.005 	 & -1.144 $\pm$ 0.048 	 & 1.906 $\pm$ 0.072 	 & 0.816 $\pm$ 0.038 	 & 0.414 $\pm$ 0.010 	 & 0.029 $\pm$ 0.009 	 & 0.421 $\pm$ 0.006 \\ 
ESO120-021 & 0.186 $\pm$ 0.006 	 & -1.099 $\pm$ 0.030 	 & 1.851 $\pm$ 0.036 	 & 0.840 $\pm$ 0.041 	 & 0.432 $\pm$ 0.011 	 & 0.025 $\pm$ 0.008 	 & 0.401 $\pm$ 0.005 \\ 
ESO145-025 & 0.195 $\pm$ 0.004 	 & -0.874 $\pm$ 0.013 	 & 1.674 $\pm$ 0.034 	 & 0.900 $\pm$ 0.013 	 & 0.427 $\pm$ 0.007 	 & 0.035 $\pm$ 0.008 	 & 0.412 $\pm$ 0.003 \\ 
ESO146-014 & 0.172 $\pm$ 0.008 	 & -0.796 $\pm$ 0.009 	 & 1.713 $\pm$ 0.028 	 & 1.195 $\pm$ 0.021 	 & 0.388 $\pm$ 0.010 	 & 0.304 $\pm$ 0.006 	 & 0.490 $\pm$ 0.005 \\ 
ESO149-001 & 0.349 $\pm$ 0.004 	 & -1.851 $\pm$ 0.094 	 & 2.955 $\pm$ 0.114 	 & 1.314 $\pm$ 0.025 	 & 0.703 $\pm$ 0.022 	 & 0.540 $\pm$ 0.013 	 & 0.505 $\pm$ 0.002 \\ 
ESO149-003 & 0.338 $\pm$ 0.014 	 & -1.618 $\pm$ 0.212 	 & 1.325 $\pm$ 0.251 	 & 0.885 $\pm$ 0.300 	 & 0.264 $\pm$ 0.061 	 & 0.222 $\pm$ 0.124 	 & 0.699 $\pm$ 0.018 \\ 
ESO150-005 & 0.240 $\pm$ 0.005 	 & -0.765 $\pm$ 0.071 	 & 2.038 $\pm$ 0.118 	 & 1.629 $\pm$ 0.032 	 & 0.791 $\pm$ 0.027 	 & 0.059 $\pm$ 0.018 	 & 0.598 $\pm$ 0.002 \\ 
ESO154-023 & 0.303 $\pm$ 0.004 	 & -1.472 $\pm$ 0.084 	 & 1.497 $\pm$ 0.074 	 & 1.931 $\pm$ 0.003 	 & 0.649 $\pm$ 0.034 	 & 0.294 $\pm$ 0.014 	 & 0.543 $\pm$ 0.002 \\ 
ESO157-049 & 0.758 $\pm$ 0.007 	 & -2.451 $\pm$ 0.145 	 & 4.224 $\pm$ 0.577 	 & 0.873 $\pm$ 0.291 	 & 0.183 $\pm$ 0.054 	 & 0.550 $\pm$ 0.036 	 & 0.525 $\pm$ 0.006 \\ 
ESO159-025 & 0.160 $\pm$ 0.004 	 & -1.003 $\pm$ 0.022 	 & 1.350 $\pm$ 0.029 	 & 0.587 $\pm$ 0.034 	 & 0.172 $\pm$ 0.003 	 & 0.051 $\pm$ 0.006 	 & 0.370 $\pm$ 0.004 \\ 
ESO187-035 & 0.167 $\pm$ 0.002 	 & -0.765 $\pm$ 0.009 	 & 1.595 $\pm$ 0.016 	 & 0.510 $\pm$ 0.039 	 & 0.231 $\pm$ 0.003 	 & 0.013 $\pm$ 0.003 	 & 0.399 $\pm$ 0.002 \\ 
ESO187-051 & 0.166 $\pm$ 0.002 	 & -0.916 $\pm$ 0.011 	 & 1.660 $\pm$ 0.016 	 & 0.463 $\pm$ 0.024 	 & 0.198 $\pm$ 0.001 	 & 0.034 $\pm$ 0.002 	 & 0.376 $\pm$ 0.001 \\ 
ESO202-035 & 0.730 $\pm$ 0.005 	 & -2.036 $\pm$ 0.183 	 & 4.413 $\pm$ 0.417 	 & 1.392 $\pm$ 0.073 	 & 0.261 $\pm$ 0.054 	 & 0.275 $\pm$ 0.041 	 & 0.529 $\pm$ 0.004 \\ 
ESO202-041 & 0.242 $\pm$ 0.007 	 & -2.099 $\pm$ 0.211 	 & 1.598 $\pm$ 0.128 	 & 1.423 $\pm$ 0.091 	 & 0.906 $\pm$ 0.060 	 & 0.018 $\pm$ 0.029 	 & 0.561 $\pm$ 0.004 \\ 
ESO234-043 & 0.209 $\pm$ 0.004 	 & -1.079 $\pm$ 0.018 	 & 1.463 $\pm$ 0.028 	 & 0.632 $\pm$ 0.057 	 & 0.255 $\pm$ 0.005 	 & 0.031 $\pm$ 0.004 	 & 0.387 $\pm$ 0.003 \\ 
ESO234-049 & 0.289 $\pm$ 0.013 	 & -1.927 $\pm$ 0.050 	 & 2.681 $\pm$ 0.077 	 & 0.565 $\pm$ 0.058 	 & 0.169 $\pm$ 0.013 	 & 0.014 $\pm$ 0.013 	 & 0.371 $\pm$ 0.009 \\ 
ESO236-039 & 0.194 $\pm$ 0.006 	 & -0.963 $\pm$ 0.018 	 & 1.884 $\pm$ 0.035 	 & 0.593 $\pm$ 0.053 	 & 0.299 $\pm$ 0.005 	 & 0.032 $\pm$ 0.003 	 & 0.393 $\pm$ 0.003 \\ 
ESO237-049 & 0.223 $\pm$ 0.005 	 & -0.690 $\pm$ 0.015 	 & 2.037 $\pm$ 0.034 	 & 0.795 $\pm$ 0.040 	 & 0.390 $\pm$ 0.006 	 & 0.282 $\pm$ 0.003 	 & 0.500 $\pm$ 0.004 \\ 

\dots	& \dots  	 & \dots  	 & \dots 	 & \dots  	 & \dots  	 & \dots 	 & \dots  \\ 
\hline
\end{tabular}
\end{center}
\label{t:ch1}
\end{table*}

\begin{table*}
\caption{The morphological parameters at 4.5 $\mu$m for the \nsamp \ $S^4G$ galaxies.  Full table is available in the {\em electronic edition}.}
\begin{center}
\begin{tabular}{llllllllllll}
Name	& Gini & $M_{20}$  & C82 &  A   & S & E & $G_M$  \\ 
\hline
\hline
ESO011-005 & 0.468 $\pm$ 0.018 	 & -2.232 $\pm$ 1.343 	 & 0.000 $\pm$ 0.780 	 & 0.626 $\pm$ 0.299 	 & 0.125 $\pm$ 0.077 	 & 0.627 $\pm$ 0.036 	 & 0.877 $\pm$ 0.008 \\ 
ESO012-010 & 0.339 $\pm$ 0.014 	 & -1.918 $\pm$ 0.169 	 & 0.000 $\pm$ 0.307 	 & 0.734 $\pm$ 0.273 	 & 0.337 $\pm$ 0.057 	 & 0.471 $\pm$ 0.023 	 & 0.563 $\pm$ 0.011 \\ 
ESO012-014 & 0.189 $\pm$ 0.012 	 & -0.420 $\pm$ 0.162 	 & 0.000 $\pm$ 0.231 	 & 1.584 $\pm$ 0.089 	 & 0.923 $\pm$ 0.021 	 & 0.117 $\pm$ 0.145 	 & 0.792 $\pm$ 0.005 \\ 
ESO013-016 & 0.450 $\pm$ 0.010 	 & -0.899 $\pm$ 0.128 	 & 0.000 $\pm$ 0.372 	 & 0.609 $\pm$ 0.189 	 & 0.184 $\pm$ 0.081 	 & 0.104 $\pm$ 0.049 	 & 0.799 $\pm$ 0.010 \\ 
ESO015-001 & 0.317 $\pm$ 0.020 	 & -1.830 $\pm$ 0.388 	 & 0.000 $\pm$ 0.791 	 & 0.652 $\pm$ 0.431 	 & 0.260 $\pm$ 0.109 	 & 0.395 $\pm$ 0.231 	 & 0.680 $\pm$ 0.016 \\ 
ESO026-001 & 0.439 $\pm$ 0.019 	 & -1.989 $\pm$ 0.180 	 & 3.542 $\pm$ 0.241 	 & 1.011 $\pm$ 0.104 	 & 0.351 $\pm$ 0.083 	 & 0.071 $\pm$ 0.039 	 & 0.584 $\pm$ 0.011 \\ 
ESO027-001 & 0.580 $\pm$ 0.012 	 & -2.458 $\pm$ 0.212 	 & 0.000 $\pm$ 0.467 	 & 1.939 $\pm$ 0.039 	 & 0.095 $\pm$ 0.049 	 & 0.178 $\pm$ 0.050 	 & 0.431 $\pm$ 0.010 \\ 
ESO027-008 & 0.528 $\pm$ 0.015 	 & -0.618 $\pm$ 0.202 	 & 0.000 $\pm$ 0.379 	 & 2.000 $\pm$ 0.001 	 & 0.066 $\pm$ 0.091 	 & 0.490 $\pm$ 0.053 	 & 0.660 $\pm$ 0.015 \\ 
ESO048-017 & 0.183 $\pm$ 0.003 	 & -0.833 $\pm$ 0.021 	 & 1.702 $\pm$ 0.037 	 & 0.774 $\pm$ 0.027 	 & 0.363 $\pm$ 0.004 	 & 0.036 $\pm$ 0.002 	 & 0.410 $\pm$ 0.002 \\ 
ESO054-021 & 0.420 $\pm$ 0.010 	 & -1.119 $\pm$ 0.137 	 & 3.085 $\pm$ 0.188 	 & 0.880 $\pm$ 0.076 	 & 0.349 $\pm$ 0.041 	 & 0.199 $\pm$ 0.050 	 & 0.612 $\pm$ 0.007 \\ 
ESO079-003 & 0.813 $\pm$ 0.004 	 & -1.763 $\pm$ 0.638 	 & 3.509 $\pm$ 0.613 	 & 2.000 $\pm$ 0.000 	 & 0.135 $\pm$ 0.039 	 & 0.079 $\pm$ 0.078 	 & 0.676 $\pm$ 0.005 \\ 
ESO079-005 & 0.448 $\pm$ 0.014 	 & -0.818 $\pm$ 0.250 	 & 0.000 $\pm$ 0.333 	 & 0.920 $\pm$ 0.254 	 & 0.322 $\pm$ 0.096 	 & 0.185 $\pm$ 0.090 	 & 0.793 $\pm$ 0.003 \\ 
ESO079-007 & 0.452 $\pm$ 0.018 	 & -1.649 $\pm$ 0.104 	 & 0.000 $\pm$ 0.306 	 & 0.600 $\pm$ 0.180 	 & 0.154 $\pm$ 0.058 	 & 0.212 $\pm$ 0.037 	 & 0.618 $\pm$ 0.012 \\ 
ESO085-014 & 0.265 $\pm$ 0.005 	 & -1.312 $\pm$ 0.073 	 & 1.109 $\pm$ 0.061 	 & 1.456 $\pm$ 0.015 	 & 0.536 $\pm$ 0.012 	 & 0.376 $\pm$ 0.006 	 & 0.513 $\pm$ 0.004 \\ 
ESO085-030 & 0.582 $\pm$ 0.009 	 & -3.001 $\pm$ 0.206 	 & 4.931 $\pm$ 0.494 	 & 0.717 $\pm$ 0.272 	 & 0.422 $\pm$ 0.078 	 & 0.076 $\pm$ 0.031 	 & 0.479 $\pm$ 0.007 \\ 
ESO085-047 & 0.172 $\pm$ 0.007 	 & -1.057 $\pm$ 0.019 	 & 1.784 $\pm$ 0.028 	 & 1.051 $\pm$ 0.047 	 & 0.622 $\pm$ 0.012 	 & 0.031 $\pm$ 0.006 	 & 0.421 $\pm$ 0.005 \\ 
ESO107-016 & 0.125 $\pm$ 0.007 	 & -0.690 $\pm$ 0.005 	 & 1.551 $\pm$ 0.006 	 & 0.409 $\pm$ 0.021 	 & 0.154 $\pm$ 0.005 	 & 0.016 $\pm$ 0.010 	 & 0.367 $\pm$ 0.006 \\ 
ESO114-007 & 0.393 $\pm$ 0.014 	 & -1.595 $\pm$ 0.678 	 & 0.000 $\pm$ 0.564 	 & 0.882 $\pm$ 0.248 	 & 0.342 $\pm$ 0.133 	 & 0.309 $\pm$ 0.089 	 & 0.873 $\pm$ 0.005 \\ 
ESO115-021 & 0.418 $\pm$ 0.023 	 & -0.809 $\pm$ 0.244 	 & 3.485 $\pm$ 0.243 	 & 2.000 $\pm$ 0.000 	 & 0.794 $\pm$ 0.046 	 & 0.488 $\pm$ 0.015 	 & 0.700 $\pm$ 0.002 \\ 
ESO116-012 & 0.395 $\pm$ 0.004 	 & -1.413 $\pm$ 0.118 	 & 1.179 $\pm$ 0.073 	 & 1.426 $\pm$ 0.021 	 & 0.459 $\pm$ 0.015 	 & 0.072 $\pm$ 0.008 	 & 0.491 $\pm$ 0.002 \\ 
ESO119-016 & 0.165 $\pm$ 0.006 	 & -0.818 $\pm$ 0.005 	 & 1.600 $\pm$ 0.016 	 & 0.626 $\pm$ 0.019 	 & 0.241 $\pm$ 0.006 	 & 0.046 $\pm$ 0.004 	 & 0.380 $\pm$ 0.005 \\ 
ESO120-012 & 0.165 $\pm$ 0.003 	 & -0.857 $\pm$ 0.019 	 & 1.700 $\pm$ 0.011 	 & 0.442 $\pm$ 0.026 	 & 0.169 $\pm$ 0.004 	 & 0.012 $\pm$ 0.007 	 & 0.373 $\pm$ 0.005 \\ 
ESO120-021 & 0.164 $\pm$ 0.004 	 & -0.926 $\pm$ 0.022 	 & 1.757 $\pm$ 0.024 	 & 0.576 $\pm$ 0.070 	 & 0.252 $\pm$ 0.004 	 & 0.041 $\pm$ 0.004 	 & 0.389 $\pm$ 0.003 \\ 
ESO145-025 & 0.167 $\pm$ 0.006 	 & -0.792 $\pm$ 0.007 	 & 1.614 $\pm$ 0.016 	 & 0.641 $\pm$ 0.028 	 & 0.239 $\pm$ 0.006 	 & 0.026 $\pm$ 0.010 	 & 0.388 $\pm$ 0.005 \\ 
ESO146-014 & 0.111 $\pm$ 0.011 	 & -0.674 $\pm$ 0.005 	 & 1.621 $\pm$ 0.017 	 & 0.753 $\pm$ 0.028 	 & 0.156 $\pm$ 0.014 	 & 0.313 $\pm$ 0.010 	 & 0.481 $\pm$ 0.006 \\ 
ESO149-001 & 0.221 $\pm$ 0.007 	 & -0.789 $\pm$ 0.015 	 & 2.457 $\pm$ 0.025 	 & 0.782 $\pm$ 0.028 	 & 0.285 $\pm$ 0.009 	 & 0.555 $\pm$ 0.005 	 & 0.526 $\pm$ 0.005 \\ 
ESO149-003 & 0.280 $\pm$ 0.009 	 & -2.113 $\pm$ 0.241 	 & 1.303 $\pm$ 0.402 	 & 1.151 $\pm$ 0.303 	 & 0.565 $\pm$ 0.075 	 & 0.067 $\pm$ 0.144 	 & 0.770 $\pm$ 0.009 \\ 
ESO150-005 & 0.232 $\pm$ 0.007 	 & -0.715 $\pm$ 0.097 	 & 2.247 $\pm$ 0.153 	 & 1.629 $\pm$ 0.058 	 & 0.950 $\pm$ 0.067 	 & 0.165 $\pm$ 0.021 	 & 0.625 $\pm$ 0.003 \\ 
ESO154-023 & 0.204 $\pm$ 0.002 	 & -1.227 $\pm$ 0.063 	 & 1.370 $\pm$ 0.062 	 & 1.941 $\pm$ 0.002 	 & 0.908 $\pm$ 0.018 	 & 0.259 $\pm$ 0.009 	 & 0.468 $\pm$ 0.001 \\ 
ESO157-049 & 0.759 $\pm$ 0.005 	 & -2.081 $\pm$ 0.576 	 & 0.000 $\pm$ 0.668 	 & 0.738 $\pm$ 0.239 	 & 0.154 $\pm$ 0.088 	 & 0.490 $\pm$ 0.079 	 & 0.626 $\pm$ 0.005 \\ 
ESO159-025 & 0.102 $\pm$ 0.004 	 & -0.919 $\pm$ 0.012 	 & 1.451 $\pm$ 0.019 	 & 0.474 $\pm$ 0.055 	 & 0.128 $\pm$ 0.003 	 & 0.054 $\pm$ 0.004 	 & 0.360 $\pm$ 0.004 \\ 
ESO187-035 & 0.095 $\pm$ 0.002 	 & -0.746 $\pm$ 0.004 	 & 1.542 $\pm$ 0.004 	 & 0.326 $\pm$ 0.025 	 & 0.127 $\pm$ 0.003 	 & 0.033 $\pm$ 0.001 	 & 0.370 $\pm$ 0.002 \\ 
ESO187-051 & 0.094 $\pm$ 0.002 	 & -0.807 $\pm$ 0.005 	 & 1.578 $\pm$ 0.008 	 & 0.310 $\pm$ 0.037 	 & 0.120 $\pm$ 0.002 	 & 0.018 $\pm$ 0.002 	 & 0.365 $\pm$ 0.001 \\ 
ESO202-035 & 0.304 $\pm$ 0.010 	 & -1.591 $\pm$ 0.060 	 & 1.880 $\pm$ 0.071 	 & 1.107 $\pm$ 0.035 	 & 0.231 $\pm$ 0.014 	 & 0.243 $\pm$ 0.005 	 & 0.473 $\pm$ 0.009 \\ 
ESO202-041 & 0.161 $\pm$ 0.002 	 & -0.783 $\pm$ 0.005 	 & 1.531 $\pm$ 0.010 	 & 0.544 $\pm$ 0.051 	 & 0.259 $\pm$ 0.002 	 & 0.007 $\pm$ 0.002 	 & 0.399 $\pm$ 0.001 \\ 
ESO234-043 & 0.147 $\pm$ 0.002 	 & -0.895 $\pm$ 0.006 	 & 1.514 $\pm$ 0.016 	 & 0.471 $\pm$ 0.022 	 & 0.160 $\pm$ 0.003 	 & 0.028 $\pm$ 0.003 	 & 0.372 $\pm$ 0.002 \\ 
ESO234-049 & 0.228 $\pm$ 0.008 	 & -1.310 $\pm$ 0.026 	 & 2.571 $\pm$ 0.045 	 & 0.561 $\pm$ 0.028 	 & 0.163 $\pm$ 0.008 	 & 0.018 $\pm$ 0.012 	 & 0.367 $\pm$ 0.008 \\ 
ESO236-039 & 0.164 $\pm$ 0.003 	 & -0.854 $\pm$ 0.013 	 & 1.700 $\pm$ 0.022 	 & 0.459 $\pm$ 0.064 	 & 0.214 $\pm$ 0.003 	 & 0.020 $\pm$ 0.004 	 & 0.385 $\pm$ 0.002 \\ 
ESO237-049 & 0.104 $\pm$ 0.009 	 & -0.643 $\pm$ 0.004 	 & 1.794 $\pm$ 0.026 	 & 0.400 $\pm$ 0.036 	 & 0.136 $\pm$ 0.011 	 & 0.310 $\pm$ 0.006 	 & 0.451 $\pm$ 0.006 \\ 
\dots	& \dots  	 & \dots  	 & \dots 	 & \dots  	 & \dots  	 & \dots 	 & \dots  \\ 

\hline
\end{tabular}
\end{center}
\label{t:ch2}
\end{table*}
\clearpage

\begin{figure*}
\begin{center}
\section{Systematics}
    \begin{minipage}{0.49\linewidth}
	\includegraphics[width=\textwidth]{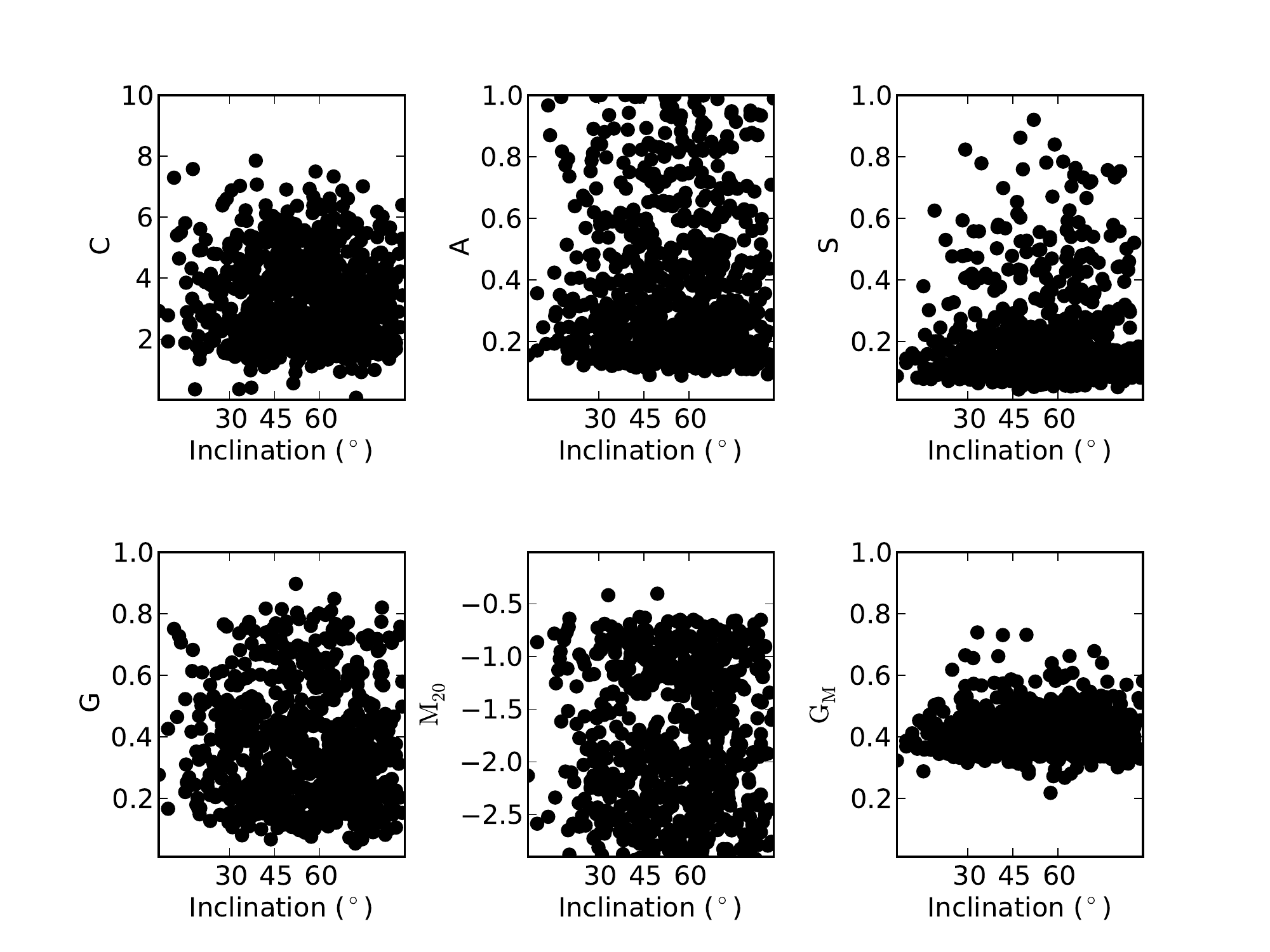}
	\caption{\label{f:incl:6par}The six morphology parameters as a function of inclination. None of the morphological parameters relate closely to the inclination, not even Concentration.}
     \end{minipage}\hfill
    \begin{minipage}{0.49\linewidth}
	\includegraphics[width=\textwidth]{./holwerda_f15a.pdf}
	\caption{\label{f:dist:6par}The six morphology parameters as a function of distance. None of the morphological parameters relate closely to the distance, not even Smoothness.}
    \end{minipage}
\end{center}
\end{figure*}

Two possible systematics are explored here in this section: any relation between inclination and the morphological parameters and the effect of smoothing due to distance on the morphological parameters. Figure \ref{f:incl:6par} show the lack of a relation between the inclination and any of the morphological parameters. One could expect a relation between disk inclination and concentration due to line-of-sight integration. However, no such relation in \s4g exists. 

Similarly, one can expect a relation between Smoothness and distance. Galaxies further away appear smoother, which is the reasoning behind the surface brightness fluctuation distance measurement method. However, there is no such relation between the distances of the \s4g galaxies and their smoothness evident in Figure \ref{f:dist:6par}. 

The Spearman rankings of the relations between inclination or distance and the \s4g morphology parameters bear out the lack of a relation (Table \ref{t:unc}), albeit at low z-value confidences.

\begin{table}
\caption{The Spearman ranking of the relation between inclination or distance and the morphological parameters in either 3.6 \mum\ or 4.5 \mum\ images for our full sample. The absolute z-values of significance for each of Spearman rankings are noted between parentheses.}
\begin{center}
\begin{tabular}{l l l l l l l l}
Type	   & C	 & A & S & $M_{20}$ & G & $G_M$ \\
\hline
\hline
3.6 $\mu$m			& & & & & & \\
\hline
Inclination ($^\circ$)   & -0.07 (1.84)         & -0.01 (0.29)         & 0.02 (0.44)   & 0.04 (1.12)   & -0.07 (1.86)         & -0.02 (0.58) \\ 
Distance (Mpc.)          & -0.01 (0.40)         & -0.02 (0.63)         & -0.01 (0.46)         & -0.01 (0.28)         & -0.01 (0.40)         & 0.00 (0.04) \\ 
\hline
\hline
4.5 $\mu$m			& & & & & & \\
\hline
Inclination ($^\circ$)   & -0.06 (1.52)         & -0.03 (0.78)         & 0.02 (0.54)   & 0.03 (0.92)   & -0.05 (1.31)         & -0.00 (0.13) \\ 
Distance (Mpc.)          & -0.01 (0.49)         & -0.02 (0.52)         & -0.04 (1.31)         & 0.02 (0.54)   & -0.03 (1.07)         & -0.03 (0.88) \\ 
\hline
\hline
\end{tabular}
\end{center}
\label{t:unc}
\end{table}%

\newpage

\begin{figure*}
\begin{center}
\section{Lopsidedness}
     \begin{minipage}{0.6\linewidth}
	\includegraphics[width=\textwidth]{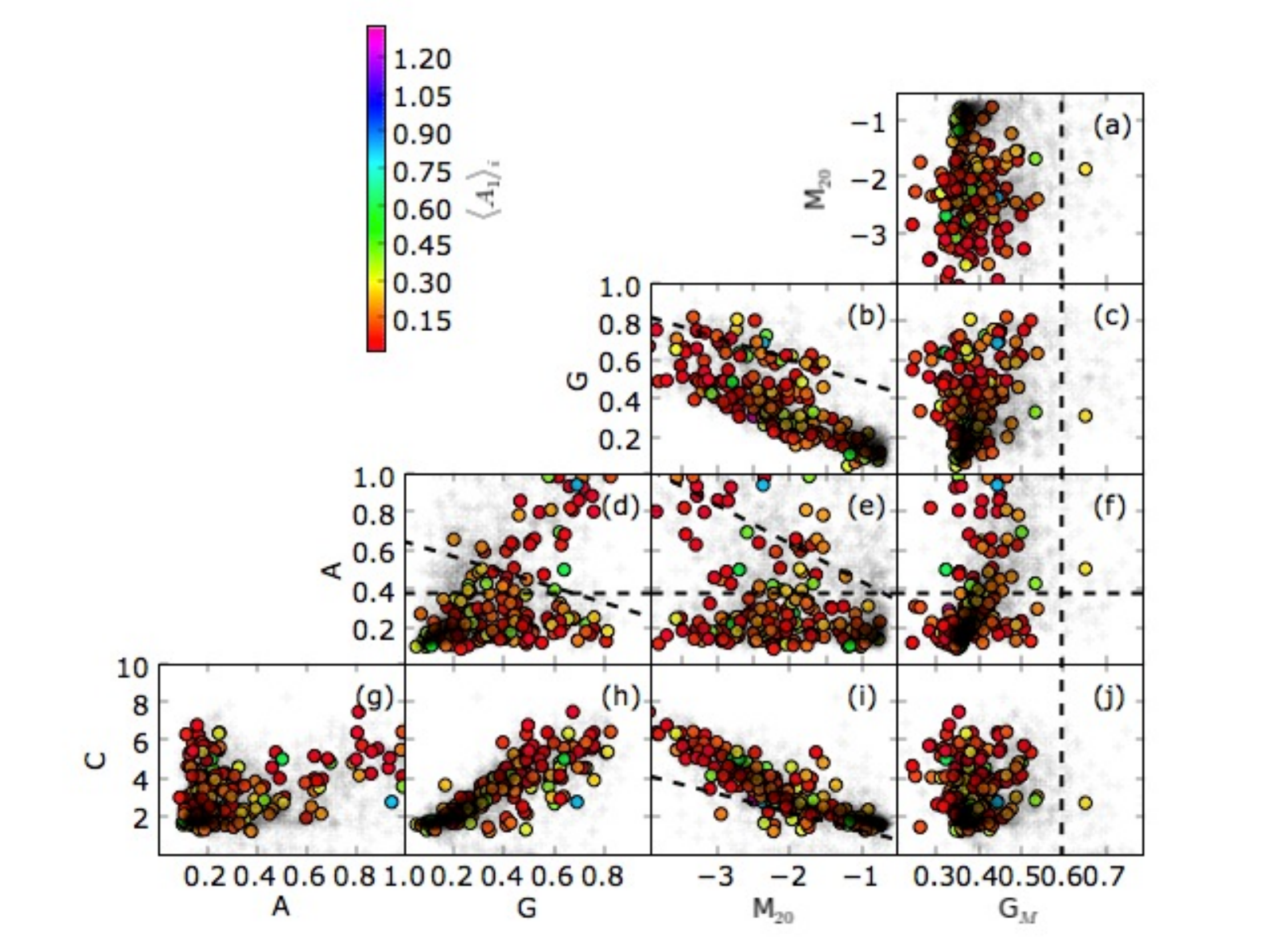}
    \end{minipage}\hfill
    \begin{minipage}{0.39\linewidth}
	\includegraphics[width=1.1\textwidth]{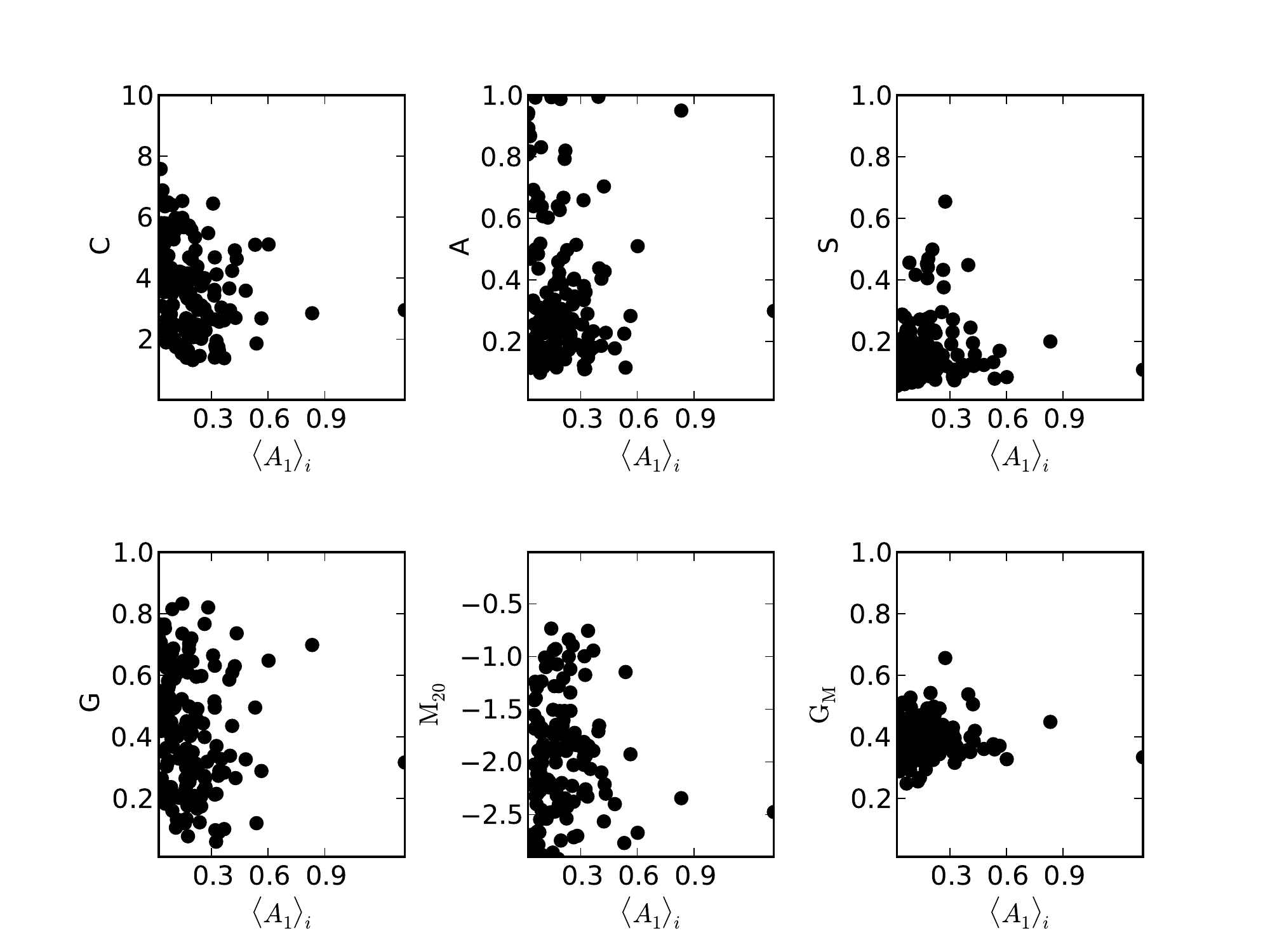}
    \end{minipage}
    \begin{minipage}{0.6\linewidth}
	\caption{\label{f:A12} The distribution of the 3.6 \mum\ morphological parameters color-coded with the $<A_{1}>_{i}$ parameter from \cite{Zaritsky13b}) where available. The full \s4g morphological sample is marked with gray crosses for reference.
Dashed lines are the merger/interaction criteria: 
Sub-panels (a), (c), (f), and (j), the \gm\ criterion (equation \ref{eq:GM})
Sub-panel (b), the G-\m20 criterion from \protect\cite{Lotz04} (equation \ref{eq:GM20},
sub-panel (d), the G-A criterion from \protect\cite{Lotz04} (equation \ref{eq:GA}), 
and sub-panels (d),(e) and (f), the horizontal line is the A$>0.38$ criterion from \protect\cite{CAS} (equation \ref{eq:AS}).
}
    \end{minipage}\hfill
    \begin{minipage}{0.39\linewidth}
	\caption{\label{f:A12:6par} The direct relation between the {\em inner} (computed between 1.5 and 2.5 scale-lengths) m=1  mode from \cite{Zaritsky13b}) and the six morphology parameters.}
    \end{minipage}
\end{center}
\end{figure*}

\begin{figure*}
\begin{center}
     \begin{minipage}{0.6\linewidth}
	\includegraphics[width=\textwidth]{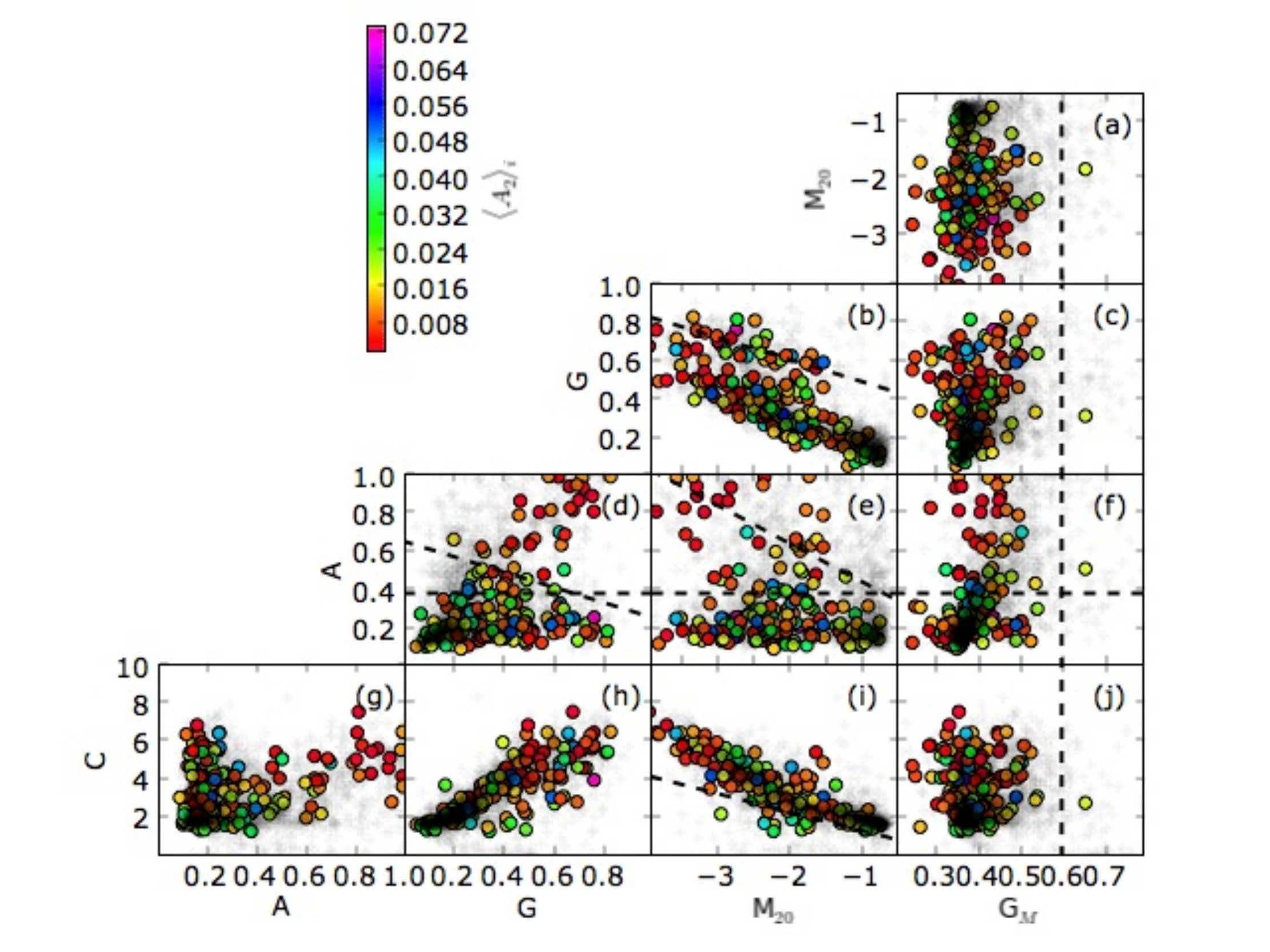}
    \end{minipage}\hfill
    \begin{minipage}{0.39\linewidth}
	\includegraphics[width=1.1\textwidth]{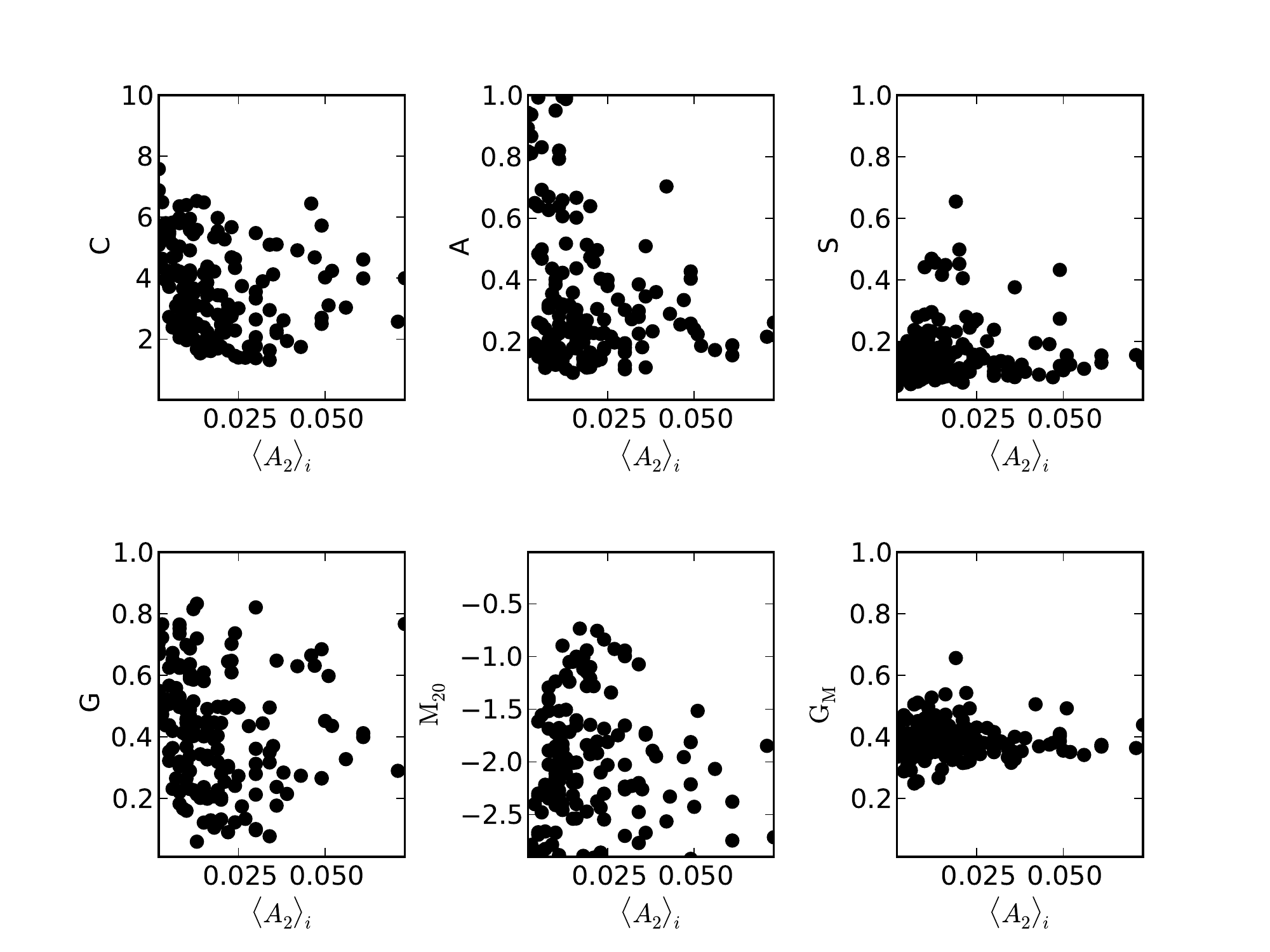}
    \end{minipage}
     \begin{minipage}{0.6\linewidth}
	\caption{\label{f:A22:6par} The direct relation between the {\em inner} (computed between 1.5 and 2.5 scale-lengths) m=2 mode from \cite{Zaritsky13b}) and the six morphology parameters.}
    \end{minipage}     
    \begin{minipage}{0.39\linewidth}
	\caption{\label{f:A22} The distribution of the 3.6 \mum\ morphological parameters color-coded with the $<A_{2}>_{i}$ parameter from \cite{Zaritsky13b}) where available. The full \s4g morphological sample is marked with gray crosses for reference. Dashed lines as in Figure \ref{f:A12}.}
    \end{minipage}
\end{center}
\end{figure*}

\begin{figure*}
\begin{center}
     \begin{minipage}{0.6\linewidth}
	\includegraphics[width=\textwidth]{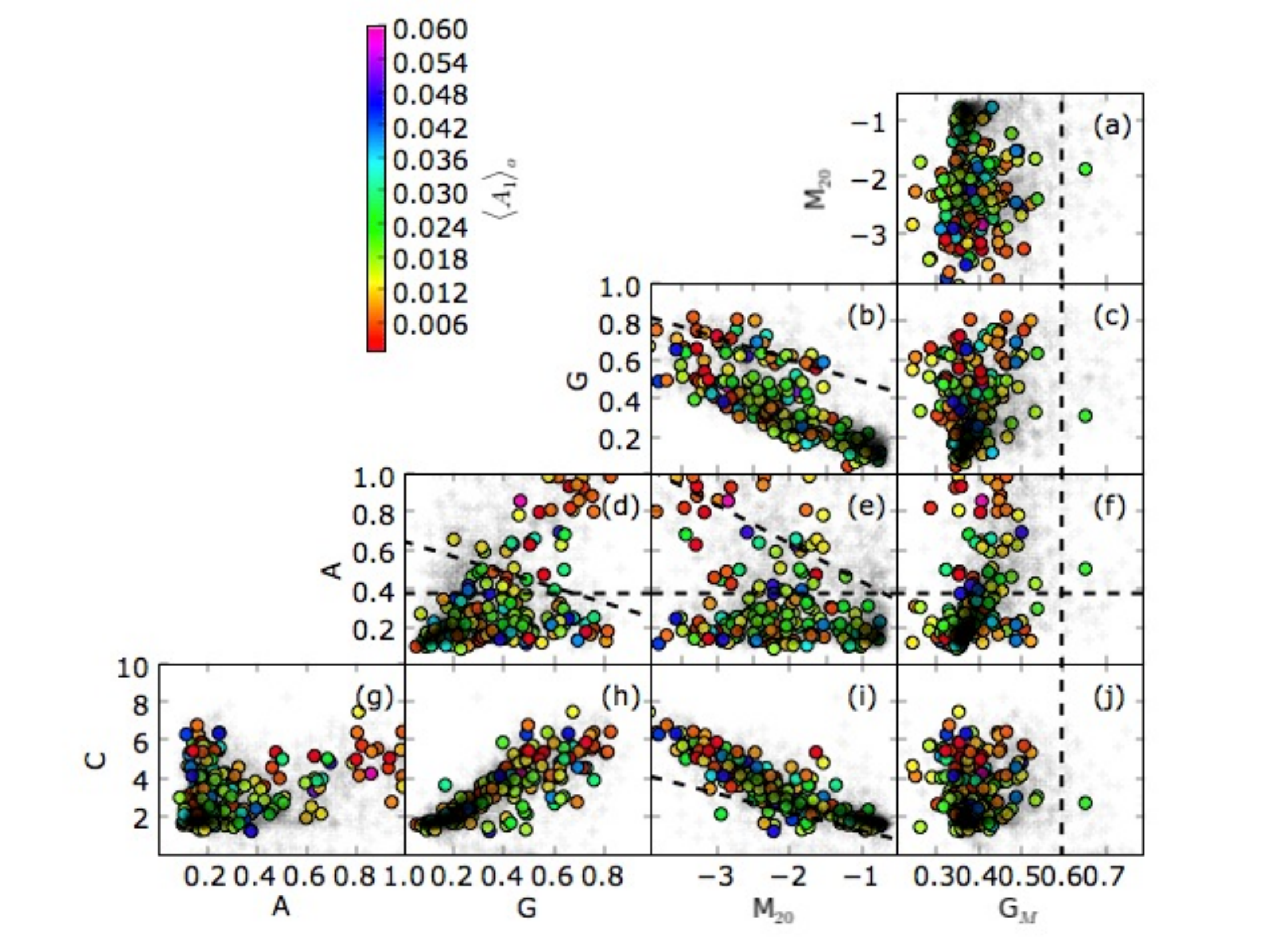}
    \end{minipage}\hfill
    \begin{minipage}{0.39\linewidth}
	\includegraphics[width=1.1\textwidth]{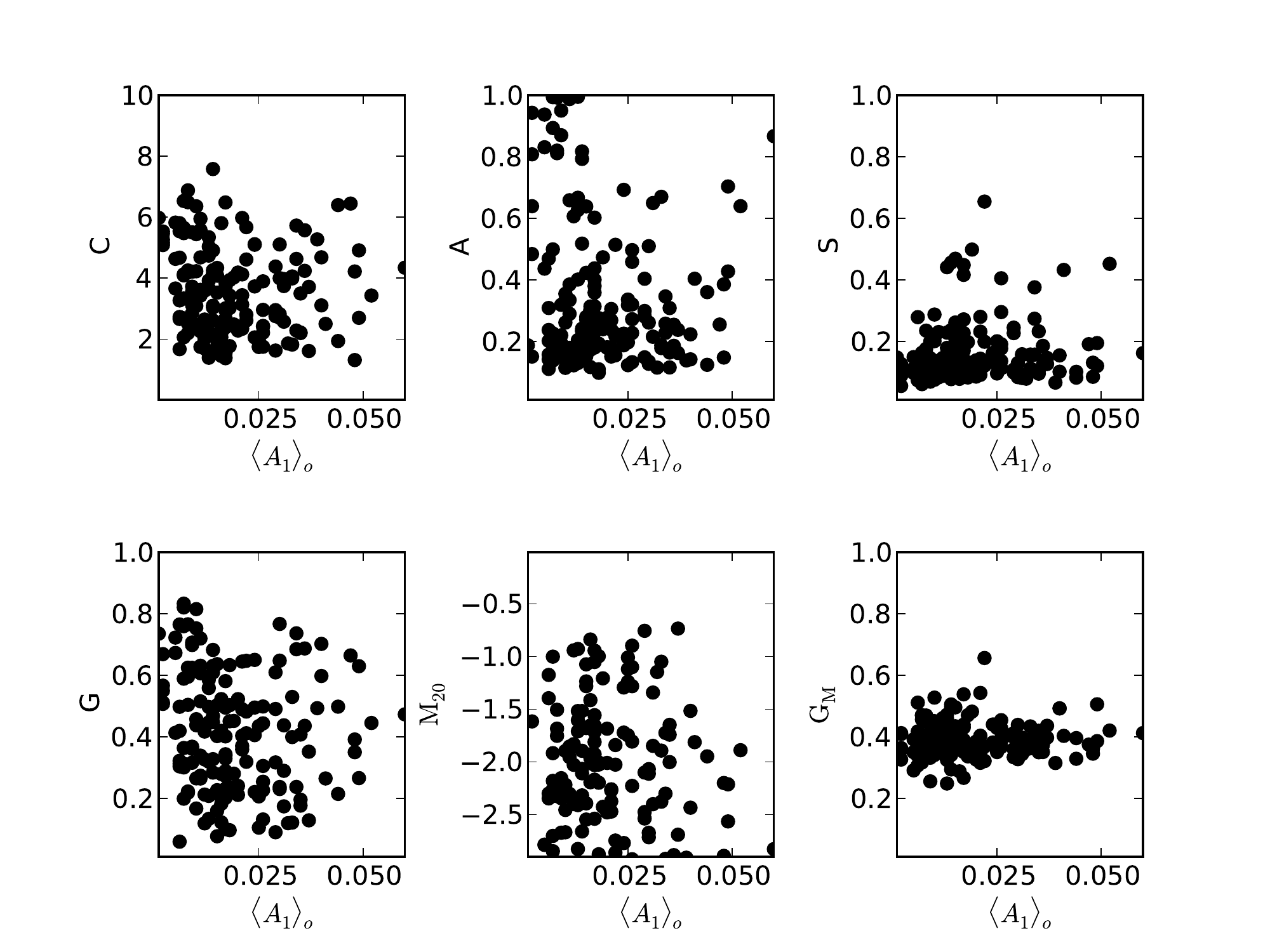}
    \end{minipage}
     \begin{minipage}{0.6\linewidth}
	\caption{\label{f:A13} The distribution of the 3.6 \mum\ morphological parameters color-coded with the $<A_{1}>_{o}$ parameter from \cite{Zaritsky13b}) where available. The full \s4g morphological sample is marked with gray crosses for reference. Dashed lines as in Figure \ref{f:A12}.}
    \end{minipage}
    \begin{minipage}{0.39\linewidth}
	\caption{\label{f:A13:6par} The direct relation between the {\em outer} (computed between 2.5 and 3.5 scale-lengths) m=1  mode from \cite{Zaritsky13b}) and the six morphology parameters.}
    \end{minipage}
\end{center}
\end{figure*}

\begin{figure*}
\begin{center}
     \begin{minipage}{0.6\linewidth}
	\includegraphics[width=\textwidth]{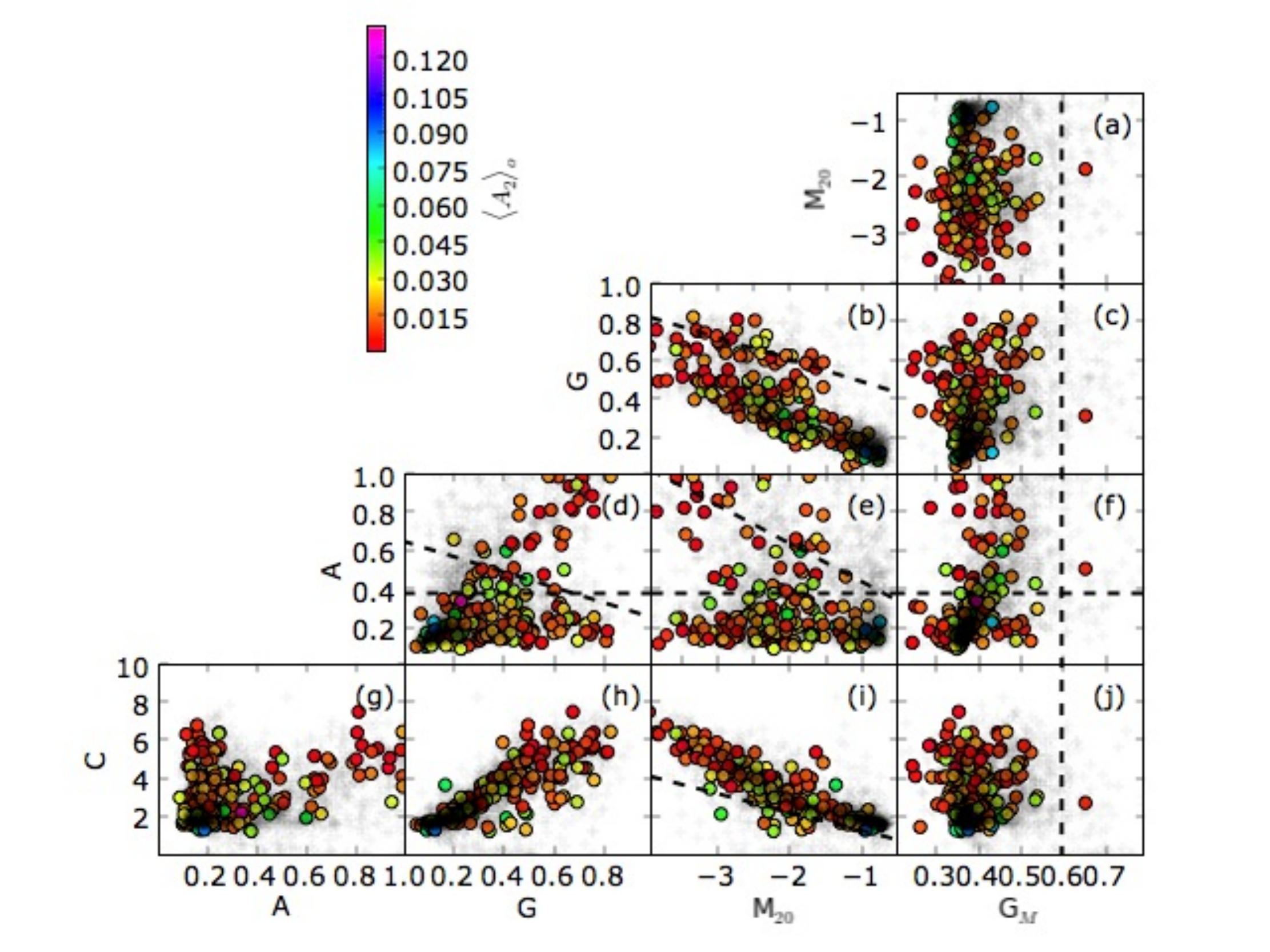}
    \end{minipage}\hfill
    \begin{minipage}{0.39\linewidth}
	\includegraphics[width=1.1\textwidth]{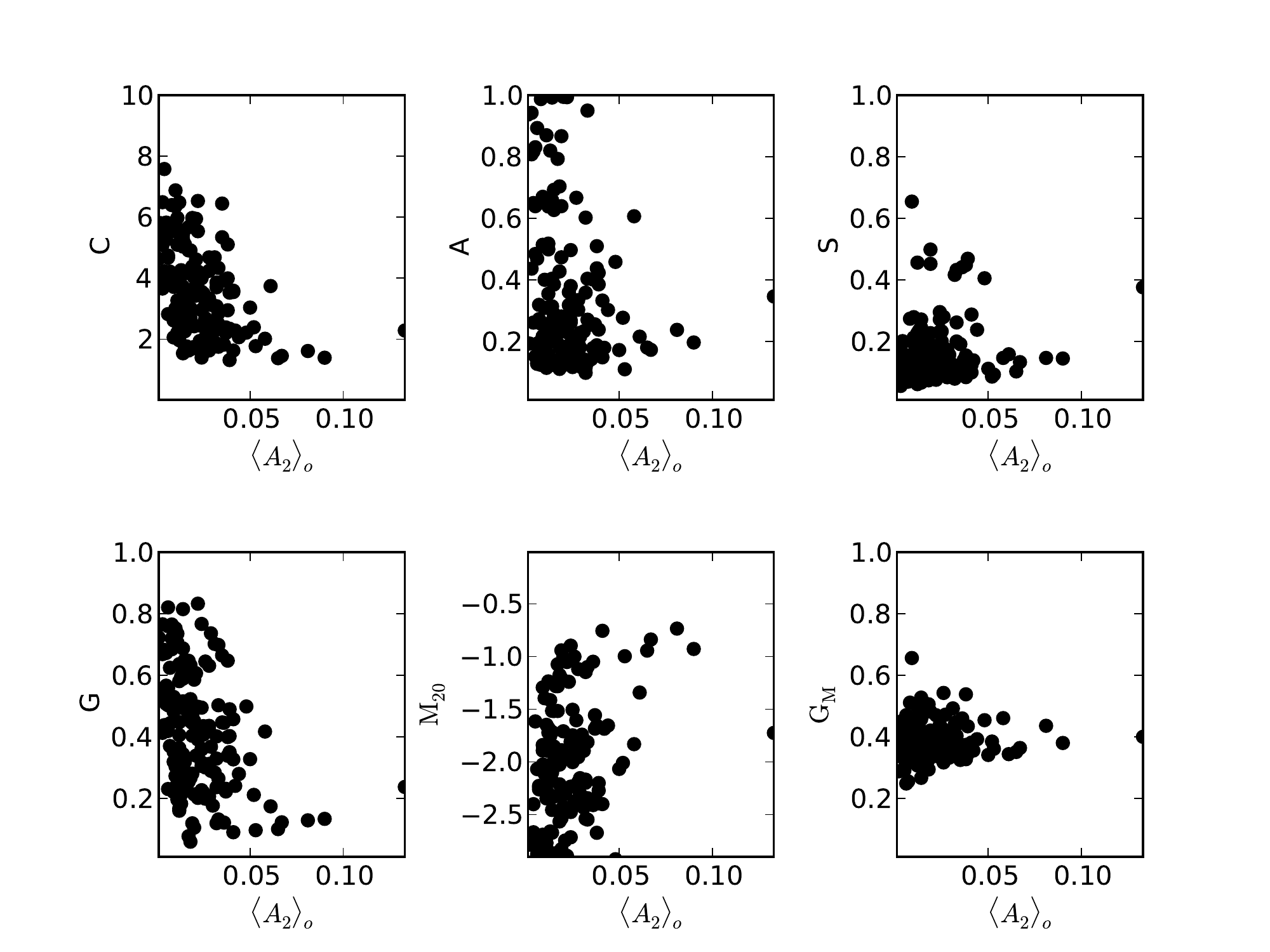}
    \end{minipage}
    \begin{minipage}{0.6\linewidth}
	\caption{\label{f:A23} The distribution of the 3.6 \mum\ morphological parameters color-coded with the $<A_{2}>_{o}$ parameter from \cite{Zaritsky13b}) where available. The full \s4g\ morphological sample is marked with gray crosses for reference. Dashed lines as in Figure \ref{f:A12}.}
    \end{minipage}
    \begin{minipage}{0.39\linewidth}
	\caption{\label{f:A23:6par} The direct relation between the {\em outer} (computed between 2.5 and 3.5 scale-lengths) m=2 mode from \cite{Zaritsky13b}) and the six morphology parameters.}
    \end{minipage}
\end{center}
\end{figure*}

\end{document}